\documentclass[11pt,secnumarabic,aps,amsmath,amssymb,amsfonts,superscriptaddress,tightenlines,nobibnotes,prd,nofootinbib,twocolumn,dvipsnames]{revtex4}

\usepackage{bm}
\usepackage[utf8]{inputenc}
\usepackage{amsmath}
\usepackage{amssymb}
\usepackage{pgfplots}

\usepackage{wasysym}
\usepackage{subeqnarray}
\usepackage{graphicx}
\usepackage{xcolor}
\usepackage{yfonts}
\usepackage[normalem]{ulem}

\usepackage{epsfig}
\usepackage[hidelinks,colorlinks=true,citecolor=Blue,linkcolor=Blue,urlcolor=BlueViolet]{hyperref}

\numberwithin{equation}{section}

\newcommand{\sumint}{\int}

\newcommand\spart{\;\raise1.0pt\hbox{/}\hskip-6pt\partial}
\newcommand\spartb{\;\overline{\raise1.0pt\hbox{/}\hskip-6pt\partial}}

\newcommand{\be}{\begin{equation}}
\newcommand{\ee}{\end{equation}}
\newcommand{\bea}{\begin{eqnarray}}
\newcommand{\eea}{\end{eqnarray}}
\newcommand{\beal}{\begin{align}}
\newcommand{\eeal}{\end{align}}
\newcommand{\beas}{\begin{subeqnarray}}
\newcommand{\eeas}{\end{subeqnarray}}
\renewcommand\[{\begin{equation}}
\renewcommand\]{\end{equation}}
\newcommand{\dd}{{\rm d}}
\newcommand{\ii}{{\rm i}}

\newcommand{\gr}[1]{\boldsymbol{#1}}
\newcommand{\ellc}{\ell_c}

\newcommand\cpsilent{}

\newcommand\gsin{\,\mathfrak{sin}}
\newcommand\gtan{\,\mathfrak{tan}}
\newcommand\gcot{\,\mathfrak{cot}}

\newcommand{\barK}{{\cal K}}
\newcommand{\myzeta}{{\zeta}}
\newcommand{\mygamma}{{\xi}}

\newcommand{\com}[1]{{}}

\begin{document}

\title{Beyond scalar, vector and tensor harmonics in maximally symmetric three-dimensional spaces}

\author{Cyril Pitrou}
\email[]{pitrou@iap.fr}
\affiliation{Institut d'Astrophysique de Paris, CNRS UMR 7095, 98 bis Bd Arago, 75014 Paris, France.}

\author{Thiago S. Pereira}
\email[]{tspereira@uel.br}
\affiliation{Departamento de Física, Universidade Estadual de Londrina, Rod. Celso Garcia Cid, Km 380, 86057-970, Londrina, Paraná, Brazil.}

\date{\today}
\begin{abstract}
We present a comprehensive construction of scalar, vector and tensor
harmonics on maximally symmetric three-dimensional spaces. Our
formalism relies on the introduction of spin-weighted spherical
harmonics and a generalized helicity basis which, together, are ideal tools to decompose harmonics into their radial and angular dependencies. We provide a thorough and self-contained set of expressions and relations for these harmonics. Being general, our formalism also allows to build harmonics of higher tensor type by recursion among radial functions, and we collect the complete set of recursive relations which can be used. While the formalism is readily adapted to computation of CMB transfer functions, we also collect explicit forms of the radial harmonics which are needed for other cosmological observables. Finally, we show that in curved spaces, normal modes cannot be factorized into a local angular dependence and a unit norm function encoding the orbital dependence of the harmonics, contrary to previous statements in the literature.  
\end{abstract}
\maketitle


\section{Introduction}
Tensor harmonics are ubiquitous tools in gravitational theories. Their applicability reach a wide spectrum of topics including black-hole physics, gravitational waves, quantum-field theory in curved spacetimes, and cosmology. In the particular context of cosmology, one is usually interested in the description of tensor harmonics over maximally symmetric manifolds, since these are the spaces in better agreement with observations. In this work we revisit the construction of scalar, vector and tensor harmonics in symmetric three-dimensional spaces with particular interest in -- but not limited to -- cosmological applications.

Scalar harmonics in symmetric spaces are well known among
cosmologists~\cite{Harrison:1967zza}, and they are defined as a
complete set of eigenfunctions of the Laplace-Beltrami
operator. Vector- and tensor-valued harmonics can be similarly
defined, and their explicit forms were gathered
in~\cite{Sandberg1978,Gerlach1978,Tomita1982}. These objects are found
nearly everywhere in cosmological applications, specially in those
related to large-scale structure and its related observables. Indeed,
on cosmological scales, where linear perturbation theory successfully
accounts for the formation of structures, perturbation modes, that is
the components in an expansion on tensor harmonics, evolve independently from one another. This fact enormously simplifies the construction of observables and the assessment of their statistics. In particular, a decomposition based on tensor harmonics is essential for the computation of cosmic microwave background (CMB) fluctuations around a maximally symmetric (but possibly curved) space~\cite{Abbott1986}. The normal modes which have been introduced in \cite{TAM1,TAM2} correspond to specific components of those of~\cite{Abbott1986}, and are consequently an equivalent presentation of them. An equivalent covariant formulation of these normal modes
is also presented in \cite{Challinor:2000as,Challinor:1999xz}.

In this article, we review the general construction of harmonics in maximally symmetric three-dimensional spaces, along with the associated normal modes, and show how they can be systematically built by recursions. In doing so, we collect all explicit expressions of the normal modes for scalar, vector
and tensor harmonics. Throughout, we choose to use a modern formulation based on spin-weighted spherical harmonics from which even and odd parts and also the general
structure is more tractable. Hence, this differs from the formulations given in \cite{Lindblom:2017maa,Dai:2012bc}.

Section~\ref{Sec21} is dedicated to definitions and notation. In particular, we define the harmonics, the helicity basis, the normal modes and the radial
functions of which many properties are collected in the
appendices. Section~\ref{SecBuild} is dedicated to the general
construction of these radial functions which characterize fully the
harmonics and most relations are collected in
Appendix~\ref{AppRecursions}. The reader interested only in the actual
expression of the harmonics can jump directly to Section~\ref{SecGatherRadial} where the explicit
expressions of the radial functions are collected, or to
Appendix~\ref{AppFlat} if interested in the flat case only. The
normalization of harmonics is discussed in Section~\ref{SecNorm}, while plane waves are built in Section~\ref{SecPlaneWaves}. 
The formalism is illustrated in Section~\ref{SecCMB} for the standard multipole expansion
of the CMB radiative transfer functions. Finally the comparison of our
results with previous references is detailed in
Appendix~\ref{SecComparison}. The tables of Appendix~\ref{SecSymbols}
gather the most important ancillary notation used throughout.

\section{Definitions}

\subsection{Maximally symmetric spaces}\label{Sec21}

We start be recalling some basic properties of maximally symmetric spaces. A nice and physicist targeted introduction can be found in~\cite{weinberg1972gravitation}.

Maximally symmetric spaces (as opposed to spacetime) are uniquely fixed by a real parameter $K$, known as constant of curvature. In three dimensions, and using standard spherical coordinates $(\chi,\theta,\phi)$, the metrics of these spaces read
\be\label{DefMetric}
g_{ij}\dd x^i \dd x^j  =  \dd  \chi^2 + r^2(\chi) [\dd \theta^2 + \sin^2 \theta \dd \phi^2]\,.
\ee
The radial coordinate $\chi$ is implicitly defined by the function $r(\chi)$, which assumes different values according to the sign of the parameter $K$:
\be\label{defrchi}
r(\chi)=\begin{cases}
\ellc \sinh(\chi/\ellc)\,, & (K<0)\,,\\
\ellc \sin (\chi/\ell_c)\,, & (K>0)\,,\\
\chi\,, & (K=0)\,.
\end{cases}
\ee
Here, ${\ellc\equiv 1/\sqrt{|K|}}$ is the curvature radius, which is related to the Ricci scalar by $R=6K$. Clearly, $K$ distinguishes between open ($K<0$), closed ($K>0$) and flat ($K=0$) spaces. When $K\neq0$ we can further use units for which $\ellc = 1$, that is, all lengths are expressed in units of the curvature radius and, in the closed case, this implies $0 \leq \chi \leq \pi$. The general case $\ellc \neq 1$ can be trivially restored from dimensional analysis if needed. To emphasize our special choice of units, let us introduce a reduced curvature parameter
\be\label{DefBarK}
\barK\equiv K \ellc^2 = K/|K|
\ee
which assumes the value $+1$ ($-1$) in the closed (open) case.

The Riemann tensor of maximally symmetric spaces can be written directly in terms of the space metric and the constant $K$:
\be
R_{ijkl} = K(g_{ik} g_{jl} - g_{il} g_{jk})\,.
\ee
This greatly simplifies identities involving commutators of covariant derivatives. One identity that we shall need is
\begin{align}\label{commut1}
[\Delta, \nabla_{k_1}\cdots \nabla_{k_n}] T_{i_1\dots i_j} & = 2 K \left[ n(n+1)/2+ nj \right]\nonumber\\ 
& \times \nabla_{k_1}\cdots \nabla_{k_n} T_{i_1\dots i_j}
\end{align}
where $\nabla_i$ is the covariant derivative associated with the metric \eqref{DefMetric} (i.e., $\nabla_k g_{ij}=0$) and $\Delta  = \nabla ^j \nabla _j $ is the Laplace-Beltrami operator. In what follows we shall adopt the unifying notation$\gsin$, $\gtan$ and $\gcot$ for trigonometric functions, defined as the usual functions when $K>0$, and as their hyperbolic counterparts when $K<0$.

\subsection{Helicity basis}

The notion of helicity (or spin) basis is more conveniently introduced in terms of an orthonormal triad of basis vectors
\beas\label{SphericalBasis1}
\gr{n}_\chi &=& \partial_\chi\,,\\
\gr{n}_\theta&=& r^{-1}(\chi)\partial_\theta\,,\\
\gr{n}_\phi&=& r^{-1}(\chi)\csc(\theta) \partial_\phi\,,
\eeas
together with its dual basis
\beas\label{SphericalBasis2}
\gr{n}^\chi &=& \dd \chi\,,\\
\gr{n}^\theta&=& r(\chi)\dd \theta\,,\\
\gr{n}^\phi&=& r(\chi)\sin(\theta) \dd\phi\,.
\eeas
From this we can form the standard helicity vector (spin 1) basis as
\begin{align}
\begin{split}\label{Defnpm}
\gr{n}_\pm &\equiv \frac{1}{\sqrt{2}}\left(\gr{n}_\theta \mp \ii \gr{n}_\phi\right), \\
\gr{n}^\pm &\equiv \frac{1}{\sqrt{2}}\left(\gr{n}^\theta \mp \ii \gr{n}^\phi\right).
\end{split}
\end{align}
Given a unit vector $\gr{n}$ at the origin ($\chi=0$), the pair $(\chi,\gr{n})$ denotes a point reached following a geodesic of length $\chi$ whose tangential direction at the origin is $\gr{n}$. It is also obvious from the spherical symmetry that the tangential vector of the geodesic at that point is $\gr{n}_\chi$. Hence it is customary to use the symbol $\gr{n}$ for both $\gr{n}_\chi$ and its dual $\gr{n}^\chi$. The helicity basis vectors $\gr{n}_\pm$ also depend on the point $(\chi,\gr{n})$ considered, but they are parallel transported along a radial curve, that is
\be\label{Parallel}
n^k \nabla_ k n^\pm _i = 0\,.
\ee
Thus, since they depend essentially only on the direction $\gr{n}$, it
is customary not to write this dependence explicitly. 

We now use the vector basis~\eqref{Defnpm} to build a suitable tensor basis (spin $s$) for symmetric trace-free (STF) tensors. For 
$0\leq |s| \leq j$, we define 
\be\label{CentralMulti}
\hat n^{\pm s}_{i_1\dots i_j} \equiv n^\pm_{\langle i_1}\dots n^\pm_{i_s} n_{i_{s+1}}\dots n_{i_j\rangle}\,,
\ee
with a similar definition when free indices are up. The angle brackets
mean that we must form the symmetric trace-free part on the enclosed
indices, and this is performed in practice with
\eqref{ExplicitSTFHelicity}. Analogously to the helicity basis, these
tensors (which are also parallel transported) depend only on the
direction $\gr{n}$ --- a dependence which will be omitted from now
on. 

In what follows, it will be convenient to introduce a multi-index notation
\be\label{Multi}
I_j \equiv i_1 \dots i_j\,,
\ee
such that the basis for STF tensors is written succinctly as $\hat n^{\pm s}_{I_j}$ or $\hat n_{\pm s}^{I_j}$. In Appendix \ref{AppYlm} we
summarize how the extended helicity basis \eqref{CentralMulti} is related to spin-weighted
spherical harmonics.

The generalized helicity basis~\eqref{CentralMulti} extends the
multi-index notation reviewed in Ref.~\cite{Thorne1980}, which is
restricted to using the tensors \eqref{CentralMulti} with $s=0$. \cpsilent{Up to
normalisation differences, it corresponds to the Legendre tensors introduced in~\cite{Zaldarriaga:1997va} for the cases $s=0,2$.} The
explicit expressions for $j \leq 3$, and a collection of properties
(which extends those already found in Appendix A of  Ref.~\cite{BlanchetDamour1986} for the case $s=0$), are given in Appendix \ref{AppPropnsIj}. The set of $\hat n^{s}_{I_j}$ with $|s| \leq j$ form a basis for STF tensors with $j$ free indices at
each point. Their normalization, used for extraction of components along that basis, is given by
\be\label{Defdjs1}
\hat n^{\pm s}_{I_j} \hat n_{\mp s'}^{I_j} = \delta_{s s'} d_{js}
\ee
where
\begin{align}
\begin{split}
d_{js}&\equiv \frac{j!}{(2j-1)!!}\frac{1}{(b_{js})^2}\,,\label{Defdjs-bjs}\\
b_{js} &\equiv \sqrt{2}^{|s|} \sqrt{\frac{(j!)^2}{(j+s)!(j-s)!}}\,.
\end{split}
\end{align}

\subsection{Decomposition of tensor fields}

Any STF tensor field in a maximally symmetric (three-dimensional) space can be decomposed onto the generalized helicity basis using spin-weighted spherical harmonics. This decomposition can be understood in two steps. At each point the helicity basis is a basis for STF tensors at that point, hence we can decompose the STF tensor field as
\be
T_{I_j}(\chi,\gr{n})  = \sum_{s=-j}^j {}_s T(\chi,\gr{n}) \hat n^s_{I_j}\,.
\ee
The spin functions ${}_s T(\chi,\gr{n})$ are then decomposed onto spin-weighted spherical
harmonics, so as to separate its radial and angular dependencies. This leads to 
\be
\label{GeneralYslmTensor}
T_{I_j}(\chi,\gr{n})  = \sum_{s=-j}^j \sum_{\ell\ge |s|}^{\infty}\sum_{M=-\ell}^\ell {}_s T_{\ell M}(\chi) {}_s Y_{\ell}^{M}(\gr{n}) \hat n^s_{I_j}\,.
\ee
Given the rotation property \eqref{RotationYslm}, this is a decomposition in irreducible components under the group of rotations.
On the left hand side, we recall that $j$ is the number of free indices given the multi-index notation \eqref{Multi}. 

The functions ${}_s T_{\ell M}$ are however constrained by the fact that the tensor fields must assume a given value at the origin of
coordinates ($\chi=0$). Let us consider the tensors ${\cal Y}^{jm}_{I_j}$ defined at the origin of the system of
coordinates, and which are explicitly given in Appendix \ref{AppYlm}, along with their properties. They form a complete basis (for STF tensors) and we can use them to decompose the value of the STF tensor at origin in the
form
\be\label{Tonchizero}
\left.T_{I_j}(\chi,\gr{n})\right|_{\chi=0} =\sum_{m=-j}^j t_{m} \,{\cal Y}^{jm}_{I_j}\,.
\ee
Therefore we find that at the origin the coefficient functions ${}_s
T_{\ell M}$ must be
\be\label{sTjlM}
\left.{}_s T_{\ell M}\right|_{\chi=0} = \delta_{j\ell} t_{M} k_{js}
\ee
where
\bea\label{Defkjs}
k_{j\,\pm s} &\equiv& (\mp 1)^s b_{js}\frac{(2j-1)!!}{j!}\nonumber\\
&=&(\mp 1)^s
(d_{js} b_{js})^{-1}\,.
\eea
This can be seen either from property \eqref{MagicSumYslm} once
\eqref{sTjlM} is replaced into \eqref{GeneralYslmTensor}, or from the
component extraction by contraction of \eqref{Tonchizero} with the
$\hat n^s_{I_j}$, and using the normalization \eqref{Defdjs1} and the
property \eqref{YlmsfromSTF}.

In the next section we define the tensor harmonics, and in the subsequent one we
shall be guided by the decomposition \eqref{GeneralYslmTensor} to
define normal modes and radial functions.

\subsection{Laplacian and harmonics}\label{Sec23}

The tensor valued eigenfunctions of the Laplacian are defined as 
\be\label{EqDelta}
(\Delta +k^2)T_{i_1\dots i_j} = 0\,.
\ee
We further ask these modes to be STF and divergence-free tensors, that is
\be\label{Divlesscondition}
\nabla^{i_1} T_{i_1\dots i_j} = 0\,.
\ee 

Solutions of \eqref{EqDelta} and \eqref{Divlesscondition} with $m$
free indices, and for a given $k$, are called \emph{harmonics of type $m$} for
the mode $k$, and are denoted as
\be
Q^{(jm)}_{i_1\dots i_j}\,,\quad j=|m|\,.
\ee
When $|m|=0,1,2$ these are called respectively scalar, vector and tensor harmonics.

We next introduce some \emph{derived harmonics} which are obtained by STF
combinations of $(j-|m|)$ derivatives of these harmonics. More precisely
they are defined as 
\be\label{DefGeneralQ}
Q_{I_j}^{(jm)} \equiv\frac{\nabla_{\langle i_1} \dots
\nabla_{i_{j-|m|}} Q^{(|m|,m)}_{i_{j-|m|+1}\dots i_j\rangle}}{k^{j-|m|}}\,,
\ee
which implies the basic relation for $j>|m|$
\be\label{STFrecursion}
Q^{(jm)}_{I_j} = \frac{1}{k} \nabla_{\langle i_j}
Q^{(j-1,m)}_{I_{j-1}\rangle }\,.
\ee
It can be checked by using~\eqref{commut1} that that they are not divergenceless and do not satisfy \eqref{EqDelta}, but they satisfy 
instead
\be
[\Delta+k^2-\barK (j-|m|)(j+|m|+1)] Q^{(jm)}_{I_j} =0
\ee
as well as
\be\label{GeneralDiv}
\nabla^{p}Q^{(jm)}_{I_{j-1} p}  = -q^{(jm)}\, Q^{(j-1,m)}_{I_{j-1}}\,,
\ee
\be\label{Defqjm}
q^{(jm)} \equiv \frac{(j^2-m^2)}{j(2j-1)} \frac{(\nu^2 - \barK j^2)}{k}\,.
\ee
Here we have introduced the notation
\be\label{Defnu}
\nu^2 \equiv k^2 + (1+|m|)\barK \,,
\ee
such that harmonics and derived harmonics can be either characterized
by the value of the mode $k$ or by the related mode $\nu$. If we fix
$k$, then $\nu$ is a function of both $|m|$ and $k$. Conversely if we
fix $\nu$, then $k$ is a function of both $\nu$ and $|m|$, and from
now on we consider this point of view.

\subsection{Comment about notation}\label{SecNotation}

For simplicity, we often omit to write the dependence of the harmonics on $k$ (or $\nu$)
to alleviate the notation. Similarly, wherever not needed, the dependence on the position on space, that is on $(\chi,\gr{n})$, is
not written explicitly. Hence, even though the full expression of an
harmonic should be $Q^{(jm)}_{I_j}(\chi,\gr{n};\nu)$, we shall use $Q^{(jm)}_{I_j}(\nu)$, $Q^{(jm)}_{I_j}(\chi,\gr{n})$ or simply $Q^{(jm)}_{I_j}$, depending on the context. Such practice will be used not only for the 
harmonics, but for any other quantities depending on $\chi$, $\gr{n}$
and $\nu$.

\subsection{Normal modes}

In order to find a decomposition of the type \eqref{GeneralYslmTensor} for the harmonics and their derivations, we follow~\cite{TAM1,TAM2} and split the radial and angular dependence through a new function
\be\label{DefGsjm}
{}_s G_\ell^{(jm)}(\chi,\gr{n};\nu) \equiv c_\ell \,\,{}_s
 \alpha_\ell^{(jm)}(\chi;\nu) \,{}_s Y_{\ell}^{m}(\gr{n})\,,
\ee
with the conventional factor
\be\label{Defcl}
c_\ell \equiv \ii^\ell \sqrt{4\pi(2\ell+1)}\,.
\ee

We insist on the fact that ${}_s G_\ell^{(jm)}$ depends on the point considered, that is, on $(\chi,\gr{n})$, while the STF basis $\hat
n^s_{I_j}$ depends on the choice of $\gr{n}$. Moreover, the radial functions $\alpha_\ell^{(jm)}(\chi;\nu)$ do not depend on $\gr{n}$, while the spin-weighted spherical harmonics ${}_s Y_{\ell}^{m}(\gr{n})$ do~\footnote{In fact, such separation between radial and angular dependence lies in the heart of the Total Angular Momentum method -- see~\cite{TAM1} for more details}. The radial functions, to be constructed in Section~\ref{sec:radial-funcs}, are conventionally normalized when $\ell=j$ as \cite{TAM1,TAM2} 
\be\label{AlphaChiZero}
\left.{}_s \alpha_\ell^{(jm)}\right|_{\chi=0} = \frac{1}{2j+1} \delta_{\ell j}\,.
\ee
Accordingly, it implies that around the origin
\be\label{NormG}
{}_s G_{\ell=j}^{(jm)} = \frac{c_j}{2j+1} {}_s Y_{\ell=j}^{m} + {O}(\chi)\,.
\ee
In general radial functions are non-vanishing only for the conditions
\be\label{GeneralRestriction}
j \geq {\rm max}(|m|,|s|)\,,\quad\ell \geq {\rm max}(|m|,|s|)\,,
\ee
and are chosen to be null functions otherwise.

We now search to build a basis for tensor harmonics (and their derivations), with $j$ free STF indices, in the form
\be\label{Qljm}
{}^\ell Q^{(jm)}_{I_j} \equiv \sum_{s=-j}^j {}_s g^{(jm)}  {}_s G_\ell^{(jm)} \hat n^s_{I_j}\,,
\ee
where the $ {}_s g^{(jm)}$ are numerical coefficients yet to be
fixed. These harmonics correspond to considering a single $(\ell,M)$
pair in the otherwise general sum of \eqref{GeneralYslmTensor}. The summation on $\ell$ will be
taken when we present the construction of plane waves in Section \ref{SecPlaneWaves}, and the summation on $M$ is needed only when considering general reference axis harmonics as detailed in Section \ref{SecRotation}.

Moving forward, let us also define
\be\label{gtildetog}
{}_s \tilde g^{(jm)} \equiv  {}_s g^{(jm)} d_{js}\,,
\ee
such that from \eqref{Defdjs1} we get the inverse relation
\be\label{QtoG}
{}^\ell Q^{(jm)}_{I_j} \hat n_{\mp s} ^{I_j} = {}_{\pm s} \tilde g^{(jm)} {}_{\pm s} G^{(jm)}_\ell \,.
\ee
From \eqref{Qljm} and \eqref{QtoG}, we see that the coefficients $
{}_s \tilde g^{(jm)} $ and $ {}_s g^{(jm)} $ are used to relate the tensors
${}^\ell Q^{(jm)}_{I_j} $ to the functions ${}_{\pm s} G^{(jm)}_\ell $ --- 
called normal modes --- and vice versa.  The normal modes are the coefficients [with a functional dependence on $(\chi,\gr{n})$] 
of the harmonics in the generalized helicity basis. Since the coefficients ${}_sg^{(jm)}$ (and thus ${}_s\tilde{g}^{(jm)}$) are yet undetermined, we can further choose that, for a given $(jm)$ pair
\begin{align}
\begin{split}\label{gtsjm}
{}_{\pm s}\tilde g^{(jm)} &= \frac{(\mp)^s}{b_{js}} {}_0 \tilde
g^{(jm)}\,,\\
{}_{\pm s}\tilde g^{(jm)} &= {}_{\pm s}\tilde g^{(j,-m)} \,,
\end{split}
\end{align}
which, given~\eqref{Defkjs}, implies that
\begin{align}
\begin{split}\label{gsjm}
{}_{\pm s}g^{(jm)} &= (\mp)^s\,b_{js} \,{}_0 g^{(jm)}\,,\\
&= {}_{\pm s} g^{(j,-m)} \,,
\end{split}
\end{align}
and, in particular
\be\label{gsgms}
{}_{-s}g^{(jm)}=(-1)^s\, {}_{s}g^{(jm)}\,,
\ee
with a similar relation for the ${}_{s}\tilde g^{(jm)}$. The choices \eqref{gsjm} [that is  ${}_{s}g^{(jm)}\propto k_{js}$ for the
dependence on $s$] ensure that 
\be\label{QToCurlyY}
\left.{}^{\ell=j} Q_{I_j}^{(jm)}\right|_{\chi=0} = {}_0
  \tilde g^{(jm)}\frac{c_j}{2j+1}{\cal Y}^{jm}_{I_j}\,,
\ee
for exactly the same reasons detailed after \eqref{Tonchizero}. 

Given the linearity of \eqref{EqDelta}, any linear combination of solutions of the type \eqref{DefGsjm} for different values of $\ell$ is also a solution. This is how plane-wave solutions are built, and we discuss this construction in \S\ref{SecPlaneWaves}. Finally, it is trivial to restore spatial dimensions ($\ellc \neq 1$) since harmonics, normal modes and radial functions are all dimensionless.

\subsection{Radial functions}\label{sec:radial-funcs}

We recall that for simplicity the dependence on $\chi$ and $\nu$ of the radial functions $_s\alpha_{\ell}^{(jm)}$ is omitted. We also  split them into even and odd parts (also called respectively electric and magnetic radial functions) as
\be\label{defeb}
{}_{\pm s} \alpha_{\ell}^{(jm)} = {}_{s}\epsilon_{\ell}^{(jm)}
  \pm \ii \,{}_s\beta_{\ell}^{(jm)} \,,
\ee
and by construction there is no odd part for $s=0$, that is
\be\label{betazero}
{}_0 \beta^{(jm)}_\ell = 0\,.
\ee
We shall check further that they also satisfy the properties
\begin{align}
\begin{split}\label{Inversionm}
{}_s \epsilon^{(j,-m)} &= {}_s \epsilon^{(jm)} \\
{}_s \beta^{(j,-m)} &= -{}_s \beta^{(jm)} \,.
\end{split}
\end{align}
In practice this means that we only need to build the radial functions for $m\geq0$.

In most cases $\nu$ is real and the electric and magnetic radial functions are real. However, when considering super-curvature modes on
open spaces~\cite{Lyth1995,Duality}, $\nu$ can be complex. In that case one
cannot deduce from \eqref{defeb} that complex conjugation on radial
functions amounts to $s \to -s$, and one must rather use \eqref{TrueConjugate}.

\section{Building harmonics}\label{SecBuild}

We now proceed to the determination of the radial functions. Indeed, they
determine the normal modes from the definition \eqref{DefGsjm}, and
subsequently the harmonics (and derived harmonics) from \eqref{Qljm}.
In the next section we first start by building the radial functions
for harmonics ($j=|m|$), and in the subsequent one we deduce the radial
functions for the derived harmonics ($j>|m|$). In all expressions, the
value of $\ell$ is general.

\subsection{Radial functions  of harmonics ($j=|m|$)}\label{Secjm}

We recall that harmonics are divergenceless. We normalize them with
\be
{}_0 \tilde g^{(j,\pm j)} = 1\,.
\ee

 We first note that [see e.g. Eq. (A.22) in \cite{Challinor:1999xz}]
\be
\Delta[{\rm curl}\,({}^\ell Q^{(j,\pm j)}_{I_j})]  ={\rm curl}[\Delta
({}^\ell Q^{(j,\pm j)}_{I_j} )]\,,
\ee
where the curl is the obvious generalization to STF tensors defined by
\be\label{Defcurl}
{\rm curl} \,T_{I_\ell} \equiv \epsilon_{jp\langle i_1}\nabla^j
T_{I_{\ell-1}\rangle}^{\phantom{I_{\ell-1}\rangle}p} \,.
\ee
Hence, the curl  of an harmonic is also an harmonic. Furthermore using the divergenceless relation~\eqref{Divlesscondition} it can be proven that [see, e.g., Eq. (3.13) of Ref.
\cite{Challinor:1999xz} for the $j=2$ case]
\be
{\rm curl}\,{\rm curl}\,({}^\ell Q^{(j,\pm j)}_{I_j}) = \nu^2
({}^\ell Q^{(j,\pm j)}_{I_{j}})\,.
\ee
Therefore, we can choose
\be\label{InitialCurl}
{\rm curl}\,({}^\ell Q^{(j,\pm j)}_{I_{j}}) = \pm \nu ({}^\ell Q^{(j,\pm j)}_{I_{j}})\,.
\ee 
The choice of sign on the right hand side (which could have been $\mp \nu$) amounts to choosing the global normalization of the odd 
radial function, and our choice is made so that we recover the flat case construction that is recalled in appendix \ref{AppFlat}.

Using the property \eqref{Divns} of the extended helicity basis, and the decomposition \eqref{Qljm} along with
the condition \eqref{gsjm}, we deduce that the divergenceless
relation \eqref{Divlesscondition} leads to the set of relations among
radial functions for $0<s<j$
\bea\label{RecursionDivLess}
&&\frac{\dd}{\dd \chi} {}_{\pm s}\alpha_\ell^{(j,\pm j)}+(j+1) \gcot \chi \,{}_{\pm
  s}\alpha_\ell^{(j,\pm j)}=\nonumber\\
&&\frac{{({}_- \lambda_\ell^s )( {}_-\lambda_j^s)}}{2(j+s)
  r(\chi)} {}_{\pm(s-1)}\alpha^{(j,\pm j)}_\ell\nonumber\\
&&+\frac{{({}_+ \lambda_\ell^s )( {}_+\lambda_j^s)}}{2
  (j-s)r(\chi)} {}_{\pm(s+1)}\alpha^{(j,\pm j)}_\ell\,,
\eea
where we defined 
\bea\label{Deflambda}
{}_{\pm} \lambda^s_\ell &\equiv& \sqrt{(\ell+1 \pm s)(\ell \mp s)}\\
&=&\sqrt{\ell(\ell+1) - s (s\pm 1)}\,.\nonumber
\eea
Condition~\eqref{RecursionDivLess}  is a special case of the divergence relation \eqref{DivergenceRelation1}.
In the special case $s=0$ it reduces to
\bea\label{Divszero}
&&\frac{\dd}{\dd \chi} {}_{0}\epsilon_\ell^{(j,\pm j)}+(j+1) \gcot
\chi \,{}_{0}\epsilon_\ell^{(j,\pm j)}\nonumber\\
&&=\frac{\sqrt{\ell(\ell+1)}}{r(\chi)}\sqrt{\frac{j+1}{j}} {}_1 \epsilon^{(j,\pm j)}_\ell,
\eea
which is a special case of \eqref{DivergenceRelation2}.

Hence, when considering the real and imaginary parts of radial functions, we see that the
divergenceless relation brings $j$ relations for the even modes and
$j-1$ relations for the odd modes (if $j\geq 1$).  Given
\eqref{betazero}, we conclude that using \eqref{RecursionDivLess} we can deduce 
all radial modes in the case $j=|m|$ (that is for all allowed values
of $s$) once we know ${}_0 \epsilon_\ell^{(j,\pm j)}$ and ${}_1 \beta_\ell^{(j,\pm j)}$.  
These terms are in turn found from the Laplace equation \eqref{EqDelta}. Again, using the decomposition \eqref{Qljm} along
with the condition \eqref{gtsjm} and the identities \eqref{Deltans},
this leads (when $s>0$) to
\bea\label{Laplaciansnonzero}
&&\frac{\dd^2}{\dd \chi^2} ({}_{\pm s}\alpha_\ell^{(j,\pm j)}) + 2 \gcot \chi \frac{\dd}{\dd \chi} ({}_{\pm
  s}\alpha_\ell^{(j,\pm j)}) \nonumber\\
&&+{}_{\pm s}\alpha_\ell^{(j,\pm j)}\gcot^2(\chi) (s^2-j(j+1))\nonumber\\
&&+{}_{\pm s}\alpha_\ell^{(j,\pm j)}\frac{1}{r^2(\chi)}(s^2-\ell(\ell+1))\nonumber\\
&&+{}_{\pm (s-1)}\alpha_\ell^{(j,\pm j)} \frac{\gcot \chi}{r(\chi)}
({}_{-} \lambda^s_j)( {}_{-} \lambda^s_\ell)\nonumber\\
&&+{}_{\pm (s+1)}\alpha_\ell^{(j,\pm j)} \frac{\gcot \chi}{r(\chi)}
({}_{+} \lambda^s_j)( {}_{+} \lambda^s_\ell)\nonumber\\
&&=-k^2 {}_{\pm s}\alpha_\ell^{(j,\pm j)}\,.
\eea
As for the $s=0$ case, it is simply
\bea\label{Laplacianszero}
&&\frac{\dd^2}{\dd \chi^2} ({}_0 \epsilon_\ell^{(j,\pm j)})+ 2 \gcot
\chi \frac{\dd}{\dd \chi} ({}_0 \epsilon_\ell^{(j,\pm j)}) \nonumber\\
&&-\left[j(j+1)\gcot^2(\chi)+\frac{\ell(\ell+1)}{r^2(\chi)}\right] {}_0\epsilon_\ell^{(j,\pm j)}\nonumber\\
&&+2 \frac{\gcot \chi}{r(\chi)}\sqrt{j(j+1)\ell(\ell+1)}{}_1
\epsilon_\ell^{(j,\pm j)} \nonumber\\
&&=-k^2 {}_0\epsilon_\ell^{(j,\pm j)}\,.
\eea
When combined with the divergenceless condition \eqref{Divszero}, this equation leads to
\bea\label{epsilon0mmeq}
&&\frac{\dd^2}{\dd \chi^2}{}_0 \epsilon_\ell^{(j,\pm j)} + 2(j+1){\gcot\chi}
\frac{\dd}{\dd \chi}{}_0 \epsilon_\ell^{(j,\pm j)} \nonumber\\
&&+{}_0 \epsilon_\ell^{(j,\pm j)}\left[k^2+
  j(j+1) \gcot^2(\chi)-\frac{\ell(\ell+1)}{r^2(\chi)}\right] \nonumber\\
&&=0 \,.
\eea
Similarly, the imaginary  part of  the relation \eqref{Laplaciansnonzero} for $s=1$, when combined with the
divergenceless condition \eqref{RecursionDivLess} at $s=1$, leads to
\bea
&&\frac{\dd^2}{\dd \chi^2}{}_1 \beta_\ell^{(j,\pm j)} + 2 j{\gcot\chi}
\frac{\dd}{\dd \chi}{}_1 \beta_\ell^{(j,\pm j)} \nonumber\\
&&+{}_1 \beta_\ell^{(j,\pm j)}\gcot^2(\chi)[j(j-1)-1]\nonumber\\
&&+{}_1 \beta_\ell^{(j,\pm j)}\frac{[1-\ell(\ell+1)]}{r^2(\chi)}\nonumber \\
&&=-k^2 {}_1 \beta_\ell^{(j,\pm j)} \,.
\eea
By comparing~\eqref{epsilon0mmeq} with equation~\eqref{EquationPsi},
we can now motivate the definition \eqref{Defnu}. Moreover, we find
that $_0\epsilon_\ell^{(j,\pm j)}\propto \Phi_\ell^\nu/r^j$, where
$\Phi_\ell^\nu$ are the hyperspherical Bessel functions -- see
Appendix~\ref{AppHyperBessel}. The normalization which
satisfies the normalization condition \eqref{AlphaChiZero} (this is
checked using \eqref{PowerLawOriginBessel}) and recovers the flat case construction of
appendix \ref{AppFlat} is
\be\label{solutionjmm}
{}_0 \epsilon_\ell^{(j,\pm j)} = 
\frac{(2j-1)!!}{\sqrt{(2j)!}}\sqrt{\frac{(\ell+j)!}{(\ell-j)!}}\frac{\mygamma_j}{k^j}\frac{\Phi_\ell^\nu}{r^j(\chi)},
\ee
with the dimensionless constants
\be\label{DefGamma}
\mygamma_m \equiv \prod_{i=1}^m \frac{k}{\sqrt{\nu^2-\barK i^2}}\,.
\ee
We also deduce that the odd radial functions must be such that ${}_1
\beta_\ell^{(j,\pm j)} \propto r^{1-j}(\chi)\Phi_\ell^\nu$. The
global normalization is deduced from the solution \eqref{solutionjmm}
using the curl condition \eqref{InitialCurl} contracted with $\hat
n^{I_\ell}$ which leads to
\be\label{betaunepsizero}
{}_1 \beta^{(j,\pm j)}_{{\ell}} =\mp\nu  r(\chi) {}_0\epsilon^{(j,\pm j)}_{{\ell}}\sqrt{\frac{j}{(j+1)(\ell+1)\ell}}\,.
\ee
Note that this is a particular case of \eqref{Curlszerojm} when $j=|m|$.

The radial functions for larger values of $s$ are found from
\eqref{Divszero} for ${}_1 \epsilon_\ell^{(j,\pm j)} $, and then from
\eqref{RecursionDivLess} for all $s>1$, and they satisfy automatically
the Laplace equation \eqref{Laplaciansnonzero}.

\subsection{Radial functions of derived harmonics ($j>|m|$)}\label{SecBuildDerivedHarmonics}

We now discuss the systematic construction of radial functions for the derived harmonics, which must be deduced using the definition \eqref{DefGeneralQ}. We start by noticing that the derived harmonics satisfy the property
\be\label{Curljm}
{\rm curl} \,{}^\ell Q^{(jm)}_{I_j} =  \frac{m \nu}{j} {}^\ell Q^{(jm)}_{I_j}
\ee
which is inherited from \eqref{InitialCurl} and the identity for STF
tensors [see e.g. Eq. (4.7) of \cite{Challinor:2000as}]
\be
{\rm curl} \,\nabla_{\langle i_{j+1}} T_{I_j \rangle} =
\frac{j}{j+1}\nabla_{\langle i_{j+1}} {\rm curl}\,T_{I_j \rangle} \,.
\ee
The derived harmonics are no more divergenceless, as was the case for the $j=|m|$ harmonics. Instead, they satisfy the relation \eqref{GeneralDiv}. As will be shown later, the normalization of the derived harmonics which is compatible with~\eqref{AlphaChiZero} requires that
\begin{align}
\begin{split}\label{gtjm}
{}_0 \tilde g^{(jm)}&\equiv\frac{(2|m|-1)!!}{(2j-1)!!} \prod_{p=|m|+1}^j
\frac{ {}_0 \kappa_p^m}{k}\,,\\
{}_0 g^{(jm)}&\equiv\frac{(2|m|-1)!!}{j!} \prod_{p=|m|+1}^j \frac{{}_0 \kappa_p^m}{k} \,,
\end{split}
\end{align}
where~\footnote{Our definition of ${}_s \kappa_\ell^m$ corresponds to
  the one of \cite{TAM2} times a factor $\nu$ such that $\nu \sqrt{1-
    \barK \ell^2/\nu^2} = \sqrt{\nu^2-\barK \ell^2}$.}
\be\label{Defkappaslm}
{}_s \kappa_\ell^m \equiv
\sqrt{\frac{(\ell^2-m^2)(\ell^2-s^2)}{\ell^2}}\sqrt{\nu^2-\barK \ell^2}\,.
\ee
The above normalization and \eqref{gtsjm} also imply the useful relations
\begin{align}
\begin{split}\label{Recursiong}
{}_{\pm s}\tilde g^{(jm)} &= {}_{\pm s}\tilde g^{(j-1,m)} \frac{1}{(2j-1)}\frac{{}_s
\kappa_{j}^m}{k}\,,\\
{}_{\pm s} g^{(jm)} &= {}_{\pm s} g^{(j-1,m)} \frac{j}{(j^2-s^2)}\,\frac{ {}_s \kappa_{j}^m}{k},\\
{}_{\pm s}\tilde g^{(jm)} &=\mp {}_{\pm(s-1)}\tilde g^{(jm)}
\sqrt{\frac{(j+s)}{2(j+1-s)}},\\
{}_{\pm s}g^{(jm)} &= \mp {}_{\pm(s-1)} g^{(jm)} \sqrt{\frac{2(j+1-s)}{(j+s)}}.
\end{split}
\end{align}
From A3 of \cite{BlanchetDamour1986}, we see that a general STF tensor obeys
\bea\label{DjTIl}
\nabla_j T_{I_\ell} &=& \nabla_{\langle j} T_{I_\ell
  \rangle}+\frac{2\ell-1}{2\ell+1} g_{j \langle i_\ell } \nabla^p
T_{I_{\ell-1}\rangle p}\nonumber\\
&&+ \frac{\ell}{\ell+1} \epsilon^p_{\,\,j\langle i_\ell} {\rm curl}
T_{I_{\ell-1}\rangle p}\,.
\eea
When combined with~\eqref{Curljm} and~\eqref{GeneralDiv}, we obtain the following relation among derived harmonics
\bea\label{GenRelHarm}
\nabla_p \left({}^\ell Q^{(jm)}_{I_j}\right) &=& k \left({}^\ell
  Q^{(j+1,m)}_{p I_{j}}\right)\nonumber\\
&&-\frac{2j-1}{2j+1} \,q^{(jm)}\, g_{p \langle i_j } \left({}^\ell
Q^{(j-1,m)}_{I_{j-1} \rangle}\right)\nonumber\\
&&+\frac{m \nu}{j+1} \epsilon^r_{\,\,p\langle i_j} \left({}^\ell
  Q^{(jm)}_{I_{j-1}\rangle r}\right)\,.
\eea

Let us now consider a given $m$ and a given $\ell\ge |m| $, and use the short notation $(j,s)$ to refer to the radial function ${}_s \alpha_\ell^{(jm)}$, since we 
want to explore the relations between radial functions with neighbor
values of $j$ and $s$. Identity \eqref{GenRelHarm} allows to derive recursive relations among radial functions in the space of the $(j,s)$ parameters, the most famous of which connects $(j,s)$ to the $(j\pm1,s)$ ones. To see how this is possible, we first need  to contract \eqref{GenRelHarm} with $n^p$ and replace the harmonics by their expansion \eqref{Qljm} while using the identities derived in Appendix~C.\ref{AppDerHelicity}. Then, a relation among radial functions is obtained by contraction with $\hat n^{I_j}_{\mp s}$ (or equivalently identification of the $\hat n^{\pm s}_{I_j}$ components), and extraction of the radial function from the orthogonality relation \eqref{OrthoYslm} of spin-weighted
spherical harmonics, along with the relations \eqref{Recursiong}. Eventually we obtain the central relation [see also Eq. (C5) in \cite{TAM2}]
\bea\label{Masteralpha1}
&&\frac{\dd}{\dd \chi} {}_s \alpha_\ell^{(jm)} =-\frac{\ii\nu m s}{j(j+1)}  {}_s  \alpha_{\ell}^{(jm)}\\
&&+\frac{1}{2j+1}\left[-{}_s \kappa_j^m  {}_s
  \alpha_{\ell}^{(j-1,m)}+{}_{s} \kappa_{j+1}^m  {}_s
  \alpha_{\ell}^{(j+1,m)}\right]\,, \nonumber
\eea
which holds for either negative or positive values of $m$ and $s$.

Since in the $(j,s)$ plane this links the $(j,s)$ radial functions with the one above [$(j-1,s)$] and the one below [$(j+1,s)$] for a given $\ell$ and $m$, we hereafter call it the North-South (NS) relation. 

Other relations can be obtained from \eqref{GenRelHarm} by contracting
with $n_\mp^p$ and then repeating the same procedure. This leads to relations connecting the $(j,s)$ radial
functions to the $(j\mp1,s)$ and $(j,s+1)$ ones. We thus call it the North-South-East (NSE) relation. Similarly, contracting instead
with $n_\pm^p$ and using the same method allows us to relate the $(j,s)$ radial functions to $(j\mp1,s)$ and $(j,s-1)$ radial functions --- a relation
that we call North-South-West (NSW). Their exact expressions are collected in Appendix \ref{AppRecursions}.

The combination of NS and NSE relations leads either to a relation
between $(j,s)$ radial functions with the $(j-1,s)$ and $(j,s+1)$
ones, which we call the North-East (NE) relation, or to a relation
between the $(j,s)$ radial functions with the $(j+1,s)$ and $(j,s+1)$
ones, which we call the South-East (SE) relation.  Similarly combining the NS and NSW relations leads either
to the North-West (NW) or the South-West (SW) relations. All these
relations are collected in Appendix~\ref{AppRecursions}.

\begin{figure}[!htb]
\includegraphics[scale=0.67]{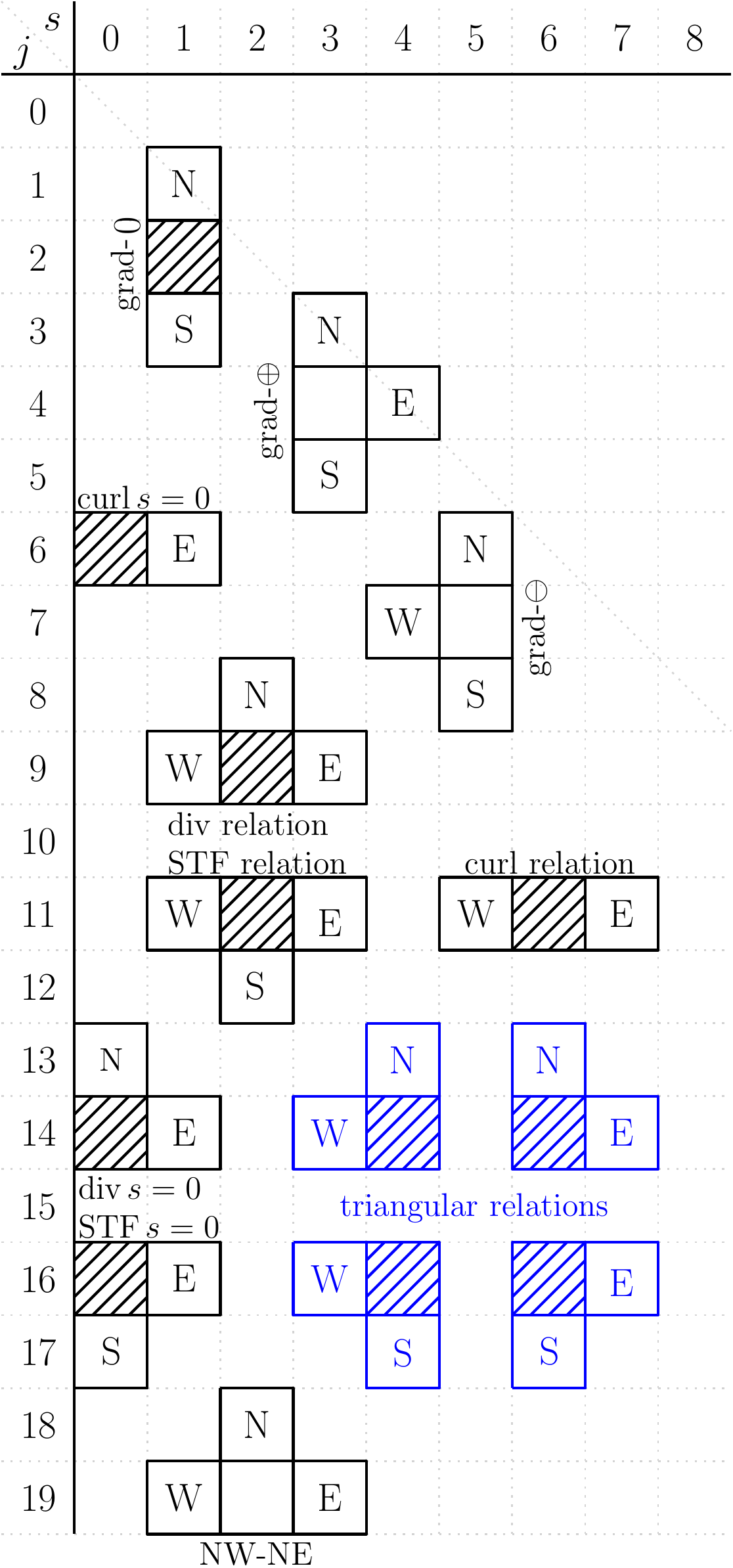}
\caption{Geographical representation of all recursion relations among
radial functions in the $(j,s)$ space of parameters. Here, `${\rm
  STF}$', `${\rm div}$' and `${\rm curl}$' denote respectively the
relations~\eqref{STFrecursion},~\eqref{GeneralDiv},
and~\eqref{Curljm}. Moreover, `grad-$0$' (resp. `grad-$\pm$') is
obtained by contraction of the gradient identity \eqref{GenRelHarm}
with $n^p$ (resp. $n_\mp^p$). The triangular relations (NW, NE, SW, and
SE) which can be formed from the grad relations are collected in
appendix \ref{AppRecursions}. Shaded squares indicate radial functions which appear with one derivative in the recursive relation. We depict only the $s \geq 0$ part in the chart as the negative $s$ are deduced from \eqref{defeb}. Only functions with $|s| \leq j$ (and $|m| \leq j$) are non-vanishing.}
\label{FigRecursions1}
\end{figure}

These triangular relations (NW, NE, SW, and SE) are the building blocks
of all sorts of recursive relations among radial function in the
$(j,s)$ space. For instance, the NS relation is a combination of the NW
and SW. It can also be found as a combination of the NE and SE
relations. Similarly the NSE relation (resp. the NSW relation) is just
the sum of the NE and SE relations (resp. the NW and SW relations).
All the recursive relations are depicted in the $(j,s)$ plane in Fig. \ref{FigRecursions1}.

There is an alternative method to obtain the triangular relations. Instead
of considering various contractions of the identity \eqref{GenRelHarm},
we can instead extract the radial functions of the divergence relation
\eqref{GeneralDiv}, the curl property \eqref{Curljm}, and the STF construction of
derived modes \eqref{STFrecursion}. Again, this proceeds by contractions
with the generalized helicity basis, extraction of the radial
functions using \eqref{OrthoYslm} and repeated use of the properties
\eqref{Recursiong}. The relations obtained are also gathered in
appendix \ref{AppRecursions2}. Combining the curl relation with the
divergence relation in two different manners leads to the NW and NE
relations. Similarly combining the curl relation with the STF relation
in two different manners leads to the SW and SE relations. While this method seems 
more appealing, it requires that we carefully separate the $s=0$ cases for which the aforementioned combinations
cannot be formed in the same manner. Instead it is found that in the
$s=0$ case, the curl relation gives the imaginary part of the NE and
SE relations. Also the $s=0$ case of the divergence relation gives the
real part of the NE relation. Finally the $s=0$ case of the STF
relation gives the real part of the SE relation. 

This indicates that the triangular relations (NW, NE, SW and SE) contain the information about recursions
in the most compact form. Their validity is only restricted by the fact that they should not produce instances 
with $s<0$ in the $(j,s)$ space, but they can be applied even if some of the radial functions vanish because of 
$s>j$. If we instead use the apparently more direct divergence  and curl
relations, we must treat the $s=0$ case separately. As an illustration, this is what has been presented in the $j=|m|$ 
case of Section~\ref{Secjm}. The solution for the $s=0$ 
case is \eqref{solutionjmm}. The divergence relation for $s=0$, given by
\eqref{Divszero}, gives only the electric function for $s=1$. One has then
to rely on the curl relation for $s=0$, which is \eqref{betaunepsizero}, to get the magnetic function
with $s=1$. In order to obtain the $s>1$ harmonics, still for $j=|m|$, one
can use the divergence relation \eqref{RecursionDivLess}, but one
could also use more directly the NE relation. Indeed, given that the north
component of the NE relations vanishes (since $j=|m|$), it gives
directly $(j=|m|,s+1)$ as a function of $(j=|m|,s)$.

Finally, we recall that all radial functions are restricted in general
to \eqref{GeneralRestriction}, and in the closed case ($\barK=1$) they are
also restricted to the integer values 
\be\label{RestrictionClosed2}
\ell \leq \nu-1\,.
\ee

\subsection{Optimal algorithm}\label{SecAlgo}

Given the plethora of recursion relations in the $(j,s)$ space for
radial functions, there are several different ways to deduce the
radial modes for the derived harmonics for increasing values
of $j$. However we can judiciously add a condition which selects one method. 
Since all radial functions are expressed in terms of
derivatives of hyperspherical Bessel functions, it is always
possible to use equation \eqref{EquadiffBessel} to reduce their form to an
expression which involves the hyperspherical Bessel function and at most its first
derivative. However, there are preferred methods for the
recursive construction of radial function, in which one never has to
rely on \eqref{EquadiffBessel} to reduce the order of derivatives. Let
us summarize one of these. Given the property \eqref{Inversionm} and the
definitions \eqref{defeb}, we only need to build harmonics for $m\geq 0$
and $s \geq 0$, and we now assume these conditions hold. 
\begin{enumerate}
\item For a given $m$, the radial function for $s=0$ and $j=m$
  is given by \eqref{solutionjmm}, and it has no derivative of
  hyperspherical Bessel functions.
\item We then use the NE recursion to obtain the $s=1$ and $j=m$
  solution, with unavoidably one derivative of Bessel
  function. However, note that this is not possible in the special case $m=0$,
  and we discuss the procedure for this case below.
\item We can then use the difference of the NE and NW relations to
  form a North-West-East relation without derivatives whose
  exact expression is \eqref{NWENoDer} and that we note NW-NE hereafter. In the
  case $j=m$, the north component vanishes so it is a relation between
  $(j=m,s-1)$, $(j=m,s)$ and $(j=m,s+1)$. Using it, one can obtain all
  radial modes for $j=m$ up to $s=j$.
\item In order to build the line with $j=m+1$, that is the radial
  functions associated with the first derived harmonics, one needs
  only to use the NSE relation to deduce ${}_0
  \alpha_\ell^{(m+1,m)}$, and then the NSW relation to deduce ${}_1
  \alpha_\ell^{(m+1,m)}$. This introduces no derivatives since the NSE and NSW relation have none. Then all ${}_s
  \alpha_\ell^{(m+1,m)}$ with $2 \leq s \leq j$ can be found either
  from the NW-NE relation \eqref{NWENoDer}, or from the use of the NSW relation. This, again, brings no 
  extra derivatives.
\item This last method is iterated to obtain all radial functions for
  increasing values of $j$.
\end{enumerate}
In the special case $m=0$, we start from the known solution ${}_0
\epsilon^{(00)}_\ell = \Phi_\ell^\nu$. Then there is no need to build
the $(j=0,s=1)$ solution since it vanishes, so we must proceed
directly by increasing the value of $j$, and build the solution for
$(j=1,s=0)$. In that case, contrary to the procedure mentioned above,
we cannot use the NSE relation since the East component vanishes, that
is, it is outside of its applicability [see \eqref{Masteralpha3}]. Instead,
we must use the NS relation \eqref{Masteralpha1} to obtain
${}_0 \alpha^{(10)}_\ell$, and this brings a derivative of a Bessel
function. Finally in order to obtain ${}_1 \alpha_\ell^{(10)}$, we can
do as in the general case, and use the NSW relation which involves no
derivative. The rest of the construction to $j\geq 2$ then proceeds
exactly like in the general case.

\begin{figure}[!htb]
\includegraphics[scale=0.86]{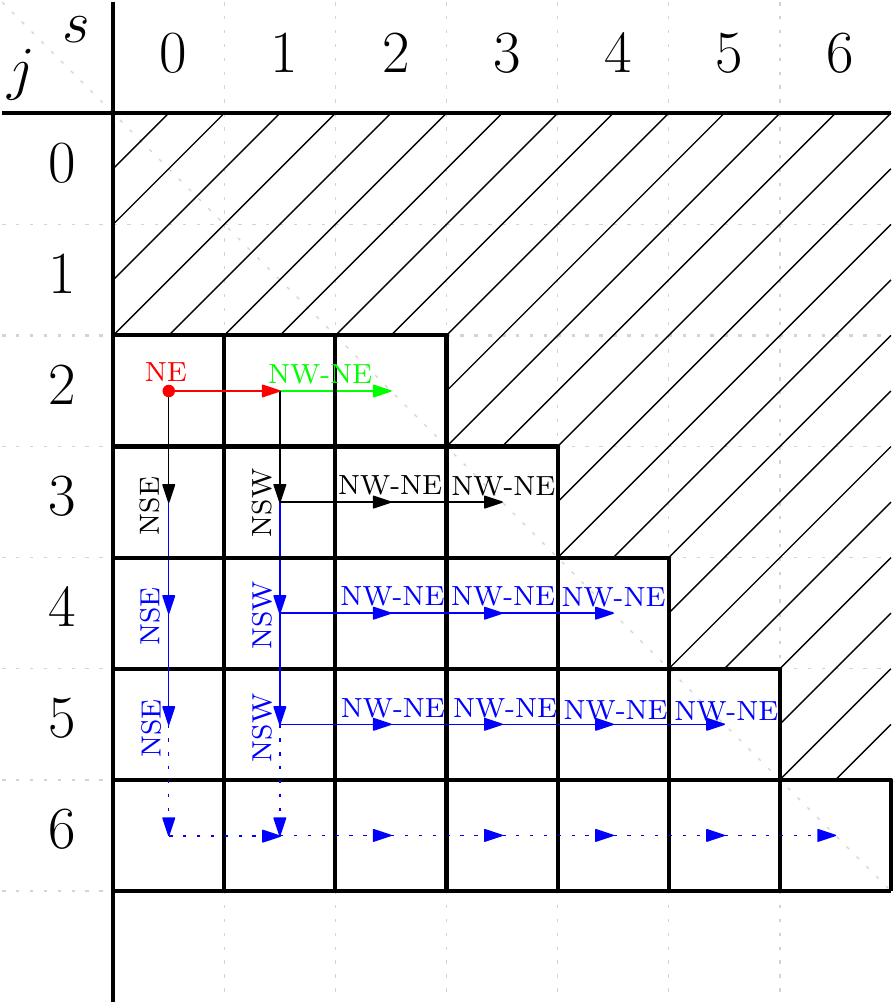}
\caption{Optimal algorithm: for a given $|m|$, the first step is to start from the solution ${}_0
  \epsilon_\ell^{(|m|,|m|)}$. Then all other radial functions are deduced
following the algorithm described in the Section \ref{SecAlgo}. The
steps $2,3,4,5$ are depicted in resp. red, green, black and blue arrows, and
the relation needed to deduce each radial function from the
previous ones is written next to the arrow.  Here we have
illustrated the case $|m|=2$, such that we have necessarily $j \geq
2$. The index $s$ must also satisfy $|s| \leq j$.  }
\label{FigRecursions2}
\end{figure}

In both cases ($m=0$ and $m>0$), there was only one step of the
procedure involving a derivative. Hence, with this method it is possible to obtain radial
functions up to any desired values of $(j,s)$ for any given $m$, as
illustrated in Fig. \ref{FigRecursions2}, and with at most one
derivative on Bessel functions, without ever having to use \eqref{EquadiffBessel} to
reduce the order of derivatives. This algorithm has been implemented
in a {\it Mathematica} notebook available at \cite{NotebookAlgorithm}.
Note that the optimal algorithm is not unique. One could for instance rely on
\eqref{Magicjjjj} to relate the $s=j$ to the $s+1=j+1$ radial functions, thus deducing the radial functions on the diagonal of Fig.~\ref{FigRecursions2}.

\subsection{Symmetry properties}

Following the same algorithm (that is the same set of recursions to travel in the $(j,s)$ space of radial function) it can
be checked that the properties
\be\label{Magicsmnu}
{}_{-s}\alpha_\ell^{(jm)}(\nu) = {}_s\alpha_\ell^{(j,-m)}(\nu) = {}_{s}\alpha_\ell^{(jm)}(-\nu)
\ee
are always satisfied. It is indeed the case for the starting radial function
\eqref{solutionjmm} of the algorithm, and it is maintained for all
values of $(j,s)$, since in all recursions for radial functions the
factors $m$ are always multiplied by the sign of $s$ and by
$\nu$. Hence from the definition \eqref{defeb} of the even and odd
parts we deduce that \eqref{Inversionm} must be satisfied.

In addition to \eqref{Magicsmnu}, there are two other symmetry properties. First, we have checked for the
first values~\footnote{In practice we checked it up to $j=4$, and for all allowed values of $s$ and $m$.} of $j$, $m$ and $s$ (but for unspecified $\ell$) that
the following property holds
\be\label{smChangejl}
{}_s \alpha_{\ell}^{(jm)}(\chi;\nu) =  {}_m \alpha_{\ell}^{(js)}(\chi;\nu)\,.
\ee
We have also checked for the first values of $m,s,j$ and $\ell$ that
\be\label{jlChangejl}
{}_s \alpha^{(jm)}_\ell =(-1)^{\ell-j} {}_s \alpha^{(\ell m)}_j\,.
\ee
Furthermore, it can be checked explicitly 
on the first values of $j$, $m$ and $s$ (but for unspecified $\ell$) that~\cite{TAM2}
\bea\label{Masteralpha}
&&\frac{\dd}{\dd \chi} {}_s \alpha_\ell^{(jm)} =-\frac{\ii\nu m s}{\ell(\ell+1)}  {}_s
  \alpha_{\ell}^{(jm)}\\
&&+\frac{1}{2\ell+1}\left[{}_s \kappa_\ell^m  {}_s\alpha_{\ell-1}^{(jm)}-{}_{s} \kappa_{\ell+1}^m  {}_s
  \alpha_{\ell+1}^{(jm)}\right]\,. \nonumber
\eea
Combined with \eqref{Masteralpha1}, this is consistent with the $j \leftrightarrow \ell$ symmetry \eqref{jlChangejl}. 
We stress that both the $m \leftrightarrow s$ and $j \leftrightarrow \ell$ symmetries are immediate in the flat case, as we demonstrate in 
Appendix~\ref{AppFlat}. 

The $m \leftrightarrow s$ symmetry is consistent with the fact that
the relations which relate radial modes with both the same $s$ and
$m$, that is Eqs. \eqref{Masteralpha1} and \eqref{Masteralpha}, are
obviously invariant under $m \leftrightarrow s$ since ${}_s \kappa_\ell^m = {}_m
\kappa_\ell^s$. Note that in the $m\leftrightarrow s$ symmetry
\eqref{smChangejl}, the same $\nu$ appears on both sides. Hence, given
the relation between $\nu$ and $k$ [see Eq.~\eqref{Defnu}], the
symmetry relates radial functions associated with different $k$, except in the flat case. The $m\leftrightarrow s$ symmetry can be used as a shortcut in the 
algorithm previously described to, for instance, calculate ${}_s\alpha^{(jm)}_\ell$ for $|m|>|s|$ from ${}_m \alpha^{(js)}_\ell$. Hence solving the radial 
functions in the plane $(j,s)$ for a given $m$ also provides automatically some of the radial
functions for larger values of $m$. Conversely, this can be left unused so as to serve as a consistency check.

\subsection{General reference axis}\label{SecRotation}

When building the harmonics and the derived harmonics, the central
relations were Eqs. \eqref{Eqdins2} and \eqref{Eqdinszero}. They depend
on $\ell$, $s$ and $j$, but not on $m$. This happens because \eqref{DefGsjm} is 
implicitly related to a special choice of axis, which is clearly not the most general construction. 
Indeed, one could perform an active rotation  $R_{\hat{\gr{\nu}}}\equiv
R(\phi_{\gr{\nu}}, \theta_{\gr{\nu}},0)$ which brings the zenith vector $\gr{e}_z$ into a general
direction $\hat{\gr{\nu}}$ with spherical coordinates $(\theta_{\gr{\nu}},\phi_{\gr{\nu}})$, that is
$R_{\hat{\gr{\nu}}}[\gr{e}_z] = \hat{\gr{\nu}}$. In order to explore this rotation, let us define the mode 
vector
\be
\gr{\nu} \equiv \nu \hat{\gr{\nu}}
\ee
which contains, at the same time, the information about the reference axis used to define harmonics, and the value of the mode $\nu$ itself. 
In Section~\ref{SecPlaneWaves} we relate $\gr{\nu}$ to the wave vector of a plane wave.

The harmonics defined with a general direction are related to the ones we have built using the zenith direction. Using the rotation rules \eqref{RotationYslm} for spherical harmonics, one finds 
\be
R[{}_s G_\ell^{(jm)} \hat n^s_{I_j}] = c_\ell\, {}_s \alpha^{(jm)}_\ell
\!\sum_{M=-\ell}^\ell\! D^\ell_{M m}(R) {}_s Y_\ell^{M} \hat n^s_{I_j}.
\ee
This naturally brings the more general definition for normal modes~\cite{Dai:2012bc}:
\begin{align}
\begin{split}
{}_s G^{(jm)}_{\ell}(\gr{\nu})&\equiv \sum_{M=-\ell}^{\ell} {}_sG^{(jm)}_{\ell M}(\nu) D^\ell_{M m}(R_{\hat{\gr{\nu}}}),\\
{}_s G^{(jm)}_{\ell M}(\nu) &\equiv c_\ell \,{}_s \alpha_\ell^{(jm)}(\nu) {}_s Y_\ell^M,
\end{split}
\end{align}
with the related more general definition for the tensor harmonics:
\bea
{}^{\ell} Q^{(jm)}_{I_j}(\gr{\nu}) &\equiv &R_{\hat{\gr{\nu}}}[{}^{\ell} Q^{(jm)}_{I_j}(\nu)]\\
 &=& \sum_{M=-\ell}^\ell {}^{\ell M} Q^{(jm)}_{I_j}(\nu)\,
 D^\ell_{Mm}(R_{\hat{\gr{\nu}}})\,,\nonumber\\
 {}^{\ell M} Q^{(jm)}_{I_j}(\nu) &\equiv& \sum_{s=-j}^j {}_s g^{(jm)}  {}_s
G_{\ell M}^{(jm)}(\nu) \hat n^s_{I_j}\,. \label{Qljmbis}
\eea
We then obtain a relation of the type \eqref{Qljm}\be
 {}^{\ell} Q^{(jm)}_{I_j}(\gr{\nu}) =\sum_{s=-j}^j {}_s g^{(jm)}\, {}_s G_\ell^{(jm)}(\gr{\nu}) \hat
 n^s_{I_j}\,.
 \ee

Evidently, we could redo the general construction of radial functions using an arbitrary reference axis (instead of our choice for the zenith). 
Provided we rotate the r.h.s of the condition Eq.~\eqref{NormG} [or Eq.~\eqref{QToCurlyY}], it would proceed exactly
through the same set of equations and steps, and one would find exactly the same radial functions. This is not a surprise, since the latter 
depend only on $\nu$.

\subsection{Conjugation and parity}\label{SecConjugationParity}

From \eqref{defeb} and \eqref{Magicsmnu} we deduce
\begin{align}
\begin{split}\label{TrueConjugate}
[{}_s \alpha_\ell^{(jm)}(\nu)]^\star &= {}_{-s}\alpha_\ell^{(jm)}(\nu^\star) \\
&=  {}_s \alpha_\ell^{(jm)}(-\nu^\star)\,.
\end{split}
\end{align}
From \eqref{Qljm} and \eqref{DefGsjm} with properties \eqref{ConjugateYslm} and \eqref{gsgms}, one then obtains the conjugation property
\be\label{MagicConjugate}
\left[{}^\ell Q_{I_j}^{(jm)}(\nu)\right]^\star =  {}^\ell Q_{I_j}^{(j,-m)}(-\nu^\star) (-1)^{(\ell+m)}\,.
\ee

Furthermore, in the special case of a rotation $R$ around the direction $\gr{e}_y$ of angle $\pi$ [that is $R_y(\pi)
\equiv R(\alpha=0,\beta=\pi,\gamma=0)$], we find from~\eqref{RotationYslmRy2}
\be\label{MagicComplex}
R_y(\pi)[{}^{\ell} Q^{(jm)}_{I_j}(\nu)] = \left[{}^{\ell} Q^{(jm)}_{I_j}(\nu^\star)\right]^\star\,,
\ee
which we can also relate to \eqref{MagicConjugate}. Rotation around
the $y$-axis by an angle $\pi$, or equivalently a parity inversion of the
$x$ and $z$ axis, is equivalent to considering the mode with $-m$ and
$-\nu$, up to a $\pm 1$ factor.

\begin{table}[!htb]
\begin{tabular}{|c|c|c|c|}
\hline
& $x \to -x$ & $y \to -y$ & $z \to -z$\\
\hline
  Factor $(-1)^m$&&yes&yes\\
  Factor $(-1)^\ell$&&&yes\\
   $Q_{I_j}^{(jm)}\to Q_{I_j}^{(j,-m)}$&yes&yes&\\
$Q_{I_j}^{(jm)}(\nu)\to Q_{I_j}^{(jm)}(-\nu) $&&&yes\\
  \hline
 \end{tabular}
\caption{Transformation rules for harmonics under the inversion of a
  single axis. \label{TableParity}}
\end{table}

We can also consider a parity transformation ${\cal P}$, which is defined on tensor fields as
\be
{\cal P}[T_{I_j}(\chi,\gr{n})] \equiv (-1)^j\,
T_{I_j}(\chi,-\gr{n})\,.
\ee
Following the same techniques using \eqref{InversionYslm} along with
\be
\label{Inversionns}
\hat n^s_{I_j}(-{\gr{n}}) = (-1)^{j+s} \,\,\hat n^{-s}_{I_j}(\gr{n})\,,
\ee
one finds
\bea\label{MagicParity}
{\cal P}[{}^{\ell} Q^{(jm)}_{I_j}(\nu)] &=& (-1)^m \left[{}^{\ell}
  Q^{(j,-m)}_{I_j}(\nu^\star)\right]^\star\nonumber\\
&=&(-1)^\ell \,\,{}^{\ell} Q^{(jm)}_{I_j}(-\nu)\,.
\eea
It is instructive to combine the previous rotation with a parity transformation. Indeed, this
corresponds to an inversion of the $y$-axis only and we find
\be\label{RP}
R_y(\pi)[{\cal P}[{}^{\ell} Q^{(jm)}_{I_j}(\nu)] ]= (-1)^m \,{}^\ell Q^{(j,-m)}_{I_j}(\nu)\,.
\ee
The factor $(-1)^m$ accounts for a rotation of angle $\pi$ around
the $z$ axis, that is, $R_z(\pi)\equiv R(\alpha=0,\beta=0,\gamma=\pi)$
which is also an inversion of the $x$ and $y$ axis. Hence, we can
deduce the transformations brought by the inversion of a single axis. The results are gathered in Table~\ref{TableParity}.

\section{Radial functions for scalars, vectors and tensors}\label{SecGatherRadial}

We now collect in this section the most common radial functions. We report the results for the even and odd components so we
can use $s \geq 0$. Furthermore we assume $m\ge 0$ since the negative values are
found from \eqref{Inversionm}. The scalar, vector and tensor cases
correspond respectively to $m=0,1,2$, with the general restrictions
\eqref{GeneralRestriction}, on which we also add the restriction
\eqref{RestrictionClosed2} in the closed case. In what follows, we only report radial functions for $j \leq 2$.

For the harmonics ($j=m$), the radial functions were already derived (even though
not formulated using spin-weighted spherical harmonics) up to $m=2$ in Ref.~\cite{Tomita1982}. Derived harmonics, that 
is with $j>m$ were reported up to $j=2$ but only in the cases $s=0$ and $s=2$ in \cite{TAM2}.
Hence this section can be used as a complete reference for radial
functions. We shall only need two particular cases of the general expression \eqref{DefGamma}:
\begin{align}
\begin{split}
\mygamma_1 &= \frac{k}{\sqrt{\nu^2 -\barK}}\\
\mygamma_2 &= \frac{k^2}{\sqrt{(\nu^2 -\barK)}\sqrt{(\nu^2 -4\barK)}}\,.
\end{split}
\end{align}
Since radial functions determine the normal modes with \eqref{DefGsjm}
and then in turn the harmonics (and derived harmonics) with \eqref{Qljm}, we also report the
values of the coefficients ${}_s g^{(jm)}$. In case one needs
contractions of the type \eqref{QtoG}, we repeat that these are
related to the ${}_s \tilde g^{(jm)}$ using \eqref{gtildetog}, and the
first few coefficients needed are
\bea
\begin{gathered}
\begin{array}{ccc}
d_{00}=1,\\
d_{10}=1, & d_{11}=1,\\
d_{20}=\frac{2}{3}, & d_{21}=\frac{1}{2}, & d_{22}=1.
\end{array}
\end{gathered}
\eea
Throughout, we abbreviate $r(\chi)$ given by \eqref{defrchi} as $r$. In
the expressions reported below, we note that the radial functions are
not invariant under $\nu \to -\nu$ in general, even though it is the
case for the hyperspherical Bessel functions of appendix
\ref{AppHyperBessel}. Indeed there is a prefactor linear in $\nu$ in
each magnetic radial function, as required by property \eqref{Magicsmnu}. 

\subsection{Scalar modes ($m=0$)}

The radial functions of the base scalar harmonics are simply the hyperspherical Bessel functions:
\be
{}_0 \epsilon^{(00)}_\ell = \Phi^\nu_\ell\,.
\ee
The radial functions for the derived harmonics are given, up to $j=2$, by
\beas
{}_0 \epsilon^{(10)}_\ell &=& \frac{\mygamma_1}{k}\frac{\dd}{\dd \chi}\Phi^\nu_\ell,\slabel{epszuz}\\
{}_1 \epsilon^{(10)}_\ell &=& \frac{\mygamma_1}{k}\sqrt{\frac{\ell(\ell+1)}{2}}\frac{\Phi^\nu_\ell}{r},\slabel{1e10}\\
{}_0 \epsilon^{(20)}_\ell &=& \frac{\mygamma_2}{2k^2}\left[3 \frac{\dd^2}{\dd \chi^2}+(\nu^2-\barK)\right]\Phi_\ell^\nu,\\
{}_1 \epsilon^{(20)}_\ell &=& \frac{\mygamma_2}{k^2}\sqrt{\frac{3\ell(\ell+1)}{2}}\frac{\dd}{\dd \chi}\left(\frac{\Phi_\ell^\nu}{r}\right), \slabel{1e20}\\
{}_2 \epsilon^{(20)}_\ell &=& \frac{\mygamma_2}{k^2} \sqrt{\frac{3(\ell+2)!}{8(\ell-2)!}}\frac{\Phi_\ell^\nu}{r^2}.\slabel{2e20}
\eeas
The constants needed to build the harmonics and derived harmonics are
\be
{}_0 g^{(00)}\mygamma_1^{-1} = {}_0 g^{(10)}=\mp{}_{\pm 1}g^{(10)}=\mygamma_1^{-1},
\ee
and
\be
{}_0 g^{(20)} =\mp\frac{\sqrt{3}}{2}{}_{\pm1}g^{20} = \sqrt{\frac{3}{2}}{}_{\pm2}g^{20} = \mygamma_2^{-1}.
\ee

\subsection{Vector modes ($m=1$)}

Similarly, for the radial functions built from the vector modes, we find
that the base harmonics are given by
\beas
{}_0 \epsilon^{(11)}_\ell &=& \frac{\mygamma_1}{k}\sqrt{\frac{\ell(\ell+1)}{2}}\frac{\Phi_\ell^\nu}{r},\slabel{0e11}\\
{}_1 \epsilon^{(11)}_\ell &=& \frac{\mygamma_1}{2k}\frac{\dd (r \Phi_\ell^\nu)}{r \dd \chi},\\
{}_1 \beta^{(11)}_\ell &=&-\frac{\nu \mygamma_1}{2 k} \Phi_\ell^\nu\,.
\eeas
Note that~\eqref{0e11} agrees with~\eqref{1e10}, which corroborates~\eqref{smChangejl}.

The radial functions for the derived harmonics are
\begin{subequations}
\begin{align}
{}_0 \epsilon^{(21)}_\ell &= \frac{\mygamma_2}{k^2}\sqrt{\frac{3\ell(\ell+1)}{2}}    \frac{\dd }{\dd \chi}\left(\frac{\Phi_\ell^\nu}{r}\right)\\,\slabel{0e21}
{}_1 \epsilon^{(21)}_\ell &=\frac{\mygamma_2}{k^2}\left[\frac{\dd^2}{\dd  \chi^2}+\gcot (\chi)\frac{\dd}{\dd \chi}\right.\nonumber\\\slabel{1e21}
  &\qquad\left.+\left(\frac{\nu^2}{2}-\frac{1}{r^2}\right)\right]\Phi_\ell^\nu, \\
{}_1 \beta^{(21)}_\ell &=-\frac{\mygamma_2 \nu}{2 k^2} \, r \frac{\dd}{\dd \chi} \left(\frac{\Phi_\ell^\nu}{r}\right),\\
{}_2 \epsilon^{(21)}_\ell &= \frac{\mygamma_2}{k^2} \frac{\sqrt{(\ell+2)(\ell-1)}}{2}\frac{\dd }{r^2 \dd \chi}(r \Phi_\ell^\nu),\\\slabel{2e21}
{}_2 \beta^{(21)}_\ell &=-\frac{\mygamma_2}{k^2} \nu  \frac{\sqrt{(\ell+2)(\ell-1)}}{2}\frac{\Phi_\ell^\nu}{r}\,. 
\end{align}\slabel{2b21}
\end{subequations}
One checks that~\eqref{0e21} agrees with~\eqref{1e20}, in agreement
with~\eqref{smChangejl}. The constants needed to build the corresponding harmonics are 
\begin{align}
\begin{split}
_{0}g^{(11)} & =\mp_{\pm1}g^{(11)}=1,\\
\frac{2}{\sqrt{3}}{}_{0}g^{(21)} & =\mp_{\pm1}g^{(21)}=\sqrt{2}_{\pm2}g^{(21)}=\frac{\xi_{1}}{\xi_{2}}.
\end{split}
\end{align}

\subsection{Tensor modes ($m=2$)}

Finally, we give the radial modes related to the base tensor harmonics. They are
\begin{subequations}
\begin{align}
{}_0 \epsilon^{(22)}_\ell &= \frac{\mygamma_2}{k^2}\sqrt{\frac{3(\ell+2)!}{8(\ell-2)!}}\frac{\Phi_\ell^\nu}{r^2},\\\slabel{0e22} 
{}_1 \epsilon^{(22)}_\ell &=\frac{\mygamma_2}{k^2} \frac{\sqrt{(\ell+2)(\ell-1)}}{2}\frac{\dd }{r^2 \dd \chi}(r \Phi_\ell^\nu), \\\slabel{1e22}
{}_1 \beta^{(22)}_\ell &=-\frac{\mygamma_2 \nu}{k^2}  \frac{\sqrt{(\ell+2)(\ell-1)}}{2}\frac{\Phi_\ell^\nu}{r}, \\\slabel{1b22}
{}_2 \epsilon^{(22)}_\ell &= \frac{\mygamma_2}{4k^2} \left[\frac{\dd^2}{\dd
    \chi^2}+4 \gcot(\chi)\frac{\dd}{\dd
    \chi}+2\gcot^2(\chi)\right.\nonumber\\
&\left.\qquad \qquad+(-\barK-\nu^2)\right]\Phi_\ell^\nu, \\
{}_2 \beta^{(22)}_\ell &=-\frac{\mygamma_2 \nu}{2 k^2}\frac{\dd}{r^2\dd
                           \chi}(r^2 \Phi_\ell^\nu).
\end{align}
\end{subequations}
Again, one checks that~\eqref{0e22} agrees with~\eqref{2e20},
\eqref{1e22} agrees with~\eqref{2e21}, and \eqref{1b22} agrees
with~\eqref{2b21}, in agreement with~\eqref{smChangejl}. There is an alternative expression for $_2\epsilon^{(22)}$, which is
\be
{}_2 \epsilon^{(22)} = \frac{\mygamma_2}{4 k^2} \left[\frac{\dd^2}{r^2 \dd
   \chi^2}(r^2 \Phi_\ell^\nu)-(\nu^2-\barK )\Phi_\ell^\nu\right].\nonumber
\ee
The constants needed to build tensor harmonics are
\be
\frac{2}{3}{}_{0}g^{(22)}=\mp\frac{1}{\sqrt{3}}{}_{\pm1}g^{(22)}=\sqrt{\frac{2}{3}}{}_{\pm2}g^{(22)}=1.
\ee

\section{Normalization}\label{SecNorm}

In this section we are going to show that expression \eqref{gtjm} is
the correct one to enforce the normalization condition \eqref{AlphaChiZero} in all cases. We will
also discuss the overall normalization of the tensor harmonics in real space.

\subsection{Normalization at origin}

Following the same algorithm as the one described in Section \ref{SecAlgo},
one can show that the radial functions scale like $\chi^{|\ell-j|}$
when $\chi \to 0$. Indeed, since $\Phi^\nu_\ell\sim\chi^\ell$ in this limit [see~\eqref{PowerLawOriginBessel}], we find from \eqref{solutionjmm} that $_0\epsilon^{(j,\pm j)}\sim\chi^{\ell-j}$. One can then check that for the various steps of the algorithm which
increase $j$ and $s$, this property is maintained. In practice, to show that
such scaling holds one needs to distinguish the cases $\ell>j$, $\ell=j$ and $\ell
< j$ when applying the algorithm. The constant in the scaling can then
be determined by $\ell-j$ iterations of \eqref{Masteralpha} (when keeping only the dominant term as $\chi \to 0$), and using \eqref{AlphaChiZero}. For  $\ell \geq j$ we find for $\chi \to 0$
\be
{}_s \alpha_\ell^{(jm)} \sim \frac{\chi^{\ell-j}  (2j-1)!!}{(\ell-j)! (2\ell+1)!!} \prod_{p=j+1}^\ell {}_s \kappa_p^m\,.
\ee
In the case $\ell \leq j$, it behaves as
\be
{}_s \alpha_\ell^{(jm)} \sim \frac{(-\chi)^{j-\ell}  (2\ell-1)!!}{(j-\ell)! (2j+1)!!} \prod_{p=\ell+1}^j {}_s \kappa_p^m\,.
\ee
This is consistent with $\ell-j$ iterations of~\eqref{Masteralpha}
(when keeping only the dominant term as $\chi \to 0$). This finally proves that \eqref{gtjm} is
the correct expression needed to enforce \eqref{AlphaChiZero}.

If we now use these results together with \eqref{NormalisationCalYContracted}, we find that the
normalization of harmonics at the origin is 
\be\label{Qatorigim}
\left.{}^{\ell=j} Q_{I_j}^{(jm)}\right|_{\chi=0} \left.{}^{\ell=j}
  Q_{I_j}^{(jm')\star}\right|_{\chi=0} = \delta_{mm'} {\cal N}_j^m
\ee
where
\bea\label{DefNjm}
{\cal N}_j^m&\equiv &({}_0 \tilde g^{(jm)})^2  \frac{(2j-1)!!}{j!}\nonumber\\
&\equiv&({}_0 g^{(jm)})^2 \frac{j!}{(2j-1)!!}\,,
\eea
and where the following contraction of indices $I_j$ was used on the left hand side of~\eqref{Qatorigim}:
\bea\label{gsjlidentities}
{}_s g^{(jm)} \hat n^s_{I_j} \,\,{}_s g^{(jm')} \hat n_s^{I_j \star}&=&
{}_s g^{(jm)} \, {}_s g^{(jm')} d_{js}\nonumber\\
&\equiv& {\cal N}_{j}^{mm'}\,,
\eea
with ${\cal N}_j^{mm} = {\cal N}_j^m$. Together with \eqref{Qljm}, the expressions above allow us to write
the contraction of harmonics at an arbitrary point. Restoring the dependence with $\nu$ of the harmonics and normal modes,
we find
\bea\label{MagicQQGG}
&&{}^{\ell}Q_{I_j}^{(jm)}(\nu) \,{}^{\ell'}
Q_{I_j}^{(jm')\star}(\nu')\\
&&\qquad={\cal N}_j^{m m'}\sum_{s=-j}^j {}_s
G_\ell^{(jm)}(\nu) \,\, {}_sG_{\ell'}^{(jm')\star}(\nu')\,.\nonumber
\eea

\subsection{Integral on space}

We checked by means of integrations by parts (on the lowest $|m|$ values)
and using \eqref{EquadiffBessel} and \eqref{Normintchi} that the harmonics satisfy the
normalization  (with again $I_j$ indices contracted)
\bea\label{NormQ}
&&\int \dd^3 V\, {}^{\ell M} Q^{(jm)}_{I_j}(\nu) \,\,
{}^{\ell' M'} Q^{(j m')
  \star}_{I_j}(\nu')= \\
&&\delta_{\ell\ell'}\delta_{m m'} \delta_{M M'} {\cal N}_{j}^{m} (2\pi)^3\frac{(2\ell+1)}{4\pi} \frac{\delta(\nu-\nu')}{(\nu^2-\barK m^2)} \nonumber
\eea
where  $\dd^3 V \equiv r^2(\chi) \dd \chi \sin \theta \dd \theta \dd
\phi$. 
The case $j>|m|$ is deduced from the case $j=|m|$ through the
construction \eqref{STFrecursion} using repeated application of
\eqref{GeneralDiv} when integrating by parts. A similar method can
be used to infer the vanishing of \eqref{NormQ} when $m \neq m'$.  Note also
that in the closed case, $\nu$ takes integer values, hence the Dirac
delta function must be understood as a Kronecker symbol instead, that
is we must read \eqref{NormQ} with $\delta(\nu-\nu') \to \delta_{\nu
  \nu'}$.

From the general definition \eqref{Qljmbis} and the identities \eqref{OrthoYslm}
and \eqref{gsjlidentities}, it follows that the normalization \eqref{NormQ} is equivalent to 
\begin{align}\label{Normalpha1}
&\sum_{s=-j}^j\int {}_s \alpha_\ell^{(jm)}(\chi;\nu)\,\, {}_s
\alpha_\ell^{(jm')\star}(\chi;\nu')  r^2(\chi) \dd
\chi\nonumber\\
&\quad=\delta_{m m'} \frac{\pi}{2}\frac{\delta(\nu-\nu')}{(\nu^2-\barK m^2)}\,.
\end{align}
This is a generalization of the normalization relation \eqref{Normintchi} which
corresponds only to scalar modes (see e.g. \cite{Lyth1995}).

Conversely, for a given $j$, a closure relation can be also
formulated, showing that we have built a complete set of basis
functions. We find 
\begin{widetext}
\bea\label{MagicClosureQ}
&&\sum_{\ell= 0}^{\infty} \sum_{M=-\ell}^\ell
\sum_{m=-j}^j\frac{4\pi}{(2\ell+1){\cal N}_j^m}\sumint  {}^{\ell M}
Q^{(jm)}_{I_j}(\chi,\gr{n};\nu)  {}^{\ell M}
Q^{(jm)\star}_{K_j}(\chi',\gr{n}';\nu) \frac{ (\nu^2 -\barK m^2) \dd
  \nu}{(2\pi)^3} \nonumber\\
&&\qquad\qquad \qquad=\frac{\delta(\chi-\chi')}{r^2(\chi)}\delta^2(\gr{n}-\gr{n}')\delta_{i_1}^{\langle k_1} \dots
\delta_{i_j}^{k_j \rangle }\,.
\eea
\end{widetext}
For open cases, the integral runs on $\nu\geq 0$, whereas in the
closed case the integrals must be understood as a discrete sum on
integer values such that $\nu \geq \ell+1$. Again, this is equivalent to
the closure relation for radial functions (which we checked for the
lowest values of $j$)
\begin{align}\label{Normalpha2}
&\sum_{m=-j}^j \sumint {}_s\alpha_\ell^{(jm)}(\chi;\nu) {}_{s'}
\alpha_\ell^{(jm)\star}(\chi';\nu)  (\nu^2-\barK m^2) \dd \nu\nonumber\\
&\qquad = \frac{\pi}{2}\delta_{s s'}\frac{\delta(\chi-\chi')}{r^2(\chi)}\,,
\end{align}
with the same convention that it is a discrete sum on $\nu \geq
\ell+1$ in the closed case. This is a generalization of the closure relation \eqref{Normintnu}
which corresponds only to scalar modes \cite{Lyth1995}. One also verifies immediately that \eqref{MagicClosureQ} is compatible
with \eqref{NormQ}, since multiplying the former by ${}^{\ell' M'}   Q^{(j m')  \star}_{I_j}(\chi,\gr{n};\nu')$ and integrating over space using the
latter, yields ${}^{\ell' M'}  Q^{(j m')\star}_{I_j}(\chi',\gr{n}';\nu')$, exactly as the r.h.s. of \eqref{MagicClosureQ} indicates. Similarly \eqref{Normalpha2} is
obviously compatible with \eqref{Normalpha1}.

\section{Plane waves}\label{SecPlaneWaves}

In flat space, a plane wave is an eigenfunction of the Laplacian which assumes constant values on planes orthogonal to a constant wavevector $\gr{k}$. For scalar functions it is simply $\exp(\ii \gr{k}\cdot\gr{x})$. This idea cannot be generalized to the curved spaces since the notion of a globally constant vector does not exist. We will nonetheless seek to build eigenfunctions of the Laplacian in curved spaces --- which we shall abusively call plane waves --- which look like flat space plane waves near the origin of coordinate system, i.e., over distances much smaller than the curvature radius.

\subsection{Zenith axis plane waves}

Plane waves are defined by summation of harmonics with different values of $\ell$. 
The most general summation on $\ell$ is of the form (restoring the explicit dependence on $\nu$)
\be\label{DefQfromsuml}
Q^{(jm)}_{I_j}(\nu) \equiv \sum_{\ell\geq |m|}^{\infty,\nu-1} \frac{\myzeta_\ell^m}{\myzeta_j^m} \,\,\,{}^\ell Q^{(jm)}_{I_j}(\nu) 
\ee
and similarly
\bea\label{GjmfromGljm}
{}_s G^{(jm)}(\nu)&\equiv& \sum_{\ell \geq (|m|,|s|)}^{\infty,\nu-1} 
\frac{\myzeta_\ell^m}{\myzeta_j^m}\,\,\,{}_s G^{(jm)}_{\ell}(\nu)	, \\
&=&  \sum_{\ell \geq(|m|,|s|)}^\infty\frac{\myzeta_\ell^m}{\myzeta_j^m}\, c_\ell\, {}_s\alpha^{(jm)}_\ell(\nu) {}_s Y_{\ell m}({\gr{n}}),\nonumber
\eea
such that
\be\label{QtogGns}
Q_{I_j}^{(jm)}(\nu) = \sum_{s=-j}^j {}_s g^{(jm)}  {}_s G^{(jm)}(\nu) \hat n^s_{I_j}\,.
\ee
The previous sums on $\ell$ run until infinity in the flat or open
case, and are limited by \eqref{RestrictionClosed2} in the closed
case. The weights $\myzeta_\ell^m$ are undetermined coefficients. From these definitions we recover, near the origin, the same behavior as in \eqref{NormG}. What we hereafter call \emph{plane waves} corresponds to the choice $\myzeta_\ell^m={\rm const.}$ (or $\myzeta_\ell^m=1$). By contrast, we name {\it pseudo plane waves} the more general case $\myzeta_\ell^m \neq {\rm const}$. A pseudo plane is thus specified both by the mode $\nu$ and by the set of
$\myzeta_\ell^m$: $Q_{I_j}^{(jm)}(\nu,\myzeta_\ell^m)$.

We chose to divide by $\myzeta_j^m$ in the definitions \eqref{DefQfromsuml}
and \eqref{GjmfromGljm} so as to maintain the normalization at origin \eqref{QToCurlyY}.
All recursion relations that were derived so far for the radial
functions in the $(j,s)$ space are, in fact, also valid for the ${}_s
G_\ell^{(jm)}$, since the coefficients of all recursions are totally
independent from $\ell$. Hence all of these recursive relations are
transposed as relations among the summed normal modes ${}_s G^{(jm)}
\myzeta_j^m$. This can be traced back to the fact that the general relation
\eqref{GenRelHarm}, from which all recursive relations originate, is satisfied for the tensors $\myzeta_j^m Q_{I_j}^{(jm)}$.

\subsection{General axis plane waves}\label{GenAxisPW}

In the previous section, we have summed the harmonics \eqref{Qljm} on
$\ell$. They correspond to a special choice where the direction used to decompose the local
structure is also the zenith direction. Hence the plane waves (or
pseudo plane waves) built in Section \ref{DefQfromsuml} correspond to a wave vector $\gr{\nu} =
\nu\gr{e}_z$. As detailed in Section \ref{SecRotation}, one can consider a general direction $\hat{\gr{\nu}}$ with the associated wave vector
$\gr{\nu} = \nu \hat{\gr{\nu}}$. The associated plane waves are built in general as
\begin{align}\label{DefQfromsumlM}
&Q^{(jm)}_{I_j}(\gr{\nu}) \equiv R_{\hat{\gr{\nu}}} [Q^{(jm)}_{I_j}(\nu) ]\\
&\qquad =\sum_{\ell\ \geq |m|}^{\infty,\nu-1} \sum_{M=-\ell}^\ell \frac{\myzeta_\ell^m}{\myzeta_{j}^{m}}
\,\,\,{}^{\ell M} Q^{(jm)}_{I_j}(\nu) D^\ell_{M m}(R_{\hat{\gr{\nu}}})\,.\nonumber
\end{align}
This form is very similar to the general decomposition of a
STF tensor field \eqref{GeneralYslmTensor} since it can also be written
more explicitly as
\bea\label{QrotatedExplicit}
Q^{(jm)}_{I_j}(\gr{\nu}) &\equiv & \sum_{\ell M s}
\frac{\myzeta_\ell^m}{\myzeta_{j}^{m}} \,c_\ell \,{}_s g^{(jm)} {}_s\alpha_\ell^{(jm)}(\chi;\nu)\nonumber\\
&&\,\,\,  \times D^\ell_{M m}(R_{\hat{\gr{\nu}}})\, {}_s Y_\ell^M ({\gr{n}}) \hat n^s_{I_j}\,.
\eea
The normal modes associated with these general axis plane waves are
\bea\label{GPW2}
{}_s{G}^{(jm)}(\gr{\nu}) &=& \sum_{\ell\geq(|m|,|s|)}^{\infty,\nu-1} \sum_{M=-\ell}^\ell  \frac{\myzeta_\ell^m}{\myzeta_j^m}\,c_\ell
\,{}_s\alpha^{(jm)}_\ell(\chi,\nu)\nonumber\\
&&\times D^\ell_{M m}(R_{\hat{\gr{\nu}}})  {}_s Y_{\ell}^{M}({\gr{n}})\,,
\eea
and we get a relation of the type \eqref{QtogGns} for the mode $\gr{\nu}$.

\subsection{Extended Rayleigh expansion}

Eq.~\eqref{QrotatedExplicit}, possibly reshaped using \eqref{RotationYslm}, is the generalized Rayleigh expansion for tensor-valued plane
waves. In the flat case, for standard plane waves (i.e., $\myzeta_\ell^m={\rm const.}$) with $j=0=m$, we recover the usual Rayleigh expansion given 
by \eqref{Rayleigh}. 

We can recast Eq.~\eqref{QrotatedExplicit} in a more covariant form as
\begin{align}\label{MasterRayleighQ}
&\hat n_{\mp s}^{I_j}(\gr{n})   Q_{I_j}^{(jm)} (\chi,\gr{n};\gr{\nu}) =\sum_{\ell \geq(|m|,|s|)}^{\infty,\nu-1}  \frac{\myzeta_\ell^m}{\myzeta_j^m} (2\ell+1)\nonumber\\
  &\frac{{}_{\pm s} \tilde   g^{(jm)} }{ {}_{\pm s} \tilde g^{(\ell m)} } \, {}_{\pm s}\alpha_\ell^{(jm)}(\chi;\nu) \,\hat n_{\mp  s}^{I_\ell}(\gr{n})   Q_{I_\ell}^{(\ell m)} (\chi=0;\gr{\nu}) \,.
\end{align}
\cpsilent{This extends Eq.~(4.13) of \cite{Challinor:1999xz} or Eq.~(4.1.47)
of \cite{Tsagas:2007yx}, which are restricted to the case $j=|m|$ and $s=0$.} Eq.~\eqref{MasterRayleighQ} can be understood essentially as
a simple Taylor expansion, since derived harmonics are precisely made of derivatives of the base harmonic. The generalized
Rayleigh expansion is essential for the computation of cosmological
observables, and we illustrate its use in Section \ref{SecOtherCosmo} and in Ref.~\cite{Duality}.

\subsection{Parity and conjugation}

The transformation properties of section \ref{SecConjugationParity}
can be extended to pseudo plane waves. For conjugation, we find
\bea\label{MagicConjugate2}
&&\left[ Q_{I_j}^{(jm)}(\nu,\myzeta_\ell^{m})\right]^\star\\
&&\quad=(-1)^{(j+m)} \times Q_{I_j}^{(j,-m)}(-\nu^\star,(-1)^\ell \myzeta_\ell^{m\star}) \nonumber\,.
\eea
For $\pi$-rotation around axis $y$, we get
\be\label{MagicComplex2}
R_y(\pi)[Q^{(jm)}_{I_j}(\nu,\myzeta_\ell^m)] = \left[Q^{(jm)}_{I_j}(\nu^\star,\myzeta_\ell^{m\star})\right]^\star\,.
\ee
Finally for parity transformations, we obtain
\begin{align}\label{MagicParity2}
&{\cal P}[Q^{(jm)}_{I_j}(\nu,\myzeta_\ell^m)] = (-1)^m
\left[Q^{(j,-m)}_{I_j}(\nu^\star,\myzeta_\ell^{m\star})\right]^\star\nonumber\\
&\qquad\qquad=(-1)^j Q^{(jm)}_{I_j}(-\nu,(-1)^\ell\myzeta_\ell^{m})\,.
\end{align}
The combination of parity and rotation (which amounts to an inversion
of the $y$-axis) is similar to \eqref{RP} and reads
\be
R_y(\pi)[{\cal P}[Q^{(jm)}_{I_j}(\nu,\myzeta_\ell^m)] ]= (-1)^m  Q^{(j,-m)}_{I_j}(\nu,\myzeta_\ell^m).
\ee
The rules for an inversion of a single axis are thus exactly the same
as in Table \ref{TableParity} for individual ${}^\ell Q_{I_j}^{(jm)}$,
except that the factor $(-1)^\ell$ manifests itself as $\myzeta_\ell^m
\to (-1)^\ell \myzeta_\ell^m$ along with a global $(-1)^j$ factor. 

\subsection{Orthogonality}

As for the special case $\hat{\gr{\nu}} = \gr{e}_z$, the plane waves (when $\myzeta_\ell^m={\rm const.}$, which we now assume) are orthogonal as we now review.
Replacing \eqref{DefQfromsumlM} in \eqref{NormQ}, and using \eqref{ClosureYslm}
and \eqref{YslmWigner}, we find that in the open or flat case, the plane waves are normalized according to
\bea\label{NormQPlane}
&&\int \dd^3 V\, Q^{(jm)}_{I_j}(\gr{\nu}) \,Q^{(j m')\star}_{I_j}(\gr{\nu}') = \delta_{m m'}(2\pi)^3\nonumber\\
&&\qquad\times{\cal N}_{j}^{m} \frac{\nu^2}{(\nu^2-\barK m^2)}\delta^3(\gr{\nu}-\gr{\nu}') \,.
\eea
Therefore, we conclude that in the open case the plane waves that we defined have
orthogonality properties very similar to the flat case plane waves, thus
justifying our abusive terminology.

However, in the closed case ($\barK=1$), the sum on $\ell$ in \eqref{DefQfromsumlM} does not extend to infinity, and
one cannot rely on \eqref{ClosureYslm}. We find instead 
\begin{align}\label{NormQPlane2}
&\int \dd^3 V\, Q^{(jm)}_{I_j}(\gr{\nu}) \, Q^{(j m')
  \star}_{I_j}(\gr{\nu}') = \delta_{\nu \nu'}\delta_{m m'}2\pi^2 {\cal N}_{j}^{m} \nonumber\\
&\qquad\times\sum_{\ell=|m|}^{\nu-1}
\frac{(2\ell+1)}{(\nu^2-\barK
  m^2)}D^\ell_{m m}(R^{-1}_{\hat{\gr{\nu}}'}  R_{\hat{\gr{\nu}}})\,.
\end{align}
In the case where the modes have the same direction  ($\hat{\gr{\nu}} = \hat{\gr{\nu}} '$), and using
\be
\sum_{\ell=|m|}^{\nu-1} (2\ell+1) = \nu^2-m^2\,,
\ee
this reduces to 
\begin{align}\label{ClosureClosedsameNudirection}
&\int \dd^3 V\, Q^{(jm)}_{I_j}(\nu\hat{\gr{\nu}}) \,Q^{(j
  m')\star}_{I_j}(\nu' \hat{\gr{\nu}})\nonumber\\
  &\qquad =2\pi^2 \delta_{\nu \nu'} \delta_{m m'} {\cal N}_{j}^{m}.
\end{align}
Eq. \eqref{NormQPlane2} shows that the plane waves as built in
\eqref{DefQfromsumlM} are not properly orthogonal in the closed
case. In that case one should work directly with ${}^{\ell
  M}Q^{(jm)}_{I_j}(\nu)$, which according to \eqref{NormQ} and
\eqref{MagicClosureQ} is a proper orthogonal basis.

In all cases, the closure relation for plane waves reads
\begin{widetext}
\be\label{MagicClosureQ2}
\sum_{m=-j}^j \int  Q^{(jm)}_{I_j}(\chi,\gr{n};\gr{\nu}) \, 
Q^{(jm)\star}_{K_j}(\chi',\gr{n}';\gr{\nu}) \frac{ (\nu^2 -\barK m^2) \dd
  \nu \dd^2 \hat{\gr{\nu}}}{(2\pi)^3 {\cal N}_j^m} =\frac{\delta(\chi-\chi')}{r^2(\chi)}\delta^2(\gr{n}-\gr{n}')\delta_{i_1}^{\langle k_1} \dots
\delta_{i_j}^{k_j \rangle }\,,
\ee
\end{widetext}
with the convention that it is a discrete sum on $\nu \geq |m|+1$
in the closed case. This relation is found from the definition \eqref{DefQfromsumlM} with \eqref{YslmWigner} to express the Wigner
 $D$-coefficients, and using the orthogonality relation \eqref{OrthoYslm} to
 handle the angular integration on $\hat{\gr{\nu}}$ so as to fall back
 onto \eqref{MagicClosureQ}. The closure \eqref{MagicClosureQ2} is
 obviously compatible with \eqref{NormQPlane} in the open of flat
 cases, and is also compatible with \eqref{NormQPlane2} even though
 there is no Dirac function on the directions of $\hat{\gr{\nu}}$ and
 $\hat{\gr{\nu}}'$ in its right hand side. This is because the factor
 $\sum_{\ell=|m|}^{\nu-1}(2\ell+1)D^\ell_{m m}(R^{-1}_{\hat{\gr{\nu}}'}
 R_{\hat{\gr{\nu}}})$ effectively plays the role of a Dirac delta when acting on
 functions with an angular structure limited to $\ell \leq \nu-1$, and
this is exactly the case for the dependence on the mode direction
$\hat{\gr{\nu}}$ of closed space plane waves as defined by \eqref{DefQfromsumlM}.

\subsection{Integral on directions}

For plane waves, and for the lowest normal modes [specifically, we checked for $j$ up to $4$, and for all allowed $m$ and $s$] we have checked that 
the following identity holds:
\bea\label{MagicC8}
&&\int {}_s
G^{(jm)}(\chi,\gr{n};\gr{\nu}) \, {}_s G^{(j'm')\star}(\chi,\gr{n};\gr{\nu})\, \dd^2 \gr{n}\nonumber\\
&&= \delta_{m m'}\delta_{j j'}\sum_\ell |c_\ell \, {}_s \alpha^{(jm)}_\ell(\chi;\nu)|^2,\nonumber\\
&&=\delta_{m m'}\delta_{j j'}\frac{4\pi}{(2j+1)} \,,
\eea
in agreement with equation C8 of \cite{TAM2}. Hence the normalization
of plane waves is such that the dependence on $\chi$,  which is there in
principle at the second line, disappears at the third line. 
In the particular case of $j=m=s=0$, and in the flat case, this relation is
proven using an addition theorem of spherical Bessel function (e.g. Eq. (A.12) of \cite{Reimberg:2015jma}).

Moreover, using properties \eqref{MagicQQGG} and \eqref{MagicC8}, we find for the plane waves harmonics \eqref{DefQfromsuml} that
\bea\label{AverageAngle}
&&\int \frac{\dd^2 \gr{n}}{4\pi} Q_{I_j}^{(jm)}(\chi,\gr{n};\gr{\nu})
Q_{I_j}^{(jm')\star}(\chi,\gr{n};\gr{\nu})\nonumber\\
&&\qquad= \delta_{m m'}{\cal N}_j^{m}\,.
\eea
If we further integrate \eqref{AverageAngle} on the measure $4\pi r^2(\chi) \dd\chi$ to complete an integration on the whole volume, we check in the closed case 
that it leads again to \eqref{ClosureClosedsameNudirection} with $\nu=\nu'$, since in the closed case $\int \dd^3 V = \int_0^\pi 4\pi r^2(\chi) \dd\chi
= 2\pi^2$.

\subsection{Discussion on general factorization of normal modes}

It is argued in Appendix C of \cite{TAM2}, and this point is recalled
in Eq.~(1.15) of \cite{Tram:2013ima} and Eq. (A9.3) of \cite{Durrer:2008eom}, that
the normal modes can be factorized in a form which separates clearly
the intrinsic angular dependence and the orbital one. Restricting the
discussion to modes $\gr{\nu} = \nu \gr{e}_z$ for simplicity, and
omitting the explicit dependence on $\nu$ on all functions, this
factorization should be of the form
\be\label{WrongFactor}
{}_s G^{(jm)}(\chi,\gr{n}) \overset{?}{=} \frac{c_j}{2j+1} {}_s Y_{j}^m(\gr{n})
F(\nu,\chi,\gr{n})
\ee
with a universal orbital function $F$ such that~\footnote{This function is often written as $e^{{\rm i}\delta(\vec{x},\vec{k})}$.}
\be\label{FunitNorm}
|F(\nu,\chi,\gr{n})|=1\,.
\ee
Translated to the plane wave harmonics using \eqref{Qljm}, this is
\be\label{WrongFactor2}
Q^{(jm)}_{I_j}\overset{?}{=} {}_0 \tilde g^{(jm)}\,\frac{c_j}{2j+1}\, {\cal Y}^{jm}_{I_j} \,F(\nu,\chi,\gr{n})\,,
\ee
with the ${\cal Y}^{jm}_{I_j}$ defined everywhere following the remark after \eqref{MagicSumYslm}.

Expressions~\eqref{WrongFactor} and~\eqref{WrongFactor2} are reminiscent of what is found in the flat case [Eqs.~\eqref{FactorFlat} and \eqref{FactorFlat2}], where the orbital function 
is a pure scalar plane-wave $\exp [\ii (k \gr{e}_z) \cdot (\chi\gr{n})]$. If this was
the case in the curved case, there would be a clear separation between orbital and angular momentum. Furthermore, the
property \eqref{MagicC8} (which is correct) would be a trivial
consequence of \eqref{OrthoYslm}. We argue in this section why this is not possible in the general
curved case, and that the factorization~\eqref{WrongFactor} does not exist. However, we insist that property \eqref{MagicC8} is still
correct since it does not imply the factorization property \eqref{WrongFactor}.

From the $j=m=s=0$ case, one infers immediately that the universal orbital function must be $F(\nu,\chi,\gr{n}) = {}_0 G^{(00)}$. 
In the flat case, $F=e^{{\rm i}\boldsymbol{k}\cdot\boldsymbol{x}}$, and since by construction we have 
\be
 {}_0 G^{(00)} = \sum_\ell \sqrt{4\pi(2\ell+1)} \ii^\ell j_\ell(\chi) Y_\ell^0 (\gr{n})
\ee
we could be tempted to deduce that the radial functions can be built exactly like in the
flat case, i.e. is using \eqref{DefGaunt}, but with the replacement $j_\ell (k r) \to
\Phi_\ell^\nu(\chi)$. In the flat case, the (usual) spherical Bessel functions can be combined by means of \eqref{DefGaunt} and 
\eqref{Recjl}, which then leads to the expressions listed in Appendix \ref{AppFlat}. If we insist on the idea of using the same combinations 
in the curved case, but with $\Phi^\nu_\ell$ in place of the usual $j_\ell$, we must also use the relations \eqref{RecurrenceBessel}. But note that these differ 
from \eqref{Recjl} by factors like $\sqrt{\nu^2 -\barK \ell^2}$ and $\sqrt{\nu^2 -\barK (\ell+1)^2}$. Thus, the results 
obtained with this method are not the radial functions reported in Section \ref{SecGatherRadial}. To be specific, let us attempt to build the 
radial function ${}_0 \alpha_\ell^{(10)}={}_0 \epsilon_\ell^{(10)}$ from the factorization \eqref{WrongFactor}. Starting from
\be
{}_0 \epsilon_\ell^{(10)}  \overset{?}{=} \frac{1}{2\ell+1}\left[\ell \Phi_{\ell-1}^{\nu}-(\ell+1)\Phi_{\ell+1}^{\nu}\right],
\ee
and using \eqref{RecurrenceBessel}, the radial function takes the form
\be
{}_0 \epsilon_\ell^{(10)}  \overset{?}{=}x_\ell \frac{\dd}{\dd
  \chi}\Phi^\nu_\ell + y_\ell \gcot{\chi}\Phi_\ell^{\nu}\,,
\ee
where the coefficients are
\bea
x_\ell &=&\frac{1}{2\ell+1}\left(\frac{\ell}{\sqrt{\nu^2 -\barK
      \ell^2}}+ \frac{\ell+1}{\sqrt{\nu^2 -\barK (\ell+1)^2}}\right)\nonumber \\
y_\ell &=& \frac{\ell(\ell+1)}{2\ell+1}\left(\frac{1}{\sqrt{\nu^2 -\barK
      \ell^2}} -  \frac{1}{\sqrt{\nu^2 -\barK (\ell+1)^2}}\right) \nonumber
\eea
and this differs from the correct expression \eqref{epszuz} since obviously $x_\ell
\neq 1/\sqrt{\nu^2 -\barK}$ and $y_\ell \neq 0$.

Another way to show that \eqref{WrongFactor} does not apply in the
curved case consists in exhibiting counterexamples. In the closed
case, the sum on $\ell$ in \eqref{GjmfromGljm} (with
$\myzeta_\ell^m={\rm const.}$) to form plane waves is a finite sum
since $0 \leq \ell \leq \nu-1$. Let us first consider the case $j=0$, $s=0$, $m=0$. If $\nu=1$, then we have only $\ell=0$ and ${}_0
\alpha^{(00)}_0(\chi;\nu=1)= F(\nu=1,\chi,\gr{n}) =1$ and there is no issue. However as soon as we
consider $\nu=2$, we have ${}_0
\alpha^{(00)}_0(\chi;\nu=2) = \cos(\chi) $ and ${}_0 \alpha^{(00)}_1(\chi;\nu=2) =
\sin(\chi)/\sqrt{3}$, and it is found that the orbital
function must be
\be\label{CounterExample1}
F(\nu=2,\chi,\gr{n}) = \cos(\chi) + \ii \sqrt{3} \cos{\theta} \sin(\chi)\,.
\ee
Hence the unit norm condition~\eqref{FunitNorm} is
not met. Of course when $\chi \ll 1$, that is for distances much smaller than the curvature scale, the norm tends to
unity, thus recovering the flat case result. Note however that
\eqref{MagicC8} still holds since $|c_0 \,{}_0 \alpha^{(00)}_0(\nu=2)|^2 =
4\pi\cos^2\chi$ and $|c_1\, {}_0 \alpha^{(00)}_1(\nu=2)|^2  = 4\pi
\sin^2\chi$. One could try to release this unit norm condition and still look for a universal orbital function. However, for $\nu=2$
but for the values $(s=0,j=m=1)$, one infers
\be
F(\nu=2,\chi,\gr{n}) = 1\,,
\ee
which is not equal to \eqref{CounterExample1}. Similarly for
$(j=1,m=s=0)$ one infers yet another orbital function, being $F(\nu=2) = \cos \chi +\ii/\sqrt{3} \sin
\chi \sec \theta$.
To conclude, not only the orbital function cannot be of unit norm, but
also it cannot be universal, that is in practice it cannot depend only on
$\nu$ and on the position in space $(\chi,\gr{n})$. At best,
\eqref{WrongFactor} can be used for a definition of $F$ for each set
of $(s,j,m)$, that is to define ${}_s F^{(jm)}$ orbital functions. In
the open case, one can also check numerically (because of the infinite sum
in $\ell$) that the orbital function cannot be of unit norm and cannot
be universal.

We thus conclude that the general factorization \eqref{WrongFactor} does not exist, and we can rely on the explicit summation
\eqref{DefGaunt} only in the flat case. In the curved case, one must determine the radial modes following the method 
used in this article (or a related one). The impossibility of the factorizations \eqref{WrongFactor} and \eqref{WrongFactor2} in the curved cases is related to  the fact that the norm squared of plane waves is not a constant, and only its average over spheres yields the constant ${\cal N}_j^m$
(independent on $\chi$) as seen on \eqref{AverageAngle}. This is different from the flat case where it is
obvious from \eqref{FactorFlat2} that the square of the norm of plane waves
is ${\cal N}_j^m$ everywhere. It is important to stress, however, that while
\eqref{WrongFactor} does not exist, its use in Ref.~\cite{TAM2} was
meant only as a heuristic motivation, and all the results are of course correct since they rely essentially
only on the property \eqref{MasterGstreaming}.

\section{Cosmological applications}\label{SecCMB}

We are now in position to discuss some physical applications of the formalism developed so far. 
We will focus on the derivation of the Boltzmann hierarchy to describe the
evolution of CMB, a key cosmological observable, following both the pioneering work of \cite{TAM2}
based on normal modes, and the approach built in \cite{Challinor:1999xz,Challinor:2000as} using STF tensors.
In the next Section we introduce harmonics and normal modes which are adapted to the
use of the propagating direction (of photons) rather than the
observing direction. In Section~\ref{SecMultDec} we summarize the angular decomposition
for CMB temperature and linear polarization. We then review the standard derivation of the Boltzmann hierarchy providing only minimal
ingredients of cosmology in Section~\ref{SecBoltzmann1}. In Section~\ref{SecOtherCosmo} we address the general method for extracting the 
multipolar decomposition of all cosmological observables when using plane wave harmonics. When using pseudo plane waves 
($\myzeta_\ell^m\neq {\rm const.}$) instead of standard plane waves
($\myzeta_\ell^m={\rm const.}$), the results are slightly different,
as detailed in \cite{Duality}.

\subsection{Relation to propagation normal modes}\label{AppPropagation}

In the context of CMB, it is often more convenient to rewrite
everything in terms of the propagating direction of photon, rather
than the observed direction of the incoming photon. Hence, let us define the propagation direction as the opposite of the
observed direction:
\be\label{barn}
\bar{\gr{n}} \equiv - \gr{n}\,.
\ee
The helicity bases associated with a direction and its opposite are
related through \eqref{Inversionns}. A given point on the manifold is either denoted by the pair
$(\chi,\gr{n})$ or the pair $(\chi,-\bar{\gr{n}})$. Plane-wave harmonics in the propagation direction are linked
with those built so far as
\be\label{DefQbar}
\overline{Q}^{(jm)}_{I_j}(\chi,-\bar{\gr{n}}) \equiv (-1)^j
\times 
Q^{(jm)}_{I_j}(\chi,\gr{n})\,.
\ee
In order to be consistent with the construction of derived harmonics we must use, instead of \eqref{STFrecursion}, the defining property
\be\label{STFrecursion2}
\overline{Q}^{(jm)}_{I_j} = -\frac{1}{k} \nabla_{\langle i_j} \overline{Q}^{(j-1,m)}_{I_{j-1}\rangle }\,.
\ee
The expansion of these new harmonics in terms of the associated normal modes is
\be\label{QG1}
\overline{Q}^{(jm)}_{I_j}(\chi,-\bar{\gr{n}}) \equiv
\sum_{s=-j}^j {}_s
g^{(jm)} {}_s \overline{G}^{(jm)}(\chi,\bar{\gr{n}}) {\hat n}^s_{I_j}(\bar{\gr{n}})\,,
\ee
whereas we recall that the harmonics built with observation directions
are expanded as \eqref{QtogGns}. The normal modes associated with plane waves are expanded in radial
functions in a similar fashion to \eqref{GjmfromGljm}, as
\bea\label{barGG}
{}_s\overline{G}^{(jm)}(\chi,\bar{\gr{n}})  = \sum_{\ell\geq |m|} \bar c_\ell\, {}_s
\bar\alpha^{(jm)}_\ell {}_s Y_{\ell m}(\bar{\gr{n}})
\eea
where
\be\label{Defbarc}
\bar c_\ell \equiv (-1)^\ell c_\ell = (-\ii)^\ell \sqrt{4\pi(2\ell+1)}\,.
\ee
Using \eqref{barGG} and \eqref{GjmfromGljm} for plane waves in \eqref{QG1} and \eqref{QtogGns}, so as to replace
in the definition \eqref{DefQbar}, we deduce from the properties
\eqref{Inversionns} and \eqref{InversionYslm} that the new radial
functions are related to the ones built with observed directions by
\be\label{Defalphabar}
 {}_s\bar\alpha^{(jm)}_\ell(\nu) =  {}_{-s}\alpha^{(jm)}_\ell(\nu)= {}_{s}\alpha^{(jm)}_\ell(-\nu)  \,.
\ee
From the decomposition \eqref{defeb} in even and odd components, we deduce
that the ones built when using propagation directions, are related
to those built using observation directions, by
\beas\label{ebpropvsebobs}
{}_s \bar \epsilon_\ell^{(jm)}(\nu) = {}_s \epsilon_\ell^{(jm)}(\nu) \\
{}_s \bar \beta_\ell^{(jm)}(\nu) =- {}_s \beta_\ell^{(jm)}(\nu) \,.
\eeas

To summarize, when using propagation direction, as is common in the
context of CMB, one needs only to add a factor $(-1)^\ell$ to
$c_\ell$, factors of $(-1)^j$ in the definition of the harmonics, and then
the radial functions are exactly the same up to a global sign for the
magnetic (odd) radial functions. Equivalently, one can use the same
radial functions but with $-\nu$ instead of $\nu$. In fact this is just the parity transformation rule \eqref{MagicParity2} since ${\cal
  P}[Q^{(jm)}_{I_j}(\gr{n}) ]= \overline{Q}^{(jm)}_{I_j}(\bar{\gr{n}}) $.

\subsection{CMB multipole decomposition}\label{SecMultDec}

At each cosmological time, the temperature fluctuation field $\Theta$ depends both on
the position in space, that is on $(\chi,\gr{n})$, and on the
propagating direction $\bar{\gr{n}}^i$ [which does not necessarily satisfy \eqref{barn}]. This dependence can be separated using a multipolar decomposition as
\be\label{Thetanprop}
\Theta= \sum_j \Theta_{i_1\dots  i_j} \bar n^{i_1}\dots \bar n^{i_j}\,,
\ee
where the STF multipoles $\Theta_{I_j}$ depend only on $(\chi,\gr{n})$ and on time. 
However, as argued in \cite{TAM1,TAM2}, a shortcut consists in fixing the propagating direction, 
\be\label{Slaven}
\bar n^i=-n^i, 
\ee
when solving for the observed CMB. This is equivalent to consider, in a given observed direction $n^i$, only the propagating directions which are observed at 
some time by the observer~\footnote{When considering first order cosmological perturbations, it is enough to consider
    the background geodesic, which is a straight line on the maximally
	symmetric spatial background. When considering higher order effects,
	time-delay and lensing corrections to the trajectory must also be considered~\cite{Hu:2000ee,Lewis:2006fu,Lewis:2017ans}.}. 

The temperature multipoles $\Theta_{I_j}$ are expanded on (general axis) plane wave harmonics as
\be\label{ThetaPlaneWave}
\Theta_{I_j} = \sum_{m=-j}^j\int \frac{\dd^3 \gr{\nu}}{(2\pi)^3}\,\frac{\Theta_j^m(\gr{\nu},\eta) }{{}_0
\tilde g^{(jm)}}\,\overline{Q}^{(jm)}_{I_j}(\gr{\nu})
\ee
The dynamical evolution for STF multipoles then translates into evolution equations for each mode components $\Theta_j^m(\gr{\nu},\eta) $.
From the choice \eqref{Slaven}, with \eqref{Thetanprop},
\eqref{ThetaPlaneWave} and \eqref{QtoG}, the temperature becomes a simple
scalar field expanded in normal modes as
\be\label{ThetaG}
\Theta = \sum_{jm}\int \frac{\dd^3 \gr{\nu}}{(2\pi)^3}\,\Theta_j^m(\gr{\nu},\eta) \,{}_0 \overline{G}^{(jm)}(\gr{\nu})\,,
\ee
where the propagating direction normal modes are related to the usual
ones as specified in \S~\ref{AppPropagation}.

The case of linear polarization is analogous, and we decompose the angular
  dependence of the Stokes parameters $Q$ and $U$ according to (see \cite{Tsagas:2007yx} or Eq. (1.67) in \cite{PitrouHDR})
\be\label{DefQUSTF}
\frac{Q \pm \ii U}{2}  = \sum_j \left[E_{I_j} \mp \ii
  B_{I_j} \right] \hat{\bar{n}}_{\mp 2}^{I_j}\,.
\ee
The electric and magnetic STF multipoles $E_{I_j}$ and $B_{I_j}$ are decomposed in terms of plane waves as
\begin{align}
\begin{split}\label{EBPlaneWave}
E_{I_j} &= \frac{1}{2}\sum_{m=-j}^j\int\!\! \frac{\dd^3
  \gr{\nu}}{(2\pi)^3}\,\frac{E_j^m(\gr{\nu},\eta) }{{}_2
\tilde g^{(jm)}}\,\overline{Q}^{(jm)}_{I_j}(\gr{\nu})\\
B_{I_j} &= -\frac{1}{2}\sum_{m=-j}^j\int\!\! \frac{\dd^3
  \gr{\nu}}{(2\pi)^3}\,\frac{B_j^m(\gr{\nu},\eta) }{{}_2
\tilde g^{(jm)}}\,\overline{Q}^{(jm)}_{I_j}(\gr{\nu})\,.
\end{split}
\end{align}
Exactly like for temperature, we then restrict the propagating direction according to \eqref{Slaven}. Hence $(Q\pm \ii U) \hat
n^{\pm 2}_{ij}$ is a tensor field on space, which is tangential to spheres
of constant $\chi$.  From \eqref{QtoG} we find that the expansion in normal modes now reads
\bea\label{EBG}
Q\pm \ii U &=& \sum_{jm}\int \frac{\dd^3 \gr{\nu}}{(2\pi)^3}\,\\
&\times& \left[E_j^m(\gr{\nu},\eta) \pm \ii   B_j^m(\gr{\nu},\eta)
\right]\,{}_{\pm 2}\overline{G}^{(jm)}(\gr{\nu})\,.\nonumber
\eea

Finally, the velocity of baryons, $V_i$, which we need to account for the
Compton collisions with electrons, is decomposed exactly as in \eqref{ThetaPlaneWave}
for $j=1$. This means that the quantity $V_i \bar n^i$ is decomposed
as in \eqref{ThetaG} with only $j=1$, which in turn defines $V^{m}(\gr{\nu},\eta)$.

In the closed case, the plane waves as defined by \eqref{DefQfromsumlM} are not
orthonormal, as shown in \eqref{NormQPlane2}. Hence we must work directly with
the ${}^{\ell M}Q^{(jm)}_{I_j}(\nu)$, which are orthogonal
according to \eqref{NormQ} and \eqref{MagicClosureQ}. The previous
expansions on harmonics can be read formally if
\be\label{Formalnu}
\gr{\nu}=(\nu,\ell,M)\,,
\ee
and then one must use the formal replacement~\cite{Challinor:1999xz,Challinor:2000as}
\be
\int \frac{\dd^3 \gr{\nu}}{(2\pi)^3} \to  \sum_{\nu=m+1}^{\infty} \sum_{\ell=m}^{\nu-1} \sum_{M=-\ell}^\ell\,.
\ee
To be clear, with the convention \eqref{Formalnu}, ${}_s
G^{(jm)}(\gr{\nu})$ refers to ${}_s G^{(jm)}_{\ell M}(\nu)$ and
$Q^{(jm)}_{I_j}(\gr{\nu})$ refers to ${}^{\ell M}Q^{(jm)}_{I_j}(\nu)$.

\subsection{Boltzmann equation}\label{SecBoltzmann1}

\subsubsection{General structure}

When using conformal cosmological time $\eta$, the general cosmological metric takes the form
\be
g^{\rm cosmo}_{\mu\nu} = a^2(\eta) (g_{\mu\nu}+ \delta g_{\mu\nu})
\ee
where $\mu,\nu$ are spacetime indices, $a(\eta)$ is the scale factor,
and where the background metric $g_{\mu\nu}$ extends the purely
spatial metric \eqref{DefMetric} with $g_{0i}=0$ and $g_{00}=-1$.

Restricting to linear cosmological perturbations, the general
Boltzmann equation dictating the evolution of the distribution
function of photons reduces to an evolution of the black body
temperature $\Theta$ which depends on $\eta$, the position on space,
and on the propagating direction $\bar n^i$. This equation possesses the general structure
\be\label{BoltzmannTheta}
\left(\partial_\eta + \bar n^i \nabla_i + \tau'\right)\Theta = {\cal C}_\Theta + {\cal G}\,.
\ee
Here, ${\cal C}$ is the collision term accounting for all processes with a final photon propagating
in the direction $\bar n^i$, whereas the term proportional to the
Compton interaction rate $\tau'\equiv a n_e \sigma_{\rm T}$ (with the
background number density of free electrons $n_e$ and
the Thomson cross section $\sigma_{\rm T}$) accounts for all
collisions with an initial photon propagating in direction $\bar
n^i$. Furthermore, ${\cal G}$ accounts for the gravitational effects which
enter when considering metric perturbations around a homogeneous and
isotropic expanding background. 

For polarization, the Boltzmann equation is even simpler since it is not affected by these gravitational effects, and one has only
\be\label{BoltzmannQU}
\left(\partial_\eta+ \bar n^i \nabla_i+ \tau'\right) (Q\pm\ii U) = {\cal C}_{Q\pm\ii U}\,.
\ee
We now discuss the individual terms in these equations in the
following sections so as to obtain a Boltzmann hierarchy in
\S~\ref{SecHierarchy}. Finally we report its formal integral solution in \S~\ref{SecSolution}.

\subsubsection{Gravitational effects}

The gravitational term selects only the scalar, vector and tensor
modes, that is $|m|\leq 2$ and in practice this implies a restriction
on the sums on $m$ in \eqref{ThetaG} and \eqref{EBG}. The effect of
the perturbed metric $\delta g_{\mu\nu}$ on temperature depends on the combinations
\be
\delta g_{00}\,,\quad \delta g_{0i} \bar n^i\,,\quad \delta g_{ij} \bar n^i \bar n^j\,,
\ee
since, from the null geodesic equation, one infers
\be\label{CalG}
{\cal G} = \frac{1}{2}\bar n^i \nabla_i \delta
g_{00}-\frac{1}{2}\delta g_{ij}' \bar n^i \bar n^j +\bar n^i \bar n^j\nabla_j \delta g_{0i} \,.
\ee
This motivates us to decompose the metric perturbations according to
\begin{align}
\begin{split}
\delta g_{00} &= -2 \int \frac{\dd^3 \gr{\nu}}{(2\pi)^3}\,
A(\gr{\nu},\eta)\,\overline{Q}^{(00)}(\gr{\nu})\\
\delta g_{0i} &= \int \frac{\dd^3 \gr{\nu}}{(2\pi)^3} \sum_{m=-1}^{1}\,
B^{(m)}(\gr{\nu},\eta)\, \overline{Q}^{(1m)}_i(\gr{\nu})\\
\delta g_{ij} &= -2 \int \frac{\dd^3 \gr{\nu}}{(2\pi)^3}
H_{L}(\gr{\nu},\eta)\, \overline{Q}^{(00)}(\gr{\nu})g_{ij}\\
&+2\int \frac{\dd^3 \gr{\nu}}{(2\pi)^3}  \sum_{m=-2}^2\,
H_{T}^{(m)}(\gr{\nu},\eta)\, \overline{Q}^{(2m)}_{ij}(\gr{\nu})\,.
\end{split}
\end{align}
It is customary to adopt a gauge in which $B^{(0)}(\gr{\nu})=0$, and we now assume this to be the case. This encompasses both the popular synchronous and Newtonian gauge.
The decomposition of \eqref{CalG} in terms of normal modes takes the form
\be\label{calGG}
{\cal G} = \sum_{jm}\int \frac{\dd^3 \gr{\nu}}{(2\pi)^3}\,
{\cal G}_j^m(\gr{\nu},\eta) \,{}_0\overline{G}^{(jm)}(\gr{\nu})\,,
\ee
and it follows from \eqref{QtoG} that factors ${}_0 \tilde
g^{(jm)}$ are brought in the expressions of the ${\cal G}_j^m$.
For completeness we report them here [see  App. C of \cite{Tram:2013ima}
which corrects \cite{TAM1,TAM2}]. The only non vanishing components
are (omitting the dependencies on $\gr{\nu}$ and $\eta$)
\begin{align}
\begin{split}
\label{Gravjm}
{\cal G}_0^0 &= H'_{L}\,,\\
{\cal G}_1^0 &= k A\,,\\
{\cal G}_2^0 &= -\,{}_0 \tilde g^{(20)}\, H_T^{(0)'},\\
{\cal G}_2^{\pm1} &= -\,{}_0 \tilde g^{(21)} \left[k\,B^{(\pm 1)}-H_T^{(\pm
    1)'}\right],\\
{\cal G}_2^{\pm 2} &= H_T^{(\pm 2)'}.
\end{split}
\end{align}
In each case, the mode $k$ is related to $\nu$ through \eqref{Defnu}, hence one must distinguish according to the value of $m$. The relevant
factors in these expressions are
\begin{align}
\begin{split}
{}_0 \tilde g^{(20)} &= \frac{2 \sqrt{k^2-3 \barK}
  }{3k} = \frac{2 \sqrt{\nu^2-4 \barK}}{3
    \sqrt{\nu^2+\barK}}\,,\\
{}_0 \tilde g^{(21)} &= \frac{\sqrt{k^2-2 \barK}}{\sqrt{3}k} =
\frac{\nu}{\sqrt{3}\sqrt{\nu^2+2 \barK}} \,.
\end{split}
\end{align}
In practice, the equations are solved in a specific gauge and not all
metric perturbations components are kept~\cite{TAM2}. The synchronous
gauge corresponds to the conditions $A = 0$ and $B^{(\pm 1)} = 0$, whereas the Newtonian gauge is found
when using $H_T^{(0)} = 0$ and $H_T^{(\pm 1)} = 0$.

In fact, the expansion in modes and multipoles of \eqref{ThetaG} \eqref{EBG}, are
exactly like Eq.~(55) of \cite{TAM1} and Eq.~(23) of \cite{TAM2}, with
the directional dependence on $\hat{\gr{\nu}}$ explicit, so we can
easily compare our results and we find that Eqs.~\eqref{Gravjm} differ
slightly from Eqs.~(35-36) of \cite{TAM2}. This is expected since our gravitational effects correspond to Eq.~C18 of \cite{Tram:2013ima} without the last term. Eqs. \eqref{Gravjm}  also  corresponds to what is obtained in \cite{Pitrou2008}, and arise when the
  observer which defines the temperature anisotropies is chosen to
  have a velocity proportional to $(\dd \eta)_\mu$. If a different
  observer is used to define anisotropies, namely, one with velocity
  proportional to $(\partial_\eta)^\mu$, then \eqref{CalG} gets  modified as we must consider the entirety of Eq.~C18 in
  \cite{Tram:2013ima}. This explains the variations found with the literature, in particular with
  \cite{TAM1,TAM2} where the contribution of $B^{(\pm
    1)}$ goes into ${\cal G}_1^{\pm 1} $ instead of
  $ {\cal G}_2^{\pm 1} $ here (see e.g. \S~4.3.1 of
  \cite{Pitrou2008} for a detailed discussion).

\subsubsection{Collisions}

The collision terms, which account for Compton collisions on electrons, depend only on the multipoles of temperature
and linear polarization, and they are expanded in multipoles with definitions following exactly the decomposition of temperature and
linear polarization of the previous section. We find \cite{Tsagas:2007yx,Pitrou2008,PitrouHDR,TAM1,TAM2}
\begin{align}
\begin{split}
{}^\Theta{\cal C}_j^m  &= \tau' \left(\delta^j_0 \delta^m_0\Theta_0^0+
  \delta^j_2 \,P^{(m)}+\delta^j_1 \, V^{(m)}\right),\\
{}^E{\cal C}_j^m &= -\tau' \sqrt{6}  \delta^j_2\,P^{(m)},\\
{}^B{\cal C}_j^m  &= 0\,, \\
P^{(m)} &\equiv \frac{1}{10}\left(\Theta_2^m-\sqrt{6}\,E_2^m\right)\,.
\end{split}
\end{align}

\subsubsection{Boltzmann hierarchy}\label{SecHierarchy}

The only non-trivial part, once the effect of gravitation and
collisions are expanded in STF multipoles, is the free streaming. 
It is sufficient to consider the case of modes aligned with
the zenith direction, and we use the expression
\be
\bar n^i \nabla_i ({}_s \overline{G}^{(jm)} \hat n^s_{I_j})=
-\frac{\dd}{\dd \chi} \left({}_s\overline{G}^{(jm)} \right) \hat n^s_{I_j}\,
\ee
as well as the recursion relation for the normal modes
\bea\label{MasterGstreaming}
&&\frac{\dd}{\dd \chi} \left({}_s \overline{G}^{(jm)} \right)  = \frac{\ii\nu m s}{j(j+1)}  {}_s \overline{G}^{(jm)} \\
&&+\frac{1}{2j+1}\left[-{}_s \kappa_j^m\,  {}_s \overline{G}^{(j-1,m)} +{}_{s} \kappa_{j+1}^m  \,{}_s \overline{G}^{(j+1,m)} \right]\nonumber
\eea
which is a consequence of \eqref{DefGsjm} and \eqref{Masteralpha1}
translated to propagating direction radial
functions~\eqref{Defalphabar}. Using this property into
\eqref{BoltzmannTheta} and \eqref{BoltzmannQU}, with \eqref{ThetaG} and
\eqref{EBG}, one finally obtains the hierarchy (again, omitting the
dependence on $\gr{\nu}$)
\begin{widetext}
\bea\label{Hierarchy}
\partial_\eta \Theta_j^{m} &=& \left[\frac{{}_0 \kappa_j^m}{2j-1}
  \Theta_{j-1}^m - \frac{{}_0 \kappa_{j+1}^m}{2j+3}
  \Theta_{j+1}^m \right] +{\cal
  G}_j^m +{}^\Theta {\cal C}_j^m -\tau' \Theta_j^m \,,\\
\partial_\eta E_j^{m} &=& \left[\frac{{}_2 \kappa_j^m}{2j-1}
  E_{j-1}^m - \frac{{}_2 \kappa_{j+1}^m}{2j+3}
  E_{j+1}^m -\frac{2m \nu}{j(j+1)} 
  B_j^m\right] +{}^E {\cal  C}_j^m -\tau' E_j^m \,,\nonumber\\
\partial_\eta B_j^{m} &=& \left[\frac{{}_2 \kappa_j^m}{2j-1}
  B_{j-1}^m - \frac{{}_2 \kappa_{j+1}^m}{2j+3}
  B_{j+1}^m +\frac{2m \nu}{j(j+1)}  E_j^m\right] +{}^B{\cal C}_j^m -\tau' B_j^m\,. \nonumber
\eea
\end{widetext}
Given the maximal symmetry of the background, the evolution of the
gravitational sources ${\cal G}_j^m$ depends only on $\nu$. However, their
initial conditions (set deep in the past) does depend fully on $\gr{\nu}$. 
Hence, in practice, the hierarchy \eqref{Hierarchy} needs to be solved only for various values of
$\nu$, and the directional dependence is simply inherited from initial
conditions. Let us also comment that we do not necessarily need to use the expansions in
normal modes \eqref{ThetaG} and \eqref{EBG} with \eqref{MasterGstreaming} to derive
the hierarchy. Indeed, this is a shortcut based on using
\eqref{Slaven}, and one might prefer using directly the expansions in
harmonics \eqref{ThetaPlaneWave} and \eqref{EBPlaneWave} along with \eqref{GenRelHarm} to compute the
effect of free streaming, as in Refs.~\cite{Challinor:1999xz,Challinor:2000as,Tsagas:2007yx,Pitrou2008}. The
hierarchy for  multipoles is eventually recovered using the orthonormality condition
\eqref{NormQPlane}.

\subsubsection{Integral solution}\label{SecSolution}

We can check using \eqref{Masteralpha} and \eqref{ebpropvsebobs} that when the gravitational effects and the collision
term can be neglected (that is when the evolution of multipoles is
only due to free streaming), the functions
\begin{align}
\begin{split}\label{SolRadial}
&(2j+1){}_0 \bar{\epsilon}_j^{(j'm)}(\eta;\nu)\,,\\
&(2j+1){}_2 \bar \epsilon_j^{(j'm)}(\eta;\nu)\,,\\
&(2j+1){}_2 \bar \beta_j^{(j'm)}(\eta;\nu)\,,
\end{split}
\end{align}
are solutions of the hierarchy \eqref{Hierarchy} for any $j'$. This
guides the general resolution of the full hierarchy when collisions
and gravitational effects are taken into account. Let us introduce the optical depth from a cosmological time $\eta$ to today ($\eta_0$)
\be
\tau(\eta,\eta_0) \equiv \int_{\eta}^{\eta_0} \dd \eta' \tau'(\eta')\,,
\ee
that we abbreviate as $\tau$. It is then straightforward to obtain the
formal solution to the full hierarchy in the integral
form~\cite{TAM1,TAM2}
\begin{align}\label{IntegralT}
\frac{\Theta_j^m(\eta_0)}{2j+1} &=  \int_0^{\eta_0}
\dd \eta {\rm e}^{-\tau} \times\\
&\sum_{j'} \left({}^\Theta{\cal C}_{j'}^m  + {\cal
    G}_{j'}^m \right){}_0\bar \epsilon_j^{(j'm)}(\chi)\,,\nonumber
\end{align}
\begin{align}
\frac{ E_j^m(\eta_0)}{2j+1} &=  \int_0^{\eta_0}
\dd \eta {\rm e}^{-\tau} \sum_{j'} {}^E{\cal
  C}_{j'}^m\,{}_2 \bar \epsilon_j^{(j'm)}(\chi)\,,\nonumber\\
\frac{B_j^m(\eta_0)}{2j+1} &=  \int_0^{\eta_0}
\dd \eta {\rm e}^{-\tau} \sum_{j'} {}^E{\cal C}_{j'}^m\,{}_2 \bar \beta_j^{(j'm)}(\chi)\,,\nonumber
\end{align}
where the argument of the radial functions is 
\be\label{SlaveChi}
\chi= \eta_0-\eta\,.
\ee

Finally, and this is crucial, it is customary to expand the directional
dependence of the observed temperature (resp. polarization) directly
in $Y_j^m$ (resp ${}_{\pm 2} Y_j^m$). Hence, to obtain the
corresponding multipoles, one must consider the normalization at
origin \eqref{NormG}, and this brings extra factors $\bar c_\ell/(2\ell+1)=(-\ii)^\ell\sqrt{4\pi/(2\ell+1)}$. Thus, for the CMB, we shall 
use
\be
{}^{\rm CMB} \Theta_\ell^m(\eta_0) = \Theta_\ell^m(\eta_0) (-\ii)^\ell\sqrt{\frac{4\pi}{2\ell+1}}\,,
\ee
with similar relations for the $E$ and $B$ modes. Eventually, one might
also prefer to use directions related to observation rather than propagation for the multipoles observed today, and this brings
extra factors of $(-1)^\ell$ for the temperature and electric-type
polarization multipoles and $(-1)^{\ell+1}$ for the magnetic-type
ones.

\subsection{Other cosmological observables}\label{SecOtherCosmo}

Quite generally, all types of cosmological observables, such as weak
lensing convergence or shear, lensing field, galaxy number counts,
redshift drifts, etc., are all of the form of an integral on the
background past light cone, which can be written formally as
\be\label{GeneralType}
{}_s {\cal O}(\gr{n}) =\int_0^{\eta_0} \dd \eta \sum_{jm} \hat n^{I_j}_{-s}\,  S^{(jm)}_{I_j}(\eta,\chi,\gr{n}) 
\ee
with \eqref{SlaveChi}. The sources $S^{(jm)}_{I_j}$ are expanded on plane waves harmonics as
\bea\label{DecSources}
&&\hat n^{I_j}_{-s}  S^{(jm)}_{I_j}(\eta,\chi,\gr{n}) \\
&&\qquad= \int \frac{\dd^3\gr{\nu}}{(2\pi)^3} \,S_j^m(\eta;\gr{\nu}) {}_s G^{(jm)}(\chi,\gr{n}; \gr{\nu})\,.\nonumber
\eea
If we decompose the observable as a sum of the effects of each harmonic as
\be\label{DecO}
{}_s {\cal O} = \int \frac{\dd^3 \gr{\nu}}{(2\pi)^3} \sum_{j m}
{}_s {\cal O}_j^m(\gr{\nu}) {}_s G^{(j m)}(\chi=0,\gr{n};\gr{\nu})\,, 
\ee
then we only need to expand the sources under the integral with the
Rayleigh expansion \eqref{MasterRayleighQ} which is equivalent to
\bea\label{MasterRayleighG}
&&{}_{\pm s} G^{(jm)}(\chi,\gr{n};\gr{\nu}) = \sum_{\ell\geq |m|}^{\infty,\nu-1}
(2\ell+1) \frac{\myzeta_\ell^m}{\myzeta_j^m} \\
&&\qquad\quad\times {}_{\pm s}\alpha_\ell^{(jm)}(\chi;\nu) {}_{\pm s} G^{(\ell m)}(\chi=0,\gr{n};\gr{\nu}) \,,\nonumber
\eea
so as to obtain the integral solutions (with $\myzeta_\ell^m = {\rm const.}$ since the decomposition \eqref{DecSources} is on plane-wave harmonics)
\be
\frac{{}_s {\cal O}_j^m(\gr{\nu})}{2j+1} =\int_0^{\eta_0} \dd
\eta \sum_{j'\geq |m|} {}_{s}\alpha_j^{(j' m)}(\chi;\nu)  S_{j'}^m(\eta;\gr{\nu})\,.
\ee
Note that in the angular decomposition \eqref{DecO} we must use the
normalization at $\chi=0$ given by \eqref{NormG} but taking into
account the rotation \eqref{GPW2}, that is~\footnote{In the closed
  case, and given the formal meaning \eqref{Formalnu}, the
  normalization at $\chi=0$ is directly ${}_s G^{(jm)}_{\ell
    M}(\chi=0;\nu) = \delta_\ell^j c_j/(2j+1) {}_s Y_j^M$.}
\be {}_s G^{(j m)}(\chi=0;\gr{\nu}) = \frac{c_j}{2j+1}\sum_{M}D^j_{M m}(R_{\hat{\gr{\nu}}}) {}_s Y_j^{M} \,.
\ee

The integral solution for the CMB multipoles arises immediately with
this method, if one notes that the Boltzmann equation
\eqref{BoltzmannTheta} is rewritten as
\be
\frac{\dd}{\dd \eta} ({\rm e}^{-\tau}\Theta) = {\rm e}^{-\tau}[ {\cal C}_\Theta + {\cal G}]
\ee
where $\dd/\dd \eta \equiv \partial_\eta + \bar n^i \nabla_i$, is the derivative along the background geodesic. Indeed, the integral form of
the type \eqref{GeneralType} is
\be
\Theta(\eta_0) = \int_0^{\eta_0} {\rm e}^{-\tau}[{\cal C}_\Theta +
{\cal G}]\dd \eta\,,
\ee
and following the aforementioned method, we then recover the solution
\eqref{IntegralT}, up to the difference that for CMB we used
propagation direction harmonics and multipoles. Even though this derivation appears much faster, one must
not forget that in the case of CMB the sources depend on the multipoles themselves, and one must rely on the Boltzmann hierarchy (which can be
found by derivation of the integral solutions with respect to $\eta_0$) to solve for their evolution. 

The physical interpretation based on this method is that free streaming builds multipoles with increasing $j$ from the initial
multipoles of sources. The effect of free streaming amounts to intersecting plane-wave harmonics with
spheres of increasing radius, and the radial functions precisely
account for the projection effects of the sources, taking into account
the local angular structure at emission.

\section{Conclusion}

Thanks to the introduction of the generalized helicity basis, we established in this work
a systematic and comprehensive construction of radial
functions, normal modes and tensor harmonics in maximally symmetric three-dimensional spaces. 
When combined with spin-weighted spherical harmonics, they provide a powerful set of tools adapted to 
the description of symmetric and trace-free tensors, and is suited for separating the radial from the angular dependencies of physical quantities. Furthermore, the developed framework allows for systematic algebraic manipulations which greatly benefits from the power of symbolic computational tools. In particular, in this work, we have made intensive use of {\it xAct} \cite{xAct}. 

In Appendix~\ref{SecComparison} our results are contrasted with earlier literature on vector and tensor harmonics around maximally symmetric curved spaces. However, our method is not restricted to vector or tensor harmonics, but can be applied to higher rank harmonics thanks to the full set of recursive relations in the $(j,s)$ space. Our results also extend to curved spaces the construction of scalar, vector
and tensor harmonics presented in \cite{Dai:2012bc}, and puts on a firmer ground the pioneering use of normal modes introduced in \cite{TAM2}. However, we stress that some of our relations were not fully demonstrated but only checked explicitly for all modes up to
reasonable values of the eigenvalue $j$ (typically $j \leq 4$ and the associated $|m| \leq j$, $|s| \leq j$), as was also the case in \cite{TAM2}. Hence, from a mathematical point of view, our formalism
would benefit from an appropriate general proof on these relations. Still, for practical physical applications which depend
only on the lowest values of $j$ (but with all allowed values of $m$
and $s$), it can be fully trusted since all relations were checked with $\ell$ being kept general, using the general properties of hyperspherical Bessel functions. Thus, in a 
restricted sense, they have been demonstrated. The relations which were checked up to $j = 4$ (and all allowed values of $m$ and
$s$) but with general values of $\ell$ are Eqs.~\eqref{Masteralpha}, \eqref{Normalpha1},
\eqref{Normalpha2} and \eqref{MagicC8}. Relation \eqref{jlChangejl}, on the other hand, was checked only for $j \leq 4$ and $\ell \leq 4$.

The radial functions have very rich properties which
fall into four categories. These are summarized as follows:
\begin{itemize}
\item Recursive relations in the space of $(j,s)$ values. They can all
  be deduced from the triangular relations (NW,NW,SW,SE) depicted in
  Fig.~\ref{FigRecursions1} and whose expressions are collected in
  appendix \ref{AppRecursions}. They allow us to deduce all radial
  functions using the algorithm described in \S~\ref{SecAlgo}. Furthermore, Eq.~\eqref{Masteralpha1}  is of direct use for the
  effect of free streaming on radiation multipoles.  
\item Sign inversions of either $m$, $s$ or $\nu$
  [Eq.~\eqref{Magicsmnu}], which are of direct use when studying the
  properties under parity transformation as in \S~\eqref{SecConjugationParity}.
\item Symmetries in the space of $(j,m,s,\ell)$ values, namely the $m \leftrightarrow s$ and the $j \leftrightarrow \ell$ exchange
  symmetries [Eqs.~\eqref{smChangejl} and \eqref{jlChangejl}].
\item Orthogonality relations \eqref{Normalpha1} and
  \eqref{Normalpha2}, which imply corresponding orthogonality relations for harmonics.   
\end{itemize}

Once knowing the radial functions, whose expressions for $j \leq 2$
are gathered in Section~\ref{SecGatherRadial} (or Appendix \ref{AppFlat} for
the flat case), the harmonics are built using \eqref{DefGsjm} and
\eqref{Qljm}, with the needed coefficients ${}_s g^{(jm)}$ given by
\eqref{gsjm} and \eqref{gtjm}, and the explicit forms of
the generalized helicity bases reported in Appendix \ref{AppPropnsIj}.

The case of a flat space is very different from the curved cases. Indeed, we have shown that the general factorization
\eqref{WrongFactor}, which can be used in the flat case to build systematically all radial functions (see Appendix \ref{AppFlat}), does
not exist in the curved case, contrary to previous statements in the literature. Our results provide a systematic algorithm to build recursively the radial functions in curved space by systematic exploration of the $(j,s)$
space of radial functions. A {\it Mathematica} notebook implementing
this algorithm is available at \cite{NotebookAlgorithm}.

The radial functions are extremely powerful for the computation of
theoretical expressions for multipoles of observables. Once an
observable is written as an integral on the background past light
cone, it is sufficient to decompose the angular structure on normal
modes, and to use the Rayleigh expansion in the form \eqref{MasterRayleighG} to
obtain the result. In practical applications, it is sometimes preferred to use harmonics
which are decomposed according to a propagation direction (e.g., photon's direction in the case 
of CMB) rather than the observation direction, and the relation between both convention is simple, as we summarized 
in section~\ref{AppPropagation}.

Finally, it is worth mentioning that radial functions (and thus harmonics) can also be defined for super-curvature
modes. They correspond to values of $\nu$ in the complex plane and rely on analytic continuations of the radial functions built here. 
In \cite{Duality} we detail how super-curvature modes can be used to described spatially anisotropic (i.e., Bianchi) space-times as super-curvature fluctuations over maximally symmetric space-times.

\acknowledgments

CP is indebted to Guillaume Faye for extensive discussions about the compendium of formulas collected in appendix \ref{AppPropnsIj}. TP thanks the Institut d'Astrophysique de Paris for its hospitality during the initial stages of this project, as well as Brazilian Funding Agencies CNPq (311527/2018-3) and Fundação Araucária (CP/PBA-2016) for their financial support. Tensorial computations in this work were performed using the tensor algebra package {\it xAct}~\cite{xAct}.

\bibliography{Harmonics}

\appendix

\section{Hyperspherical Bessel functions}\label{AppHyperBessel}

Hyperspherical Bessel functions are derived in detail in Refs.~\cite{Harrison:1967zza,Tomita1982,Abbott1986}. For the convenience of the reader, though, we present some key properties in this appendix.

The Hyperspherical Bessel functions $\Phi^\nu_\ell$ are solutions of the following differential equation
\bea\label{EquadiffBessel}
&&\frac{1}{r^2(\chi)} \frac{\dd}{\dd \chi} r^2(\chi) \frac{\dd}{\dd
	\chi} \Phi_\ell^\nu \nonumber\\
&&+ \left[\nu^2-\barK  -\frac{\ell(\ell+1)}{r^2(\chi)}\right]\Phi_\ell^\nu=0\,.
\eea
They satisfy the following recurrence relations
\begin{align}
\begin{split}\label{RecurrenceBessel}
\frac{\dd}{\dd \chi} \Phi_\ell^\nu &=\frac{\ell}{2\ell+1}\sqrt{\nu^2 - \barK \ell^2}\Phi_{\ell-1}^\nu \\
& -\frac{(\ell+1)}{2\ell+1}\sqrt{\nu^2 - \barK (\ell+1)^2}\Phi_{\ell+1}^\nu\,,\\
\gcot \chi \,\Phi_\ell^\nu &= \frac{1}{2\ell+1} \sqrt{\nu^2-\barK \ell^2}\Phi_{\ell-1}^\nu\\
&+ \frac{1}{2\ell+1}\sqrt{\nu^2 - \barK (\ell+1)^2}\Phi_{\ell+1}^\nu\,,
\end{split}
\end{align}
with
\be
\Phi_0^\nu \equiv \frac{\sin \nu \chi}{\nu \gsin \chi}\,.
\ee
A closed expression for a general $\ell$ is~\cite{Tomita1982,Lyth1995}
\bea
\Phi_\ell^\nu &=& \frac{1}{\nu^2}\left(\prod_{i=1}^\ell \frac{1}{\sqrt{\nu^2-\barK
      i^2}}\right) \\
&&\times \gsin^\ell \chi\left(\frac{-1}{\gsin \chi} \frac{\dd}{\dd \chi}\right)^{\ell+1}
\cos(\nu \chi)\,.\nonumber
\eea
Near the origin ($\chi \to 0$), they are power-law suppressed (except
for $\ell=0$) as
\be\label{PowerLawOriginBessel}
\Phi_\ell^\nu \sim \chi^\ell \prod_{i=1}^\ell \frac{\sqrt{\nu^2 - \barK i^2}}{(2i+1)}\,.
\ee

In the closed case, $\gsin=\sin$ and the variable $\nu$ must take positive integer values constrained by
\be
0 \leq \ell < \nu\,.
\ee
The lowest value $\nu=1$ corresponds to a constant global perturbation since there
is only $\Phi_0^{\nu=1} = 1$. The $\nu=2$ mode allows only for global
dipolar modulations since $\Phi_0^{\nu=2} = \cos \chi$ and $\Phi_1^{\nu=2} = (\sin \chi)/\sqrt{3}$.

The hyperspherical Bessel functions are normalized similarly to usual
spherical Bessel functions [Eq. (\ref{Normjl})]
\begin{align}
&\int \Phi_\ell^\nu(\chi) \Phi_\ell^{\nu'}(\chi) r^2(\chi) \dd \chi =
\frac{\pi}{2} \frac{\delta(\nu-\nu')}{\nu^2}\,,\label{Normintchi}\\
&\int \Phi_\ell^\nu(\chi) \Phi_\ell^{\nu}(\chi') \nu^2 \dd \nu = \frac{\pi}{2} \frac{\delta(\chi-\chi')}{r^2(\chi)}. \label{Normintnu}
\end{align}
In the closed case, the integral on $\nu$ must be understood as a discrete sum
on $\nu \geq \ell +1$. 

A class of related functions is given by
\be
\Psi_\ell^{\nu,n} \equiv \frac{\Phi_\ell^\nu}{r^n(\chi)}\,.
\ee
One can check that these functions satisfy the differential equation
\bea\label{EquationPsi}
&&\frac{\dd^2}{\dd \chi^2} \Psi_\ell^{\nu,n} + 2(1+n)\gcot\chi \frac{\dd}{\dd
  \chi} \Psi_\ell^{\nu,n}\nonumber\\
&&+\left[\nu^2-\barK(1+n)-\frac{\ell(\ell+1)}{r^2(\chi)}\right.\nonumber\\
&&\quad\left.+ n(n+1) \gcot^2(\chi)\right]\Psi_\ell^{\nu,n}=0\,.
\eea

In practical numerical computations, hyperspherical Bessel functions are challenging to compute. The reader interested in fast and 
accurate implementations can check Refs.~\cite{Lesgourgues:2013bra,Tram:2013xpa}.

\section{Spherical Harmonics and helicity basis}\label{AppYlm}

In this appendix we work in the flat (Euclidean) three-dimensional space, also identified with the tangent space at the origin of the
coordinates ($\chi=0$) of curved spaces. The unit direction vector is $\gr{n}$, and we also use the helicity vectors \eqref{SlaveChi} along with the general helicity basis \eqref{CentralMulti} and the multi-index notation \eqref{Multi}.

\subsection{Spherical harmonics}

Spherical harmonics are defined as functions on the unit sphere:
\be
Y_{\ell}^{m}(\theta,\varphi) = \sqrt{\frac{2 \ell+1}{4
    \pi}\frac{(\ell-m)!}{(\ell+m)!}} e^{\ii m \varphi} P_\ell^m(\cos \theta)\,,
\ee
with the associated Legendre polynomials being given by
\bea
P_\ell^m(z) &=& (-1)^m
(1-z^2)^{m/2}\nonumber\\
&&\times\frac{\dd^{\ell+m}}{\dd z^{\ell+m}}(z^2-1)^\ell\,.
\eea
Let us define, for any pair of functions $A(\gr{n})$ and $B(\gr{n})$, the Hermitian inner product
\be\label{DefABProduct}
\left\{ A | B \right\} = \left\{ B | A \right\}^\star \equiv \int \dd^2 \gr{n} A^\star(\gr{n}) B(\gr{n})\,.
\ee
The spherical harmonics are orthonormal
\be
\left\{ Y_{\ell}^{m} | Y_{\ell'}^{m'}\right\} = \delta_{\ell \ell'}\delta_{m
  m'}\,,\label{OrthoYlm}
\ee
and complete
\be\label{ClosureYlm}
\sum_ {\ell=0}^\infty \sum_{m=-\ell}^\ell  Y_\ell^m(\gr{n}) Y_\ell^{m\star}(\gr{n}') = \delta^2(\gr{n}-\gr{n}')\,.
\ee

Given the spherical harmonics, which form a basis for scalar functions on the sphere, one can define spin-weighted spherical harmonics as basis for spin functions on the sphere~\cite{Goldberg:1966uu}. These are defined as
\beas
{}_s Y_{\ell}^{m} &=& \sqrt{\frac{(\ell-s)!}{(\ell+s)!}}
\spart^s Y_{\ell m}\,, \quad s\ge 0\,,\nonumber\\
{}_s Y_{\ell}^{m} &=& \sqrt{\frac{(\ell+s)!}{(\ell-s)!}}
(-1)^s \spartb^{-s} Y_{\ell m}\,, \quad s\le 0\,,\nonumber
\eeas
or by induction as 
\beas
{}_{s+1} Y_{\ell}^{m}\! &=&\! \frac{1}{\sqrt{(\ell-s)(\ell+s+1)}} \spart {}_s Y_{\ell}^{m}\,, \nonumber\\
{}_{s-1} Y_{\ell}^{m}\! &=&\!- \frac{1}{\sqrt{(\ell+s)(\ell-s+1)}} \spartb {}_s Y_{\ell}^{m}\,,\nonumber
\eeas
where the spin-raising ($\spart$) and spin-lowering ($\spartb$) operators are
\beas
\spart &\equiv& -(\sin \theta)^s \left(\partial_\theta +\frac{\ii}{\sin
    \theta} \partial_\varphi \right) (\sin\theta)^{-s}\,, \nonumber\\
\spartb &\equiv& -(\sin \theta)^s \left(\partial_\theta -\frac{\ii}{\sin
    \theta} \partial_\varphi \right) (\sin\theta)^{-s}\,. \nonumber
\eeas
These operators are related to the covariant derivative ${\bf D}$ on the sphere. Indeed, it is found that
\beas
\spart &=& - \sqrt{2} D_{+} = -\sqrt{2} \gr{n}_- \cdot {\bf D}\,,\\
 \spartb &=& - \sqrt{2} D_{-} = -\sqrt{2} \gr{n}_+ \cdot {\bf D}\,.
\eeas
This allows us to derive a central relation for the computation of covariant derivatives on the sphere 
\bea
D_j \left({}_{\pm s} Y_{\ell}^{m} \hat n^{\pm s}_{I_s}\right) &=&\mp
\frac{({}_\pm \lambda_\ell^s)}{\sqrt{2}} {}_{\pm(s+1)} Y_{\ell}^{m} n^\pm_j
\hat n^{\pm s}_{I_s} \nonumber\\
&&\pm \frac{({}_\mp \lambda_\ell^s)}{\sqrt{2}} {}_{\pm(s-1)} Y_{\ell}^{m} n^\mp_j \hat n^{\pm s}_{I_s} \,,\nonumber
\eea
where the coefficients ${}_\pm\lambda_\ell^s$ where introduced in~\eqref{Deflambda}. 

An explicit form of the spin weighted spherical harmonics is
\bea
&&{}_s Y^{m}_\ell = e^{\ii m \varphi }\sqrt{\frac{2 \ell+1}{4
    \pi}\frac{(\ell+m)!(\ell-m)!}{(\ell+s)!(\ell-s)!}}\nonumber\\
&&\sum_{r={\rm max}(0,m-s)}^{{\rm min}(\ell-s,\ell+m)}\binom{\ell-s}{r}
\binom{\ell+s}{r+s-m}\nonumber\\
&&\times (-1)^{\ell+m-r-s} \frac{(\cos \frac{\theta}{2})^{2r+s-m}}{(\sin \frac{\theta}{2})^{2r+s-m-2\ell}}\,.
\eea
Useful properties are
\beas
{}_s Y_{\ell}^{m\,\star}(\gr{n})\! &=&\! (-1)^{m+s} {}_{-s} Y_{\ell}^{-m}(\gr{n}),\slabel{ConjugateYslm}\\ 
{}_s Y_{\ell}^{m}(-\gr{n})\! &=&\! (-1)^\ell {}_{-s}Y_{\ell}^{m}(\gr{n}),\slabel{InversionYslm}\\
{}_{-s} Y_{\ell}^{m}(\gr{e}_z)\!  &=&\! \delta_{m s}(-1)^m \sqrt{\frac{2 \ell + 1}{4 \pi}},
\eeas
and the $s \leftrightarrow m $ interchange property
\be\label{msChange}
(-1)^s {\rm e}^{\ii s \phi} {}_s Y_\ell^m = (-1)^m {\rm e}^{\ii m \phi} {}_m Y_\ell^s\,.
\ee
Finally, we also find the orthogonality relation
\be\label{OrthoYslm}
\left\{ {}_s Y_{\ell}^{m} | {}_s Y_{\ell'}^{m'}\right\} = \delta_{\ell \ell'}\delta_{m m'}\,,
\ee
as well as the closure relation, which generalizes \eqref{ClosureYlm}
\be\label{ClosureYslm}
\sum_ {\ell=|s|}^\infty \sum_{m=-\ell}^\ell {}_s Y_\ell^m(\gr{n}) {}_s Y_\ell^{m\star}(\gr{n}') = \delta^2(\gr{n}-\gr{n}')\,.
\ee

\subsection{Relation with helicity basis}

In this section, we detail how spherical harmonics and spin-weighted spherical harmonics are related to the generalized helicity basis~\eqref{CentralMulti}. This extends the results already collected  in appendix D of \cite{PitrouHDR}. First, from the general rule for the integration over products of $\gr{n}$~\cite{Thorne1980} 
\bea
\int \frac{\dd^2 n}{4\pi} n^{I_{2\ell+1}}&=&0\,,\\
\int \frac{\dd^2 n}{4\pi}n^{I_{2\ell}}&=&\frac{1}{2\ell+1}\delta^{(i_1 i_2} \dots
\delta^{i_{2\ell-1} i_{2\ell})}\nonumber\,,
\eea
where the parentheses mean full symmetrization on enclosed indices, it is possible to show that
\bea 
\left\{ \hat n^{I_\ell} | \hat n_{J_\ell}
\right\} &=& \Delta_\ell\delta_{j_1}^{\langle i_1} \dots
\delta_{j_\ell}^{i_\ell \rangle }\,, \label{OrthoSTF}\\
\Delta_\ell &\equiv& \frac{4\pi \ell!}{(2\ell+1)!!}\,. \label{DefDeltaell}
\eea
$\hat n^{I_\ell}$ is a special case of \eqref{CentralMulti} with
multi-index notation, and is simply the STF product of $\ell$ unit direction
vectors (see e.g. \cite{FBI2014, Thorne1980}). Eq.~(\ref{OrthoSTF}) is a particular case of Eq.~(C2) in \cite{FBI2014}. 

If we define
\be\label{DefCurlyY}
{\cal Y}_{\ell m}^{I_\ell} \equiv \Delta_\ell^{-1}\left\{ \hat n^{I_\ell} | Y_{\ell}^{m}\right\}\,,
\ee
we can relate the spherical harmonics to the generalized helicity basis (with $s=0$) as
\be\label{YlmfromSTF}
Y_{\ell}^{m}(\gr{n}) = \Delta_\ell^{-1}\hat n_{I_\ell}
\left\{ \hat n^{I_\ell} | Y_{\ell}^{m}\right\} = \hat n_{I_\ell} {\cal
  Y}_{\ell m}^{I_\ell}\,.
\ee
The inverse relation is
\bea\label{MagicSTFYlm}
\hat n^{I_\ell} &=& \sum_{m=-\ell}^\ell Y_{\ell}^{m}(\gr{n})\left\{ Y_{\ell m} | \hat n^{I_\ell} \right\}\,,\\
&=&\sum_{m=-\ell}^\ell\Delta_\ell Y_{\ell}^{m}(\gr{n}) {\cal  Y}_{\ell m}^{\star I_\ell}\,. \nonumber
\eea
From the identity
\be
\sum_{m=-\ell}^\ell Y_{\ell}^{m}(\gr{n}) Y_{\ell}^{m\,\star}(\gr{n}) = \frac{2\ell+1}{4\pi}\,,
\ee
we get the closure relation
\be\label{closureCalY}
\sum_{m=-\ell}^\ell {\cal Y}_{\ell m}^{I_\ell} {\cal Y}^{\ell m\,\star}_{J_\ell}=\Delta_\ell^{-1}\delta_{j_1}^{\langle i_1} \dots
\delta_{j_\ell}^{i_\ell \rangle }\,.
\ee
Explicitly the ${\cal Y}_{\ell m}^{I_\ell}$ are given by (for $m>0$) 
\bea
&&{\cal Y}_{\ell m}^{I_\ell} = C_{\ell m}
\sum_{j=0}^{[\tfrac{(\ell-m)}{2}]}\left(\delta_1^{i_1}+\ii
  \delta_2^{i_1}\right)\dots \left(\delta_1^{i_m}+\ii
  \delta_2^{i_m}\right)\nonumber\\
&&a_{\ell m j} \delta_3^{i_{m+1}}\dots\delta_3^{i_{\ell-2j}}\delta^{{\tiny\ell-2j+1}\,\,
{\tiny \ell-2j+2}}\dots\delta^{i_{\ell-1}\,i_\ell}\,,\nonumber
\eea
where
\bea
C_{\ell m} &\equiv& (-1)^m \left[\frac{2\ell+1}{4\pi}\frac{(\ell-m)!}{(\ell+m)!}\right]^{1/2}\,,\nonumber\\
a_{\ell m j} &\equiv& \frac{(-1)^j (2\ell - 2j) !}{2^\ell j! (\ell-j)!(\ell-m-2 j)!}\,.\nonumber
\eea
Since we used a Cartesian basis in a Euclidean space, we also define
${\cal Y}_{\ell m}^{I_\ell} ={\cal Y}^{\ell m}_{I_\ell} $ and we have the property $ {\cal Y}_{\ell m}^{\star\,I_\ell} = (-1)^m {\cal Y}_{\ell
  \,-m}^{I_\ell} $ which extends the definition for negative $m$. The ${\cal Y}_{\ell m}^{I_\ell}$ satisfy the orthogonality property
\be\label{NormalisationCalYContracted}
{\cal Y}_{\ell m}^{I_\ell}{\cal Y}^{\ell m' \star}_{I_\ell}=\Delta_\ell^{-1}\delta_m^{m'}\,.
\ee
They also allow us to build spin-weighted spherical harmonics, in close analogy to~\eqref{YlmfromSTF}:
\be\label{YlmsfromSTF}
{}_{\pm s}Y_{\ell}^{m}(\gr{n})=(\mp)^s b_{\ell s}\,{\cal Y}^{\ell m}_{I_\ell}
\hat n_{\mp s}^{I_\ell}\,.
\ee
This relation is inverted as
\be\label{Magiceplusns}
\hat n_{\mp s}^{I_\ell}=\frac{(\mp)^s\Delta_\ell}{b_{\ell s}}
\sum_{m=-\ell}^\ell {}_{\pm s}Y_{\ell}^{m}(\gr{n})
{\cal Y}^{\star I_\ell}_{\ell m}\,.
\ee
Using~\eqref{OrthoYslm} and~\eqref{closureCalY} we deduce immediately the useful orthogonality
condition for the generalized helicity basis
\bea\label{OrthoAngularGeneralHelicity}
\left\{ \hat n_{\pm s}^{I_\ell} | \hat n^{\pm s}_{J_{\ell'}}
\right\} &=& \delta_{\ell \ell'}\frac{\Delta_\ell}{(b_{\ell s})^2}\delta_{j_1}^{\langle i_1} \dots
\delta_{j_\ell}^{i_\ell \rangle }\\
&=&\delta_{\ell \ell'} d_{\ell s}\frac{4\pi}{(2\ell+1)}\delta_{j_1}^{\langle i_1} \dots
\delta_{j_\ell}^{i_\ell \rangle }\,.\nonumber
\eea
Furthermore, since the generalized helicity basis is a complete basis
for STF tensors at each point, we also find the closure relation
\be\label{ClosureHelicity}
\sum_{s=-\ell}^\ell (d_{\ell s})^{-1}\hat n_{\pm s}^{I_\ell}  \hat n^{\mp
    s}_{J_\ell} = \delta_{j_1}^{\langle i_1} \dots
\delta_{j_\ell}^{i_\ell \rangle }\,.
\ee

In the construction of harmonics of this paper, and more specifically
in \eqref{QToCurlyY}, we are not working in a Euclidean space. However
we can still use the object \eqref{DefCurlyY} if it is understood that
it is defined in the tangent space at the origin.

Finally note that the ${\cal Y}_{\ell m}^{I_\ell} $ are related to the
generalized helicity basis in the zenith direction Since in this special direction ($\theta=0$) $\phi$ is not defined, we choose
the convention 
\be
\gr{n}_\theta=\gr{e}_x,\quad \gr{n}_\phi=\gr{e}_y,\quad \gr{n}_\pm =
\frac{1}{\sqrt{2}}(\gr{e}_x \mp \ii \gr{e}_y)
\ee
which implies that at any point $(\theta,\phi)$ of the unit sphere,
the helicity basis is obtained by a rotation of angle $\theta$ around the $y$
axis and a rotation of angle $\phi$ around the $z$ axis from the basis
at the zenith direction. With this choice we get in particular for
$m\ge 0$
\be\label{YslmToGenHelicity}
{\cal Y}_{\ell\, \pm m}^{I_\ell} =(\mp)^m(\Delta_\ell d_{\ell m})^{-1/2}\left.\hat n_{\mp m}^{I_\ell}\right|_{\rm zen}\,\,.
\ee
Note that we can recast the value at $\chi=0$ of harmonics given in
\eqref{QToCurlyY}. We find
\bea\label{QToCurlyY2}
\left.{}^{\ell} Q_{I_j}^{(j\,\pm m)}\right|_{\chi=0} &=& \delta_\ell^j   (\mp
\sqrt{2})^m \,\ii^j\, \frac{(2m-1)!!}{\sqrt{(2m)!}} \nonumber\\
&&\times\frac{\mygamma_m}{\mygamma_j} \left.\hat n_{\mp m}^{I_j}\right|_{\rm zen}\,.
\eea

With the formulation \eqref{YslmToGenHelicity}, Eq. \eqref{YlmsfromSTF} can be recast as 
\be
{}_{\pm s} Y_\ell^{\pm m} = (\mp1)^s (\mp 1)^m
\sqrt{\frac{4\pi}{2\ell+1}} \hat{n}^{\mp s}_{I_\ell }\left.\hat n_{\mp m}^{I_\ell}\right|_{\rm zen}
\ee
which obviously leads to \eqref{ConjugateYslm} after complex conjugation. Using
that the generalized helicity basis is a complete basis for STF
tensors, we also obtain by decomposing the generalized helicity
basis in the zenith direction (considered as a constant tensor)
\be\label{MagicSumYslm}
\sum_{s=-j}^j (k_{js})\, {}_s Y_j^m \hat n^s_{I_j } = {\cal Y}^{jm}_{I_j}\,.
\ee
with the factors $k_{js}$  defined by \eqref{Defkjs}. 

In flat case, the definition of the ${\cal Y}^{jm}_{I_j}$ in the tangent space at $\chi=0$ can be trivially extended to any point by
simple translations. In the curved case, one uses the relation
\eqref{YslmToGenHelicity} and extend it to any point by parallel
transport along the radial geodesic reaching this point. Since the
generalized helicity basis is also parallel transported along radial
geodesics, the properties \eqref{YlmsfromSTF} and \eqref{MagicSumYslm} are also valid when
${\cal Y}^{jm}_{I_j}$ and $\hat n^s_{I_j}$ are evaluated at a general point with $\chi \neq 0$.

\subsection{Rotations}

Let us consider an active rotation of angles
$R(\alpha,\beta,\gamma)$. With the Euler angle notation, it consists in actively rotating around the
$z$-axis by an angle $\gamma$, then actively rotating around the
$y$-axis by an angle $\beta$, and finally rotating around the $z$-axis
by an angle $\alpha$. The rotated spherical and spin-weighted
spherical harmonics are related to the original ones by (see
e.g. appendix A of \cite{Challinor:2005jy} or appendix D.3 of \cite{Pitrou:2015iya})
\be\label{RotationYslm}
R[{}_s Y_\ell^m \hat n^s _{I_j}] = \sum_{m'} D^\ell_{m' m}(R) {}_s Y_\ell^{m'} \hat n^s
_{I_j} 
\ee
where the Wigner $D$-coefficients are related to spin spherical
harmonics through
\be\label{YslmWigner}
{}_s Y_\ell^m(\beta,\alpha) = \sqrt{\frac{2\ell+1}{4\pi}}(-1)^m {\rm
  e}^{\ii s \gamma} D^\ell_{-m\, s}(\alpha,\beta,\gamma)\,.
\ee
In particular, when considering only a rotation $R_y(\pi)$ around the $y$-axis by
an angle $\pi$ (that is $\alpha=0,\beta=\pi,\gamma=0$)
\be\label{RotationYslmRy2}
D^\ell_{m'm}(R_y(\pi)) = \delta_{m\,-m'} (-1)^{\ell+m'}\,,
\ee
and
\be\label{RotationYslmRy}
R_y(\pi)[{}_s Y_\ell^m \hat n^s _{I_j}] = (-1)^{\ell-m} {}_s Y_\ell^{-m} \hat n^s
_{I_j}\,. 
\ee

\section{A compendium of useful formulae}\label{AppPropnsIj}

\subsection{Explicit expression of the generalized helicity basis}

From the general formula to extract the STF part from a symmetric
tensor (see e.g. Eq. (2.2) of \cite{Thorne1980}) one infers the general expression
for the generalized helicity basis which is
\begin{align}\label{ExplicitSTFHelicity}
\hat n^{\pm s}_{I_\ell} &= \sum_{n=0}^{[(\ell-s)/2]}\, {}_s a^\ell_n\, g_{(i_i i_2}\dots
  g_{i_{2n-1} i_{2n}} \nonumber\\
  &\qquad n^\pm_{i_{2n+1}} \dots n^\pm_{i_{2n+s}} n_{i_{2n+s+1}} \dots n_{i_\ell)}\,,
\end{align}
where the parentheses mean full symmetrization on enclosed indices, and
with the coefficients
\be
{}_s a_n^\ell \equiv \frac{(-1)^n (2 \ell- 2n -1)!! (\ell-s)!}{(2 \ell
  -1)!! (2n)!! (\ell-2n-s)!}\,.
\ee

It is instructive to write down explicitly the first few terms of the
generalized helicity basis~\eqref{CentralMulti}. For $j=1$, we have by
convention $\hat n_i = n_i$ and $\hat n^{\pm 1}_i =
n^\pm_i$. For two and three indices, we find respectively
\begin{align}
\begin{split}
\hat n_{ij} &= n_i n_j - \tfrac{1}{3}g_{ij},\\
\hat n^{\pm 1}_{ij} &= \frac{1}{2}(n^\pm_i n_j+n_i n^\pm_j),\\
\hat n^{\pm 2}_{ij} &=  n^\pm_i n^\pm_j,
\end{split}
\end{align}
and
\begin{align}
\hat n_{ijk}&=n_i n_j n_k -\frac{1}{5}(g_{ij}n_k + g_{jk}n_i + g_{ki} n_i),\nonumber\\
\hat n^{\pm 1}_{ijk}&=\frac{1}{3}(n^\pm_i n_j n_k+ n_i n^\pm_j n_k+n_i n_j n^\pm_k),\nonumber\\
&\quad-\frac{1}{15}(g_{ij}n^\pm_k + g_{jk}n^\pm_i + g_{ki} n^\pm_i),\nonumber\\
\hat n^{\pm 2}_{ijk}&=\frac{1}{3}(n^\pm_i n^\pm_j n_k+ n^\pm_i n_j n^\pm_k+ n_i n^\pm_j n^\pm_k),\nonumber\\
\hat n^{\pm 3}_{ijk}&=n^\pm_i n^\pm_j n^\pm_k.
\end{align}

\subsection{Products and contractions}

In the process of obtaining recursive relations among radial functions, we encounter a series of products and contractions of
generalized helicity basis elements which we now collect. The contractions
\bea
\hat n_{\pm s}^{I_{\ell-1} p} n_{p} &=& \hat n_{\pm s}^{I_{\ell-1}}\frac{(\ell^2-s^2)}{\ell(2\ell-1)}\label{C1}\\
\hat n^{I_{\ell-1} p} n^\pm_{p} &=& -\frac{\ell-1}{2\ell-1}\hat n_\pm^{I_{\ell-1}} 
\eea
generalize Eq. (A23) of \cite{BlanchetDamour1986}. 
For $s\geq0$ we also find
\be
\hat n_{\pm s}^{I_{\ell-1} p} n^{\pm}_{p} = -\frac{(\ell-s)(\ell-s-1)}{\ell(2\ell-1)}\hat n_{\pm(s+1)}^{I_{\ell-1}} \,.
\ee
For $s >0$ we obtain
\be\label{C4}
\hat n_{\pm s}^{I_{\ell-1} p} n^{\mp}_{p} = \frac{(\ell+s)(\ell+s-1)}{2\ell(2\ell-1)}\hat n_{\pm(s-1)}^{I_{\ell-1}} \,.
\ee
Repeated application of \eqref{C1}-\eqref{C4} allows to prove the orthogonality property \eqref{Defdjs1}.

Defining the Levi-Civita tensor on the spheres as
\be
\epsilon_{ij} = n^k \epsilon_{kij}\,,
\ee
we also have parity inverting relations
\be
\epsilon^p_{\,\,\langle i_\ell} \hat n^{\pm s}_{I_{\ell-1}\rangle p} = \pm \frac{\ii
  s}{\ell } \hat n^{\pm s}_{I_\ell}\,.
\ee

Let us now collect relations related to products of the generalized
helicity basis. Applying A3 of \cite{BlanchetDamour1986}, we get 
\bea\label{UsefulBlanchetPitrou}
\hat n_{\pm s}^{I_\ell} n^j &=& \hat n_{\pm s}^{\langle I_\ell j \rangle} +
\frac{(\ell-s)(\ell+s)}{\ell(2\ell+1)}\hat n_{\pm s}^{\langle
  I_{\ell-1}}g^{i_\ell \rangle j} \nonumber\\
&&\pm\frac{\ii s}{\ell+1} \epsilon_p^{\,\,\,j \langle i_\ell}
\hat n_{\pm s}^{I_{\ell-1} \rangle p}\,.
\eea
For $s \geq 0$ we also find
\begin{align}\label{ProductHelicities1}
\hat n_{\pm s}^{I_\ell} n_\pm^j &= \hat n_{\pm(s+1)}^{I_\ell j}\mp\frac{\ii (\ell-s)}{\ell+1} \epsilon_p^{\,\,\,j \langle i_\ell}
\hat n_{\pm(s+1)}^{I_{\ell-1} \rangle p}\nonumber\\
&-\frac{(\ell-s)(\ell-s-1)}{\ell(2\ell+1)}\hat n_{\pm(s+1)}^{\langle
  I_{\ell-1}}g^{i_\ell \rangle j} \,.
\end{align}
For $s>0$ it reads instead
\bea
\hat n_{\pm s}^{I_\ell} n_\mp^j &=& -\frac{1}{2}\hat n_{\pm(s-1)}^{I_\ell j}\mp\frac{\ii (\ell+s)}{2(\ell+1)} \epsilon_p^{\,\,\,j \langle i_\ell}
\hat n_{\pm(s-1)}^{I_{\ell-1} \rangle p}\nonumber\\
&+&\frac{(\ell+s)(\ell+s-1)}{2\ell(2\ell+1)}\hat n_{\pm(s-1)}^{\langle
  I_{\ell-1}}g^{i_\ell \rangle j} \,.
\eea
Note that the missing case $s=0$ for this relation is in fact given by \eqref{ProductHelicities1} evaluated in $s=0$.

Finally, note that the STF part of products of helicity vectors are
the expected relation (for $s\geq 0$)
\be\label{STFprod1}
\hat n_{\pm s}^{\langle I_\ell} n_\pm^{j\rangle } = \hat n_{\pm  (s+1)}^{I_\ell j}
\ee
but one should be careful that for  $s\geq1$ we also find
\be\label{STFprod2}
\hat n_{\pm s}^{\langle I_\ell} n_\mp^{j\rangle } = -\frac{1}{2}\hat n_{\pm
  (s-1)}^{I_\ell j}\,.
\ee

\subsection{Derivatives of helicity basis}\label{AppDerHelicity}

\subsubsection{Simple derivative}\label{AppSimpleDerHelicity}

In this section, we collect relations related to derivatives of the
generalized helicity basis. We work on the maximally symmetric curved
space with metric \eqref{DefMetric} whose associated covariant derivative is
$\nabla_i$. We first find
\bea
\nabla_p \hat n^{\pm s}_{I_j} &=& \frac{\pm \ii s}{r(\chi)}
\cot \theta e^\phi_p \hat n^{\pm s}_{I_j}\nonumber\\
&+& (j-s)\gcot \chi\left[g_{p\langle i_j} \hat n^{\pm
    s}_{I_{j-1}\rangle} -n_p \hat n^{\pm    s}_{I_j}\right]\nonumber\\
&-&s \gcot \chi \,n^\pm_p \hat n^{\pm (s-1)}_{I_j}\,.\label{Eqdins0}
\eea
The first line might seem peculiar at first sight, but it can be
absorbed when considering instead the derivative of products of
spin-weighted spherical harmonics multiplied by the associated
helicity basis, which is precisely what we always have in all expressions. Indeed, for $s>0$ we get
\bea\label{Eqdins2}
&&\nabla_p ({}_{\pm s}Y_{\ell}^{m}\hat n^{\pm s}_{I_j}) =\\
&&+ (j-s)\gcot \chi\left[g_{p\langle i_j} \hat n^{\pm
    s}_{I_{j-1}\rangle} -n_p \hat n^{\pm    s}_{I_j}\right]{}_{\pm s}Y_{\ell}^{m}\nonumber\\
&&-s \gcot \chi \,n^\pm_p \hat n^{\pm (s-1)}_{I_j} {}_{\pm s}Y_{\ell}^{m}\nonumber\\
&&\mp \frac{({}_+ \lambda_\ell^s)}{\sqrt{2}r(\chi)}  {}_{\pm
  (s+1)}Y_{\ell}^{m} n^\pm_p \hat n^{\pm s}_{I_j}\nonumber\\
&&\pm \frac{({}_- \lambda_\ell^s)}{\sqrt{2}r(\chi)}  {}_{\pm (s-1)}Y_{\ell}^{m} n^\mp_p \hat n^{\pm s}_{I_j}\,,\nonumber
\eea
and in the particular case $s=0$ we find simply
\bea\label{Eqdinszero}
&&\nabla_p (Y_{\ell}^{m}\hat n_{I_j}) =\\
&&+ j\gcot \chi\left[g_{p\langle i_j} \hat n_{I_{j-1}\rangle} -n_p \hat n_{I_j}\right]Y_{\ell}^{m}\nonumber\\
&&-\frac{1}{r(\chi)} \sqrt{\frac{\ell(\ell+1)}{2}} {}_{+}Y_{\ell}^{m} n^+_p \hat n_{I_j}\nonumber\\
&&+\frac{1}{r(\chi)} \sqrt{\frac{\ell(\ell+1)}{2}} {}_{-} Y_{\ell}^{m} n^-_p \hat n_{I_j}\,.\nonumber
\eea
From \eqref{STFprod1} and \eqref{STFprod2}, it is obvious to consider only the STF part of these relations.

\subsubsection{Divergence of helicity basis}

Furthermore, by contraction with $g^{p i_j}$ of the expressions in the
previous section, we obtain relations
associated with the divergence of an helicity basis. For $s>0$ it is
\bea\label{Divns}
&&\nabla^{p} ({}_{\pm s}Y_{\ell}^{m}\hat n^{\pm s}_{I_{j-1} p}) = \frac{(j^2-s^2)}{j(2j-1)}\\
&&\times \left\{(j+1) \gcot \chi\, {}_{\pm s}Y_{\ell}^{m}\hat n^{\pm s}_{I_{j-1}}\right.\nonumber\\
&&\pm \frac{({}_+ \lambda_\ell^s)}{\sqrt{2}r(\chi)}
\frac{(j-s-1)}{(j+s)} {}_{\pm  (s+1)}Y_{\ell}^{m}\hat n^{\pm (s+1)}_{I_{j-1}}\nonumber\\
&&\left.\pm \frac{({}_- \lambda_\ell^s)}{\sqrt{2}r(\chi)} \frac{(j+s-1)}{2(j-s)} {}_{\pm  (s-1)}Y_{\ell}^{m}\hat n^{\pm (s-1)}_{I_{j-1}}\right\}\nonumber
\eea
whereas in the $s=0$ case it is

\bea\label{Divnszero}
&&\nabla^{p} (Y_{\ell}^{m}\hat n_{I_{j-1}p}) = \frac{j(j+1)}{2j-1} \gcot \chi\,Y_{\ell}^{m}\hat n_{I_{j-1}}\nonumber\\
&&+ \frac{1}{r(\chi)}\sqrt{\frac{\ell(\ell+1)}{2}}
\frac{(j-1)}{(2j-1)} {}_{+1}Y_{\ell}^{m}\hat n^{+1}_{I_{j-1}}\nonumber\\
&&-\frac{1}{r(\chi)}\sqrt{\frac{\ell(\ell+1)}{2}} \frac{(j-1)}{(2j-1)} {}_{-1}Y_{\ell}^{m}\hat n^{-1}_{I_{j-1}}\,.\nonumber
\eea

\subsubsection{Curl of helicity basis}

The curl is also deduced by contraction with the Levi-Civita tensor of
the expressions in section \ref{AppSimpleDerHelicity}. If $s>0$ we get
\bea
&&{\rm curl}\left( {}_{\pm s} Y_{\ell}^{m}\hat n^{\pm s}_{I_j} \right)=\pm \ii \frac{s }{j} \gcot \chi\,
{}_{\pm s} Y_{\ell}^{m} \hat n^{\pm s}_{I_j}\nonumber\\
&&+\ii \frac{({}_+ \lambda_\ell^s)}{2 r(\chi)} \frac{(j-s)}{j} {}_{\pm (s+1)} Y_{\ell}^{m} \hat n^{\pm (s+1)}_{I_j}\nonumber\\
&&-\ii \frac{({}_- \lambda_\ell^s)}{2 r(\chi)} \frac{(j+s)}{2j} {}_{\pm (s-1)} Y_{\ell}^{m} \hat n^{\pm (s-1)}_{I_j}
\eea
and for $s=0$ we get simply
\bea
&&{\rm curl}\left(Y_{\ell}^{m}\hat n_{I_j} \right)=\\
&&\ii \frac{\sqrt{\ell(\ell+1)}}{\sqrt{2} r(\chi)} \left({}_{+1} Y_{\ell}^{m} \hat n^{+1}_{I_j}+ {}_{-1} Y_{\ell}^{m} \hat n^{-1}_{I_j}\right)\,.\nonumber
\eea

\subsubsection{Laplacian}

For a generic function $f(\chi)$, and using twice \eqref{Eqdins2} we
find for $s>0$
\bea\label{Deltans}
&&\Delta(f {}_{\pm s}Y_{\ell}^{m} \hat n_{\pm s}^{I_j})=\\
&&+\left[f''+2\gcot\chi\,f' \right]{}_{\pm s}Y_{\ell}^{m} \hat n_{\pm s}^{I_j} \nonumber\\
&&+\left[ f \gcot^2(\chi) (s^2 -  j(j+1))\right]{}_{\pm s}Y_{\ell}^{m} \hat n_{\pm s}^{I_j} \nonumber\\
&&+\left[\frac{f}{r^2(\chi)}(s^2-\ell(\ell+1))\right]{}_{\pm s}Y_{\ell}^{m} \hat n_{\pm s}^{I_j} \nonumber\\
&&\mp\frac{\gcot
  \chi}{r(\chi)}(j-s)\sqrt{2} ({}_+\lambda_\ell^s) {}_{\pm (s+1)}Y_{\ell}^{m} \hat n_{\pm (s+1)}^{I_j} \nonumber\\
&&\mp \frac{\gcot
  \chi}{r(\chi)}\frac{(j+s)}{\sqrt{2}}({}_- \lambda_\ell^s) {}_{\pm (s-1)}Y_{\ell}^{m} \hat n_{\pm (s-1)}^{I_j} \,.\nonumber
\eea
In the $s=0$ case we find 
\bea\label{Deltanszero}
&&\Delta(f Y_{\ell}^{m} \hat n^{I_j})=\\
&& \left[f''+2\gcot \chi\,f' - f \gcot^2(\chi) j(j+1)\right]Y_{\ell}^{m} \hat n^{I_j} \nonumber\\
&&- \frac{f}{r^2(\chi)}\ell(\ell+1) Y_{\ell}^{m} \hat n^{I_j} \nonumber\\
&&- \frac{\gcot
  \chi}{r(\chi)} j\sqrt{2\ell(\ell+1)}{}_{+1}Y_{\ell}^{m} \hat n_{+1}^{I_j} \nonumber\\
&&+ \frac{\gcot
  \chi}{r(\chi)} j\sqrt{2\ell(\ell+1)}{}_{-1}Y_{\ell}^{m} \hat n_{-1}^{I_j} \,.\nonumber
\eea

\section{Geographical recursion relations}\label{AppRecursions}
We collect recursive relations among radial functions in the $(j,s)$
plane. We illustrate in Fig.~\ref{FigRecursions1} the radial functions
which are related by these individual relations.

The NSW relation, whose derivation is summarized in section
\eqref{SecBuildDerivedHarmonics} is ( for $0\leq s \leq j$) 
\bea\label{Masteralpha2}
&&\frac{({}_+\lambda_\ell^s) ({}_+\lambda_j^s)}{(j+s+1) r(\chi)}  {}_{\pm s}\alpha^{(jm)}_\ell =
\nonumber\\
&&(j-s)\gcot \chi  \,\,{}_{\pm (s+1)}\alpha^{(jm)}_\ell \nonumber\\
&&+\frac{ {}_{(s+1)}\kappa_{j+1}^m }{(2j+1)}\,{}_{\pm
  (s+1)}\alpha^{(j+1,m)}_\ell\nonumber\\
&&+\frac{(j-s)\, {}_{ (s+1)}\kappa_{j}^m}{(2j+1)(j+1+s)} \, {}_{\pm (s+1)}\alpha^{(j-1,m)}_\ell\nonumber\\
&&\pm \ii \frac{(j-s) m \nu}{j(j+1)}   {}_{\pm (s+1)}\alpha^{(jm)}_\ell\,.
\eea
It is understood that when $j=s$, this is a relation linking ${}_{\pm
  s}\alpha^{(jm)}_\ell$ to ${}_{\pm
  (s+1)}\alpha^{(j+1,m)}_\ell$only. Indeed the coefficients in front
of these two terms contain $\sqrt{j-s}$, but the coefficients in front
of the other terms contain $(j-s)$ (and also multiply radial functions
that no longer satisfy $|s| \leq j$). Hence it must be understood that
we must divide first by $\sqrt{j-s}$ before evaluating in $j=s$ and we
get
\begin{align}\label{Magicjjjj}
&\frac{({}_+\lambda_\ell^j) }{\sqrt{2} r(\chi)}  {}_{\pm j}\alpha^{(jm)}_\ell
                = {}_{\pm (j+1)}\alpha^{(j+1,m)}_\ell \times\\
& \sqrt{\frac{(j+1)^2-m^2}{(j+1)(2j+1)}}\sqrt{\nu^2-\barK (j+1)^2} \,.\nonumber
 \end{align} 
A recursive application of this relation allows to deduce ${}_{s=j}
\alpha^{(j,0)}_\ell$ from ${}_0 \epsilon^{(00)}_\ell = \Phi^\nu_\ell$. Using \eqref{smChangejl}, we then recover \eqref{solutionjmm}.

As for the NSE relation, it is ($0< s \leq j$)
\bea\label{Masteralpha3}
&&\frac{({}_-\lambda_\ell^s) ({}_-\lambda_j^s)}{(j-s+1) r(\chi)} {}_{\pm s}\alpha^{(jm)}_\ell =
\nonumber\\
&&(j+s)\gcot \chi  \,\,{}_{\pm (s-1)}\alpha^{(jm)}_\ell \nonumber\\
&&+\frac{ {}_{ (s-1)}\kappa_{j+1}^m }{(2j+1)}\,{}_{\pm
  (s-1)}\alpha^{(j+1,m)}_\ell\nonumber\\
&&+\frac{(j+s)\, {}_{ (s-1)}\kappa_{j}^m}{(2j+1)(j+1-s)} \,{}_{\pm (s-1)}\alpha^{(j-1,m)}_\ell\nonumber\\
&&\mp \ii \frac{(j+s) m \nu}{j(j+1)}   {}_{\pm (s-1)}\alpha^{(jm)}_\ell\,.
\eea

Combining relations NS \eqref{Masteralpha1} with the NSE or NSW leads to the a set of four triangular relations (see interpretation of this denomination on Fig. \ref{FigRecursions1})
\begin{itemize}
\item NW relation  (for $0<s \leq j$) :
\bea\label{AngleRelation1}
&&\frac{({}_-\lambda_\ell^s) ({}_-\lambda_j^s)}{(j+s) r(\chi)}   {}_{\pm (s-1)}\alpha^{(jm)}_\ell =\frac{\dd}{\dd \chi} {}_{\pm s}\alpha^{(jm)}_\ell  \\
&&+(j+1-s)\gcot \chi  \,\,{}_{\pm s}\alpha^{(jm)}_\ell \nonumber\\
&&+\frac{{}_{s}\kappa_{j}^m}{(j+s)} \, {}_{\pm s}\alpha^{(j-1,m)}_\ell\pm \ii \frac{m \nu}{j}   {}_{\pm s}\alpha^{(jm)}_\ell\,.\nonumber
\eea

\item NE relation (for $0 \leq s < j$) :
\bea\label{AngleRelation2}
&&\frac{({}_+\lambda_\ell^s) ({}_+\lambda_j^s)}{(j-s) r(\chi)} {}_{\pm (s+1)}\alpha^{(jm)}_\ell =\frac{\dd}{\dd \chi} {}_{\pm s}\alpha^{(jm)}_\ell  \nonumber\\
&&+(j+1+s)\gcot \chi  \,\,{}_{\pm s}\alpha^{(jm)}_\ell \\
&&+\frac{{}_{s}\kappa_{j}^m}{(j-s)} \, {}_{\pm s}\alpha^{(j-1,m)}_\ell\mp \ii \frac{m \nu}{j}   {}_{\pm s}\alpha^{(jm)}_\ell\,.\nonumber
\eea

\item SW relation (for $0< s \leq j+1$) :
\bea\label{AngleRelation3}
&&\frac{({}_-\lambda_\ell^s) ({}_-\lambda_j^s)}{(j+1-s) r(\chi)}   {}_{\pm (s-1)}\alpha^{(jm)}_\ell =-\frac{\dd}{\dd \chi} {}_{\pm s}\alpha^{(jm)}_\ell  \nonumber\\
&&+(j+s)\gcot \chi  \,\,{}_{\pm s}\alpha^{(jm)}_\ell \\
&&+\frac{{}_{s}\kappa_{j+1}^m}{(j+1-s)} \, {}_{\pm s}\alpha^{(j+1,m)}_\ell\pm \ii \frac{m \nu}{j+1}   {}_{\pm s}\alpha^{(jm)}_\ell\,.\nonumber
\eea
In the case $s=j+1$, it reduces to a relation between$ {}_{\pm
  (s-1)}\alpha^{(jm)}_\ell $ and $ {}_{\pm s}\alpha^{(j+1,m)}_\ell$.
However, both terms contain a factor $1/\sqrt{j+1-s}$, so it must be understood that the expression is to be
multiplied by $\sqrt{j+1-s}$ before being evaluated in $s=j+1$, and we
recover \eqref{Magicjjjj}.

\item SE relation is (for $0\leq s < j$) :
\begin{align}\label{AngleRelation4}
&\frac{({}_+\lambda_\ell^s) ({}_+\lambda_j^s)}{(j+1+s) r(\chi)}   {}_{\pm (s+1)}\alpha^{(jm)}_\ell =-\frac{\dd}{\dd \chi} {}_{\pm s}\alpha^{(jm)}_\ell  \nonumber\\
&+(j-s)\gcot \chi  \,\,{}_{\pm s}\alpha^{(jm)}_\ell \\
&+\frac{{}_{s}\kappa_{j+1}^m}{(j+1+s)} \, {}_{\pm s}\alpha^{(j+1,m)}_\ell\mp \ii \frac{m \nu}{j+1}   {}_{\pm s}\alpha^{(jm)}_\ell\,.\nonumber
\end{align}
\end{itemize}

Let us also stress that by taking the difference of the NW and NE
relations, we obtain a useful NW-NE relation which involves no
derivative, which is (for $0<s< j$)
\begin{align}\label{NWENoDer}
&\frac{({}_+ \lambda_\ell^s ) ({}_+ \lambda_j^s )}{2r(\chi)(j-s)}{}_{\pm
  (s+1)}\alpha_\ell^{(jm)} =\\
& \frac{({}_- \lambda_\ell^s ) ({}_- \lambda_j^s )}{2r(\chi)(j+s)}{}_{\pm
  (s-1)}\alpha_\ell^{(jm)}\mp \ii \frac{m \nu}{j}  {}_{\pm
  s}\alpha_\ell^{(jm)}\nonumber\\
&+s \gcot \chi\,{}_{\pm s}\alpha_\ell^{(jm)} +\frac{s}{(j^2-s^2)}\, {}_s \kappa_j^m\, {}_{\pm
  s}\alpha_\ell^{(j-1,m)}\,.\nonumber
\end{align}
Similarly, one could combine the SW and SE relations to get a SW-SE
relation without derivatives.

\section{Divergence, curl and STF recursions}\label{AppRecursions2}

Following the method of section \ref{SecBuildDerivedHarmonics}, we can
obtain recursive relations among radial functions in the $(j,s)$
space, from the divergence relation \eqref{GeneralDiv}, the curl property \eqref{Curljm}, and the STF construction of
derived modes \eqref{STFrecursion}.

The divergence relation leads for $0<s\leq j$ to 
\bea\label{DivergenceRelation1}
&&\frac{\dd}{\dd \chi}{}_{\pm s}\alpha_\ell^{(jm)} + (j+1)\gcot \chi\,
{}_{\pm s}\alpha_\ell^{(jm)} \\
&&+\frac{j }{j^2-s^2} \, {}_s \kappa_j^m \,{}_{\pm
  s}\alpha_\ell^{(j-1,m)}\nonumber\\
&&=\frac{({}_+ \lambda_\ell^s ) ({}_+ \lambda_j^s )}{2r(\chi) (j-s)}{}_{\pm
  (s+1)}\alpha_\ell^{(jm)}\nonumber\\
&&+\frac{({}_- \lambda_\ell^s ) ({}_- \lambda_j^s )}{2r(\chi) (j+s)}{}_{\pm (s-1)}\alpha_\ell^{(jm)}\,.\nonumber
\eea
In the $s=0$ case it reduces to
\bea\label{DivergenceRelation2}
&&\frac{\dd}{\dd \chi}{}_0\epsilon_\ell^{(jm)} + (j+1)\gcot \chi
\,{}_0 \epsilon_\ell^{(jm)} +\frac{ {}_0 \kappa_j^m }{j}\,{}_{0}\epsilon_\ell^{(j-1,m)}\nonumber\\
&&=\frac{\sqrt{\ell(\ell+1)(j+1)}}{\sqrt{j} r(\chi)
}{}_{1}\epsilon_\ell^{(jm)}\,.
\eea
We can check that this latter case corresponds to the real part of the
NE relation \eqref{AngleRelation2}.

The curl relation among radial functions is for $0<s\leq j$

\begin{align}\label{Curlrelation}
&s\left[\frac{\dd}{\dd \chi}{}_{\pm s}\alpha_\ell^{(jm)} + \gcot \chi\,
{}_{\pm s}\alpha_\ell^{(jm)}\right] \pm \ii m \nu  {}_{\pm
  s}\alpha_\ell^{(jm)}\nonumber\\
&=-\frac{({}_+ \lambda_\ell^s ) ({}_+ \lambda_j^s )}{2r(\chi)}{}_{\pm
  (s+1)}\alpha_\ell^{(jm)}\nonumber\\
&\,\,+\frac{({}_- \lambda_\ell^s ) ({}_- \lambda_j^s )}{2r(\chi)}{}_{\pm (s-1)}\alpha_\ell^{(jm)}\,.
\end{align}
In the $s=0$ case it reduces to
\bea\label{Curlszerojm}
{}_1 \beta_\ell^{(jm)} \frac{\sqrt{\ell (\ell+1)j(j+1)}}{r(\chi)} = -m\nu {}_0 \epsilon_\ell^{(jm)}\,.
\eea
This latter relation is exactly the imaginary part of the NE relation \eqref{AngleRelation2} or the SE relation \eqref{AngleRelation4}.
Note also that combining the curl relation \eqref{Curlrelation} and the div
relation \eqref{DivergenceRelation1} allows to remove the derivative
of the radial function and leads also the NW-NE relation without
derivative \eqref{NWENoDer}.

Finally the STF construction of derived harmonics brings the relation
(for $0<s\leq j$)
\bea
&&\frac{\dd}{\dd \chi}{}_{\pm s}\alpha_\ell^{(jm)} - j \gcot \chi\,
{}_{\pm s}\alpha_\ell^{(jm)}\\
&&-\frac{(j+1)}{(j+1)^2-s^2} \,{}_s \kappa_{j+1}^m \,{}_{\pm s}\alpha_\ell^{(j+1,m)} \nonumber\\
&&=-\frac{({}_+ \lambda_\ell^s ) ({}_+ \lambda_j^s )}{2r(\chi) (j+1+s)}{}_{\pm
  (s+1)}\alpha_\ell^{(jm)}\nonumber\\
&&-\frac{({}_- \lambda_\ell^s ) ({}_- \lambda_j^s )}{2r(\chi) (j+1-s)}{}_{\pm (s-1)}\alpha_\ell^{(jm)}\,.\nonumber
\eea
The case $j=s+1$ can also be considered with method similar to those
detailed after \eqref{AngleRelation3}, and it also reduced to \eqref{Magicjjjj}.

In the $s=0$ case it is
\bea
&&\frac{\dd}{\dd \chi}{}_0\epsilon_\ell^{(jm)} - j \gcot \chi\,
{}_0\epsilon_\ell^{(jm)}- \frac{{}_0 \kappa_{j+1}^m}{j+1} \,{}_{0}\epsilon_\ell^{(j+1,m)} \nonumber\\
&&=-\frac{1}{r(\chi)} \sqrt{\frac{\ell(\ell+1)j}{(j+1)}} {}_1 \epsilon_\ell^{(jm)}\,.
\eea
This latter relation is nothing but the real part of the SE relation \eqref{AngleRelation4}.

We then check that combinations of all the relations of this section can be used to form
the triangular relations (NW, NE, SW and SE relations). Hence this is an
alternative derivation for all recursions among radial functions in
the $(j,s)$ space. However, the fact that we need to separate
explicitly the real and imaginary parts of the triangular relations in the
cases $s=0$ makes this derivation less direct and we prefer the method
based on the various projections of \eqref{GenRelHarm}.

\section{Radial functions in flat space}\label{AppFlat}

In the flat case, there is a complete separability between the angular
and the spatial dependencies. The spatial dependence is the same for all
modes, and thus the same as for scalar harmonics, that is, it is a pure
Fourier mode. We choose to align the wavevector $\gr{k}$ of the Fourier
mode with the zenith direction $\gr{e}_z$. From this separability,
the plane-wave normal modes are all of the form \cite{TAM1,Dai:2012bc}
\be\label{FactorFlat}
{}_s G^{(jm)}(r,\gr{n}) = \frac{c_j}{2j+1} {}_s Y_{j}^m(\gr{n})
{\rm e}^{\ii k \gr{e}_z \cdot \gr{r}}\,,
\ee
where $\gr{r} = r \gr{n}$, and $r$ is now the radial coordinate, 
corresponding to plane waves harmonics
\be\label{FactorFlat2}
Q_{I_j}^{(jm)}(r,\gr{n})  = {}_0 \tilde g^{(jm)} \,\frac{c_j}{2j+1}\,{\cal Y}^{jm}_{I_j} \,{\rm e}^{\ii k \gr{e}_z \cdot \gr{r}}\,.
\ee
We do not use $\chi$ which was in units of curvature, since now the curvature
length $\ellc$ is infinite. Note that the Fourier mode magnitude $k$
is also no more in units of inverse curvature length. \cpsilent{In practice, the trigonometric functions $\gsin(\chi)$, $\gtan(\chi)$ and $\gcot(\chi)$ need also to be replaced respectively by $r$, $r$ and $1/r$ in all expressions.} Using the Rayleigh expansion 
\be\label{Rayleigh}
{\rm e}^{\ii k \gr{e}_z \cdot \gr{n}} =\sum_\ell \sqrt{4\pi(2\ell+1)}
\ii^\ell j_\ell (k  r ) Y_{\ell}^{0}(\gr{n})
\ee
the decomposition of the plane-wave normal modes is then given by
\be
{}_s G^{(jm)}(r,\gr{n}) = \sum_\ell c_\ell \,{}_s \alpha_\ell^{(jm)}(k r) {}_s Y_{\ell}^{m}(\gr{n})
\ee
with the radial functions built as
\bea\label{DefGaunt}
{}_s \alpha_{\ell}^{(jm)} (x)&\equiv& \sum_{L} {}^s C^{m 0 m}_{\ell L j}
j_L(x)  \ii^{L+j-\ell} \nonumber\\
&&\times\sqrt{\frac{(4 \pi)(2 L +1)}{(2 \ell + 1)(2
    j +1)}}\,,
\eea
with the coefficients ${}^s C^{m 0 m}_{\ell L j}$ defined in~\eqref{C_Gaunt}.

The $j_\ell$ are the usual spherical Bessel functions satisfying the relations
\bea\label{Recjl}
j_\ell'(x) &=& \frac{1}{2\ell+1}\left[\ell
  j_{\ell-1}(x)-(\ell+1)j_{\ell+1}(x)\right]\nonumber\\
\frac{j_\ell(x)}{x} & = & \frac{1}{2\ell+1}\left[j_{\ell-1}(x)+j_{\ell+1}(x)\right]
\eea
with $j_0(x) \equiv \sin(x)/x$. They also satisfy the differential equation
\be\label{EquadiffBesselFlat}
\frac{1}{x^2 } \frac{\dd}{\dd x} \left(x^2 \frac{\dd}{\dd
  x} j_\ell \right)+ \left[1 -\frac{\ell(\ell+1)}{x^2}\right]j_\ell=0\,,
\ee
and the normalization condition
\be\label{Normjl}
\int j_\ell(a x) j_\ell (b x )x^2 \dd x=\frac{\pi}{2}\frac{\delta(a-b)}{a^2}\,.
\ee
The constants in \eqref{DefGaunt} are the so-called Gaunt coefficients, and are defined as
\bea
\label{C_Gaunt}
&&{}^s C^{m_1 m_2 m_3}_{\ell_1  \ell_2 \ell_3} \equiv \int \dd^2 \Omega
\,({}_s Y^{m_1\,\star}_{\ell_1})\,(
Y_{\ell_2}^{m_2})\,( {}_s Y_{\ell_3}^{m_3})\nonumber\\
&&= (-1)^{m_1+s}\sqrt{\frac{(2\ell_{1}+1)(2\ell_{2}+1)(2\ell_{3}+1)}{4\pi}}\nonumber \\
&& \times\left(\begin{array}{ccc}
\ell_{1} & \ell_{2} & \ell_{3}\\
s & 0 & -s
\end{array}\right)\left(\begin{array}{ccc}
\ell_{1} & \ell_{2} & \ell_{3}\\
-m_{1} & m_{2} & m_{3}
\end{array}\right)\,,
\eea
where the $3\times2$ matrices on the third line are the well-known
Wigner 3-j symbols. From the symmetry properties of the 3-j symbols, we deduce that
\be
 {}^s C^{m 0 m}_{\ell L j} = {}^m C^{s 0 s}_{\ell L j}\,,\quad {}^s C^{m 0 m}_{\ell L j} ={}^s C^{m 0 m}_{j L \ell} 
\ee
which with \eqref{DefGaunt} proves rigorously the properties
\eqref{smChangejl} and \eqref{jlChangejl} in the flat case. 

Let us now collect the explicit forms of the radial functions in flat
space. We recover results of \cite{TAM1,TAM2} for $s=0$ or $s=2$ up to the global sign inversion for odd modes since we collect results when
defining harmonics with respect to the observed direction (see
section~\ref{AppPropagation} for propagation direction harmonics). We also use the compact notation $x\equiv
kr$ and we recall that in the flat case $\mygamma_n =1$ for all $n$ since $\nu=k$, so the constants ${}_s g^{(jm)}$ are directly read from
those reported in section~\ref{SecGatherRadial}. The first radial functions are
\beas
{}_0 \epsilon^{(00)}_\ell &=& j_\ell(x),\\
{}_0 \epsilon^{(10)}_\ell &=& j_\ell'(x),\\
{}_1 \epsilon^{(10)}_\ell &=& \sqrt{\frac{\ell(\ell+1)}{2}}\frac{j_\ell(x)}{x},\\
{}_0 \epsilon^{(20)}_\ell &=& \frac{1}{2}\left[3 j_\ell''(x)+j_\ell(x)\right],\\
{}_1 \epsilon^{(20)}_\ell &=& \sqrt{\frac{3\ell(\ell+1)}{2}}\frac{\dd}{\dd x}\left(\frac{j_\ell(x)}{x}\right),\\
{}_2 \epsilon^{(20)}_\ell &=& \sqrt{\frac{3(\ell+2)!}{8(\ell-2)!}}\frac{j_\ell(x)}{x^2}\,,
\eeas

\beas
{}_0 \epsilon^{(11)}_\ell &=& \sqrt{\frac{\ell(\ell+1)}{2}}\frac{j_\ell(x)}{x},\\
{}_1 \epsilon^{(11)}_\ell &=& \frac{1}{2}\frac{\dd (x j_\ell(x))}{x \dd x},\\
{}_1 \beta^{(11)}_\ell &=&-\frac{1}{2} j_\ell(x)\,,
\eeas
\beas
{}_0 \epsilon^{(21)}_\ell &=& \sqrt{\frac{3\ell(\ell+1)}{2}}
\frac{\dd }{\dd x}\left(\frac{j_\ell(x)}{x}\right),\\
{}_1 \epsilon^{(21)}_\ell &=&j_\ell''(x) + \frac{j_\ell'(x)}{x}-\frac{j_\ell(x)}{x^2}+\frac{j_\ell(x)}{2},\nonumber\\
{}_1 \beta^{(21)}_\ell &=&-\frac{1}{2} \, x \frac{\dd}{\dd x} \left(\frac{j_\ell(x)}{x}\right), \\
{}_2 \epsilon^{(21)}_\ell &=& \frac{\sqrt{(\ell+2)(\ell-1)}}{2}\frac{1}{x^2}\frac{\dd }{\dd x}[x j_\ell(x)],\nonumber\\
{}_2 \beta^{(21)}_\ell &=&- \frac{\sqrt{(\ell+2)(\ell-1)}}{2}\frac{j_\ell(x)}{x},
\eeas

\beas
{}_0 \epsilon^{(22)}_\ell &=& \sqrt{\frac{3(\ell+2)!}{8(\ell-2)!}}\frac{j_\ell(x)}{x^2},\\
{}_1 \epsilon^{(22)}_\ell &=&\frac{\sqrt{(\ell+2)(\ell-1)}}{2}\frac{1}{x^2}\frac{\dd }{\dd x}[x j_\ell(x)],\nonumber\\
{}_1 \beta^{(22)}_\ell &=&-\frac{\sqrt{(\ell+2)(\ell-1)}}{2}\frac{j_\ell(x)}{x}, \\
{}_2 \epsilon^{(22)}_\ell &=& \frac{1}{4} \left[j_\ell''(x)-j_\ell(x)+4\frac{j_\ell'(x)}{x}+2\frac{j_\ell(x)}{x^2}\right],\nonumber\\
{}_2 \beta^{(22)}_\ell &=&-\frac{1}{2 x^2}\frac{\dd}{\dd
  x}[x^2 j_\ell(x)].
\eeas

\section{Comparison with literature}\label{SecComparison}

The harmonics built in this paper can be related to the scalar, vector
and tensor harmonics derived in \cite{Tomita1982} and
\cite{Lindblom:2017maa} in the closed case, and expressed in the usual
orthonormal spherical basis \eqref{SphericalBasis1} rather than with the helicity basis. In these references, the
harmonics and derived harmonics are separated into their electric
(even parity) and odd parity by considering the contributions
\be
{}^\ell Q^{(j,|m|)}_{I_j} \pm  {}^\ell Q^{(j,-|m|)}_{I_j} \,.
\ee
From the property \eqref{Inversionm}, we see that the plus sign selects only the
contribution of the electric (even parity) radial modes, whereas the
negative sign selects the magnetic (odd parity) radial modes. To be
specific the three vector harmonics defined in Eqs. (12-14) of
\cite{Lindblom:2017maa} are proportional to respectively the $m=0$ harmonics
(necessarily of even type), the
$m=1$ magnetic harmonics, and the $m=1$ electric harmonics, where the
notation used is $k \equiv \nu-1$, such that it takes positive integer
values. Similarly the tensor harmonics of Eqs. (26-30) are
successively proportional to the $m=1$ magnetic harmonics, the $m=1$ electric harmonics, the $m=0$
harmonics (necessarily of even type), the $m=2$ magnetic harmonics and
the $m=2$ electric harmonics.

The spectrum of eigenvalues of the Laplacian can also be compared for
scalar and vector harmonics with the exterior calculus approach of
\cite{Achour:2015zpa} in the closed case, and we now detail our agreement. We still work
in units such that $\ellc=1$. The Laplace-de Rahm operator is defined as $\tilde \Delta \equiv
-(\dd\delta + \delta \dd)$. For scalar functions it matches exactly the
Laplace-Beltrami operator \eqref{EqDelta}. In the closed case the set of
eigenvalues for scalar harmonics ($j=m=0$) is the set of $k^2 = \nu^2
-1 = L (L+2)$ where $L \geq 0$ and $\nu \geq 1$ are integers. For the derived vector valued harmonics ($j=1$ and $m=0$) which correspond to exact forms, we find that the
spectrum is the same since
\be
\tilde \Delta \nabla_i Q^{(00)} = \nabla_i \Delta Q^{(00)} = - k^2 \nabla_i Q^{(00)} \,.
\ee 
However for vector harmonics ($j=m=1$), which correspond to co-exact
forms since they are divergenceless, we find
\be
\tilde \Delta Q^{(11)}_i = (\Delta -2 \barK) Q^{(11)}_i = -
(k^2 +2 \barK) Q^{(11)}_i\,. \nonumber
\ee
The spectrum of $\tilde \Delta$ in that case is the set of $k^2+2 =
\nu^2$ with the integer values $\nu\geq2$, in agreement with \cite{Achour:2015zpa}.


\section{Tables of symbols}\label{SecSymbols}

We gather in the Tables below the most commonly used symbols of this work.

\begin{table}[!htb]
\begin{tabular}{|l|r|}
\hline
Variable & Definition \\
\hline
 $r(\chi)$ &\eqref{defrchi} \\
$\barK$ &\eqref{DefBarK}\\
 $\nu $&\eqref{Defnu}\\
   $b_{js}$ &\eqref{Defdjs-bjs} \\
  $d_{js}$ & \eqref{Defdjs-bjs} \\
  $k_{js}$ & \eqref{Defkjs} \\
${}_s g^{(jm)}$ & \eqref{Qljm}\\
${}_s \tilde g^{(jm)}$ & \eqref{gtildetog}\\
$c_\ell, \bar c_\ell$ &\eqref{Defcl}, \eqref{Defbarc}\\
$\mygamma_n$& \eqref{DefGamma}\\
   $q^{(jm)}$ & \eqref{Defqjm}\\
   ${\cal N}_{j}^m$ & \eqref{DefNjm}\\
   $\gsin,\gtan,\gcot$ & \S~\ref{Sec21} \\
  ${}_s \kappa_\ell^m$&\eqref{Defkappaslm}\\
${}_\pm \lambda_\ell^s$ & \eqref{Deflambda}\\
$\myzeta_\ell^m$&\eqref{DefQfromsuml}\\
$\Delta_\ell$&\eqref{DefDeltaell}\\
\hline
 \end{tabular}
\caption{Main symbols used in the construction of harmonics. \label{TableDefinitions}}
\end{table}

\begin{table}[!htb]
\begin{tabular}{|l|r|}
\hline
Function & Definition \\
\hline
$\hat n^s_{I_j}$ & \eqref{CentralMulti}\\
${}_s \alpha_\ell^{(jm)}$&\eqref{DefGsjm}\\
${}_s \epsilon_\ell^{(jm)}$, ${}_s \beta_\ell^{(jm)}$&\eqref{defeb}\\
${}_s G_\ell^{(jm)}$&\eqref{DefGsjm}\\
${}_s G^{(jm)}$&\eqref{GjmfromGljm}\\
${}^\ell Q^{(jm)}_{I_j} $&\S~\ref{Sec23}\\
$Q^{(jm)}_{I_j} $&\eqref{DefQfromsuml}\\
${}_s G^{(jm)}_{\ell M} $,  $ {}_s{G}^{(jm)}_\ell(\gr{\nu})$ & \S~\ref{SecRotation}\\
$ {}^{\ell M} Q^{(jm)}_{I_j} $, ${}^\ell Q^{(jm)}_{I_j} (\gr{\nu})$& \S~\ref{SecRotation}\\
$ {}_s{G}^{(jm)}(\gr{\nu})$,&\S~\ref{GenAxisPW}\\
$Q^{(jm)}_{I_j} (\gr{\nu})$&\S~\ref{GenAxisPW}\\
${\cal Y}_{\ell m}^{I_\ell}$&\eqref{DefCurlyY}\\
\hline
 \end{tabular}
\caption{Main functions and tensors used in building harmonics. The barred version of these
    functions are related to the propagating direction, and are
    defined in \S~\ref{AppPropagation}. \label{TableDefinitions2}}
\end{table}

\end{document}